\documentclass[12pt,a4paper]{article}
\usepackage[T1]{fontenc}
\usepackage[utf8]{inputenc}
\usepackage{authblk}
\usepackage{bbding} 
\usepackage{soul}

\usepackage{amsmath}
\usepackage{graphicx,float,color}
\usepackage{hyperref}
\hypersetup{colorlinks = true, linkcolor = blue, urlcolor  = blue, citecolor = blue, anchorcolor = blue}
\usepackage{multirow}
\usepackage[numbers,sort&compress]{natbib}
\usepackage{braket}
\usepackage{amsfonts}
\usepackage{amssymb} 
\usepackage{epic}
\usepackage{eepic}
\usepackage{epsfig}
\usepackage{latexsym}
\usepackage{url}  
\usepackage{caption}
\usepackage[caption=false]{subfig}
\usepackage{bm}
\usepackage{geometry}
\usepackage{slashed}
\usepackage[detect-all]{siunitx}
\usepackage{lipsum}
\makeatletter
\def\ps@pprintTitle{%
 \let\@oddhead\@empty
 \let\@evenhead\@empty
 \def\@oddfoot{}%
 \let\@evenfoot\@oddfoot}
\makeatother
\usepackage{etoolbox}

\begin{document}
\title{\vspace{-1cm} \textbf{Diffeomorphism Invariance Breaking in Gravity and Cosmological Evolution \vspace{0.5cm}
}}    
\author[]{Ufuk Aydemir\thanks{uaydemir@metu.edu.tr} }
\affil[]{\vspace{0.1cm} \small Department of Physics, Middle East Technical University, Ankara 06800, T\"urkiye}

\author[]{Mahmut Elbistan\thanks{mahmut.elbistan@bilgi.edu.tr } }
\affil[]{\vspace{0.1cm}\small Department of Energy Systems Engineering, Istanbul Bilgi University, Istanbul 34060, T\"urkiye}
\date{}
\maketitle

 
\vspace{-0.2cm} 
\begin{abstract}
Diffeomorphism invariance breaking has been investigated in the literature in several contexts, including emergent General Relativity (GR). If GR emerges from an underlying theory without diffeomorphism invariance, there may be small violations of this symmetry at low energies. Since such small violations should not cause instabilities in cosmological evolution, it is a suitable framework for examining such symmetry-breaking effects. In this paper, the cosmological evolution with broken diffeomorphism invariance is investigated in the (modified) 
 FLRW spacetime in the effective theory framework. The GR Lagrangian is augmented with all diffeomorphism-breaking but Lorentz-invariant terms in the leading order, namely, those involving two derivatives. The magnitudes of (minor) violations are kept general modulo the conditions arising in the linearized theory. The analytic solutions of the scale factor in the full non-linear theory for the single-component universes are attempted; the radiation and vacuum solutions are found analytically, whereas the matter solution is worked out numerically since an analytic solution does not exist in the required form. It is observed that the solutions smoothly connect to those of GR  in the limit of vanishing symmetry-breaking. The more realistic, two-component, and three-component universes are numerically studied, and no signs of singular behavior are observed: minor diffeomorphism-violating modifications to GR at the level of two derivatives do not cause instabilities in the basic cosmological evolution.    
\end{abstract}

\newpage
{
  \hypersetup{linkcolor=black}
}

\section{Introduction\label{sec:intro}}
Gauge symmetries play an essential role in establishing new theories. The common practice is to invoke larger gauge symmetries at higher energies beyond those respected at low energies. However, it is also common in nature, particularly in condensed matter physics~\cite{Volovik:2003fe,Volovik:2008dd}, that the symmetries with new degrees of freedom emerge as lower-energy manifestations of the underlying theory, and that the underlying theory does not respect these symmetries of the low-energy world. The symmetries may gradually emerge in the infrared from a non-symmetric, more fundamental theory~\cite{Nielsen:1978is,Forster:1980dg,Wetterich:2016qee,Chadha:1982qq}.
  
The emergent gauge symmetry approach has been a direction to seek answers regarding
 the fundamental principles of nature~\cite{Forster:1980dg,Nielsen1983,Bjorken:2001pe,Donoghue:2010qf,Barcelo:2016xhp,Wetterich:2016qee,Witten:2017hdv,Barcelo:2021idt,Bass:2021wxv,Bass:2023ece,Carlip:2012wa,Betzios:2020sro,Barcelo:2021ryh}, alternative to the idea of unification. 
In the emergent gravity models with background independence~\cite{Gaul:1999ys,Ambjorn:2006hu,Ambjorn:2006jf,Lee:2006gp,Konopka:2008hp}, the diffeomorphism invariance is always respected. Still, there exist cases, some of which are motivated by condensed matter analogies, in which the underlying theory does not have this feature~\cite{Gu:2009jh,Horava:2009uw,Weinfurtner:2007dnb,Liberati:2009uq,Volovik:1999cn,Zaanen:2011hm,Barcelo:2021ryh,Betzios:2020sro,Linnemann:2017hdo,Sindoni:2011ej}. (See Refs.~\cite{Volovik:2000ua,Klinkhamer:2005tz,Barcelo:2005fc,Sindoni:2009fc,Sindoni:2011nh,Sindoni:2011ej} as some of the other works regarding the analog gravity approach.)~Although there are challenges such as the Weinberg-Witten theorem~\cite{Weinberg:1980kq,Loebbert:2008zz,Jenkins:2009un} and Marolf's theorem~\cite{Marolf:2014yga}, there are ways to circumvent them~\cite{Carlip:2012wa,Betzios:2020sro,Barcelo:2021ryh,Barcelo:2021idt}.  The diffeomorphism breaking in gravity has been studied in many other contexts including the unimodular gravity~\cite{vanderBij:1981ym,Unruh:1988in,Buchmuller:1988wx,Alvarez:2006uu,Alvarez:2007nn,Lopez-Villarejo:2010uib,Bonder:2025njt}, massive gravity~\cite{Hinterbichler:2011tt,deRham:2014zqa}, theories with local Lorentz symmetry violation such as; the Standard Model Extension (SME)~\cite{Kostelecky:2003fs,Kostelecky:2011qz,Bluhm:2019ato}, Bumblebee models~\cite{Bluhm:2004ep,Bertolami:2005bh,Bluhm:2007bd}, Ho\v{r}ava gravity~\cite{Horava:2009uw,Charmousis:2009tc,Sotiriou:2010wn,Barvinsky:2019rwn,Chaichian:2015asa}, and Einstein-aether gravity~\cite{Jacobson:2007veq}. There have also been some discussions about Lorentz symmetry violation in string theory~\cite{Kostelecky:1988zi,Kostelecky:1989jp,Kostelecky:1989jw}.
 
 In the emergent symmetry concept, where the symmetry may gradually emerge in the infrared, a perfectly symmetric low-energy theory is not ensured. Therefore, if diffeomorphism-invariant gravity is an emergent phenomenon, one may expect small violations of this symmetry in the low-energy effective theory, as reminiscences of the underlying theory.  In this context, a phenomenological study was conducted in Ref.~\cite{Anber:2009qp}, where the coefficient of a sample diffeomorphism breaking operator (referred to as $\mathcal{L}_3$, given in our Eq.~(\ref{diffLs})) was constrained through the Parametrized Post Newtonian (PPN) formalism, and found it to be extremely suppressed, bounded to be smaller than $\sim10^{-20}$. Other than constraining the parameters through the post-Newtonian effects, there are other important points to be examined. We know from experience that breaking a gauge symmetry can lead to a variety of problems. For instance, in Pauli-Fierz massive gravity, there is an issue known as the vDVZ discontinuity~\cite{Zakharov:1970cc,vanDam:1970vg}, where the broken theory does not reduce to GR in the limit where the symmetry-breaking parameter vanishes.~\cite{Vainshtein:1972sx,Porrati:2000cp,Kogan:2000uy,Grisa:2009yy,Hinterbichler:2011tt,deRham:2014zqa})\footnote{Note that we give an example here. The model we work with does not have a vDVZ discontinuity~\cite{Anber:2009qp}.}. Furthermore, it is known that breaking diffeomorphism invariance may lead to ghost-like degrees of freedom in the linearized theory, depending on certain conditions~\cite{Anber:2009qp}.  Another commonly discussed issue regards explicit breaking of the symmetry; as we will note below, this requires certain constraint equations to be satisfied.

In this paper, we adopt a cosmological approach;
we consider the \textit{entire} Lagrangian with diffeomorphism-violating Lorentz invariant terms at the leading order and investigate the cosmological evolution.\footnote{ This subject  was discussed for the first time in Ref.~\cite{Aydemir:2012gda} in a limited context. Recently, in Ref.~\cite{Bello-Morales:2023btf}, diffeomorphism breaking was considered in the cosmological framework based on the above Lagrangian introduced in Ref.~\cite{Anber:2009qp}, yet the authors of Ref.~\cite{Bello-Morales:2023btf} interpret the effects of the symmetry breaking in the context of non-standard fluids, which is quite different from what we do in this paper.} We investigate whether there are well-behaved solutions of the scale factor that are continuously connected to GR. Namely, we aim to determine whether minor violations of diffeomorphism invariance yield approximate solutions to the GR scale factor and whether the solutions reduce to those of GR in the diffeomorphism-invariant limit.  We emphasize that answers to these questions are nontrivial. If, for instance, there are solutions for the scale factor that have different time dependence than the GR solution (as will be the case of the vacuum-only universe) and/or solutions that have singular behavior in the limit of vanishing symmetry breaking parameter (e.g. $1/\alpha$ dependence), even small terms can lead to drastic differences from the GR solution and will contradict the observed evolution of the universe. This would also raise concerns about the reliability of the perturbative effective action and could pose challenges for the underlying theory in admitting a low-energy effective description. Furthermore, we will investigate whether the ghost-like instabilities arising in the linear theory might play a role in the full nonlinear theory regarding cosmological evolution.

Note that in this work, we \textit{do not} propose a diffeomorphism-breaking theory, \textit{nor do} we advocate for or against the diffeomorphism breaking; we just do a neutral analysis as a follow-up to Ref.~\cite{Anber:2009qp}.  There is no question regarding the effective diffeomorphism invariance of gravity; the question is whether this situation can emerge from an underlying theory without this symmetry. If this is the case, there may be minor violations at low energies, small enough to avoid direct observations. If such small violations can mathematically yield instabilities in the basic evolution of the universe, this would provide strong evidence against such a possibility of emergence. This is the motivation of this paper.

Explicit diffeomorphism breaking has been heavily studied in the literature~\cite{Kostelecky:2003fs,Kostelecky:2011qz,Bluhm:2015dna,Bluhm:2016dzm,Bluhm:2017pje,Bonder:2018asb,Bluhm:2019ato,Bonder:2020fpn,Kostelecky:2020hbb,Kostelecky:2021tdf,Bluhm:2021lzf,Boehmer:2021aji,Boehmer:2023fyl,Boehmer:2023lpb,Bello-Morales:2023btf,Jensko:2024bee,Bailey:2024zgr,Lafkih:2024qva}, most of which are in the context of the local Lorentz symmetry violation. Our case is less severe; we maintain the local Lorentz symmetry while breaking the general covariance in order to focus on the diffeomorphism-breaking effects in the Lorentz-invariant theory.~Just like a Lorentz-violating theory introduces a preferred direction in spacetime due to a Lorentz-violating background, the diffeomorphism-invariance breaking in the theory fixes the geometry to one that satisfies a certain consistency equation.

This brings us to another point of discussion in the literature: spontaneous vs explicit diffeomorphism invariance breaking. This is rather a discussion more related to the possible underlying theory, not necessarily the effective theory analysis at low energies. Regarding the low-energy phenomenology, the symmetry-breaking terms in the low-energy effective theory can always be treated explicitly, regardless of how the symmetry is broken in the underlying theory~\cite{Kostelecky:2003fs,Anber:2009qp,Donoghue:2010qf,Bluhm:2013mu,Bello-Morales:2023btf}. This is what we are interested in this paper, the low-energy phenomenology without referring to a possible underlying theory in which the symmetry might be broken either way, depending on the theory.  It has been discussed in the literature that the explicit diffeomorphism-invariance-breaking terms might lead to some geometrical inconsistencies~\cite{Kostelecky:2003fs,Bluhm:2014oua}. Yet, satisfying certain consistency equation(s) (see Eq.~(\ref{cons5}) below), resolves the issue~\cite{Anber:2009qp,Bluhm:2016dzm,Bluhm:2021lzf}. 
 
The rest of the paper is organized as follows. In section~\ref{sec:theory}, we give a brief account of the formulation of the theory and establish the cosmological set-up. In section~\ref{sec:analytical}, we investigate the single-component universes, where we find analytical solutions for radiation-only and vacuum-only cases and numerical solutions for the matter-only case. In sections~\ref{sec:radmatnumerical} and \ref{sec:matvacnumerical}, we look at the two-fluid systems, radiation-matter and matter-vacuum cases, respectively. In section~\ref{sec:mixnumerical}, we address the most realistic scenario where we have all three components. In section~\ref{sec:summary}, we summarize our results. Finally, we conclude in section~\ref{sec:conclusion}, while leaving some computational details to our Appendices \ref{depapp}, \ref{expMapp}, \ref{Appconstraint} and \ref{EinsteinApp}.


\section{The diffeomorphism violating construction \label{sec:theory}}

\subsection{Formulation \label{sec:Formulation}}
We begin by laying down the basics of the theory. The zeroth-order term in the energy expansion in the effective theory of gravity is just the cosmological constant. The next order terms are the ones involving two derivatives of the metric. At this order, in the diffeomorphism invariant theory (i.e., GR), there is only a single term that can be written, the Ricci scalar, $R=g^{\sigma \nu} \partial_\mu \Gamma_{\nu \sigma}^\mu-g^{\sigma \nu} \partial_{\nu} \Gamma_{\mu \sigma}^\mu+g^{\sigma \nu} \Gamma_{\mu\lambda}^\mu \Gamma_{\nu \sigma}^\lambda-g^{\sigma \nu} \Gamma_{\nu \lambda}^\mu \Gamma_{\mu \sigma}^\lambda$, which is invariant under diffeomorphisms.  

Let us briefly discuss diffeomorphisms. Under the coordinate change $x^\mu \rightarrow x^{\prime \mu}$,  the metric transforms as 
\begin{equation}
\label{passive}
g_{\mu \nu}(x) \rightarrow g_{\mu \nu}^{\prime}\left(x^{\prime}\right)=\frac{\partial x^\rho}{\partial x^{\prime \mu}} \frac{\partial x^\sigma}{\partial x^{\prime \nu}} g_{\rho \sigma}(x)\;.
\end{equation}
 Considering the infinitesimal diffeomorphisms generated by an infinitesimal vector field $\xi^\mu(x)$, namely $x^\mu \rightarrow x^{\prime \mu}= x^\mu+\xi^\mu(x)$,  Eq.~(\ref{passive}), up to first order in $\xi^\mu(x)$, yields
\begin{eqnarray} 
\label{metrictransformation}
 g_{\mu \nu}^{\prime}=g_{\mu \nu}-g_{\rho \nu} \partial_\mu \xi^\rho-g_{\mu \sigma} \partial_\nu \xi^\sigma-\xi^\lambda \partial_\lambda g_{\mu \nu}\;,
\end{eqnarray}
where all the terms are valued at the same point on the manifold, say $x^\mu$. This can be expressed in the common form: $g_{\mu \nu}^{\prime}=g_{\mu \nu}-\nabla_{\mu} \xi_\nu-\nabla_\nu \xi_\mu=g_{\mu \nu}-\mathcal{L}_{\xi} g_{\mu \nu}$, where $\mathcal{L}_{\xi}$ is the Lie derivative.\footnote {These transformations are sometimes called passive diffeomorphisms, or general coordinate transformations, where we change the coordinates from $x$ to $x^{\prime}$ and evolve the fields accordingly at $x^{\prime}$ (then relabel them as $x$ for simplicity). If instead one adopts the active diffeomorphism perspective, where the coordinates remain fixed and the metric is transformed by the pullback of the diffeomorphism generated by the vector field $\xi^\mu$, the change in the metric is given by its Lie derivative with the plus sign: $g_{\mu \nu}^{\prime}=g_{\mu \nu}+\mathcal{L}_{\xi} g_{\mu \nu}$~\cite{Carroll:2004st}. In our context, the passive and active perspectives are dual to each other.}
 
 Considering the identities $\partial_\nu g^{\rho \gamma}=-g^{\rho \alpha} \Gamma_{\nu \alpha}^\gamma-g^{\beta \gamma} \Gamma_{\beta \nu}^\rho$ and $\partial_\mu \sqrt{-g}=\sqrt{-g}\; \Gamma_{\mu \nu}^\nu$, we can express the Einstein-Hilbert term $\sqrt{-g} R$ using terms that involve two factors of Christoffel connections up to surface terms as
\begin{equation}
\label{Ricci}
\sqrt{-g} R= \sqrt{-g} \left(g^{\mu \nu} \Gamma_{\mu \lambda}^\alpha \Gamma_{\nu \alpha}^\lambda -g^{\mu \nu} \Gamma_{\mu \nu}^\alpha \Gamma_{\lambda \alpha}^\lambda\right)\quad +\quad  \mathrm{surface\;  terms}\;,
\end{equation}
which is  $\sqrt{-g}(\mathcal{L}_2-\mathcal{L}_1)$ in the notation below.  Namely, a specific combination of terms composed of two Christoffel connections makes up the Einstein-Hilbert up to surface terms, which is the only allowed term at the quadratic order in metric derivatives in the diffeomorphism invariant theory. Once we give up the diffeomorphism invariance and take the form of the Lagrangian the same as above, namely in the form of~\footnote{ Note that if we break (local) diffeomorphisms down to the global Lorentz group, there could be an infinite number of terms with extra powers of  $\sqrt{-g}$ at the same, two derivative, order since $\sqrt{-g}$ is scalar under the latter group. However, here, as emphasized in Ref.~\cite{Bello-Morales:2023btf}, diffeomorphisms can be broken down to global $ GL(4,\mathbb R)$, which contains the global Lorentz group $SO(3,1)$ as a subgroup (this is distinct from the global $SO(3,1)$ contained in the local unbroken Lorentz gauge symmetry, which acts fiberwise on the orthonormal frame bundle, namely on the tetrads $e^{a}{}_{\mu}$, the spin connection, etc.). Under $ GL(4,\mathbb R)$,  $\sqrt{-g}$ transforms as a scalar density of weight $-1$. Therefore, other than a single factor of $\sqrt{-g}$, which, together with $ d^4 x$, will constitute a scalar under $ GL(4,\mathbb R)$, no other factors of $\sqrt{-g}$ are allowed in the action. Recall that the connection coefficients $\Gamma_{\mu \lambda}^\alpha$ transform as (1,2) tensors under this linear group.}  $\sqrt{-g}\;\times$ (terms without any power of $\sqrt{-g}$), we should include in the Lagrangian all terms that are composed of two factors of Christoffel connections, but with arbitrary coefficients. Other terms involving two derivatives such as $\sqrt{-g} \partial_\gamma g^{\alpha \beta} \partial^\gamma g_{\alpha \beta}$ and $\sqrt{-g} \partial_\alpha \partial_\beta g^{\alpha \beta}$, can be expressed as a linear combination of terms with two factors of Christoffel connections up to surface terms, in similar to $g^{\rho\mu}\partial_\mu \Gamma^\nu_{\nu \rho}$ and $g^{\nu\rho} \partial_\mu \Gamma^\mu_{\nu\rho}$ terms, as mentioned while getting rid of these terms in obtaining Eq.~(\ref{Ricci}) (see  Appendix \ref{depapp}).  
Then, the most general diffeomorphism-breaking action in the form of $\sqrt{-g}\;\times$ (involving terms up to quadratic order in metric derivatives) becomes 
\begin{equation}
\label{action}
\mathcal{S}=\int d^4 x \sqrt{-g} \;\mathcal{L}\;,
\end{equation}
 where the Lorentz invariant effective Lagrangian is given as
\begin{eqnarray}
\label{eqn:TotalLagrangian}
\mathcal{L}=\frac{1}{16 \pi G}\left[R-2\Lambda+\sum_{a=1}^5 \alpha_a \mathcal{L}_a\right]+\mathcal{L}^m\;.
\end{eqnarray}
Here, in addition to the usual GR terms (coupled with matter), i.e., the Einstein-Hilbert term, cosmological constant, and the matter Lagrangian $\mathcal{L}^m$, we also have the diffeomorphism-violating terms given as~\cite{Anber:2009qp}
\begin{eqnarray}
\label{diffLs}
 \mathcal{L}_1&=&-g^{\mu \nu} \Gamma_{\mu \lambda}^\alpha \Gamma_{\nu \alpha}^\lambda, \quad \mathcal{L}_2=-g^{\mu \nu} \Gamma_{\mu \nu}^\alpha \Gamma_{\lambda \alpha}^\lambda, \quad  \mathcal{L}_3=-g^{\alpha \gamma} g^{\beta \rho} g_{\mu \nu} \Gamma_{\alpha \beta}^\mu \Gamma_{\gamma \rho}^\nu,\nonumber \\
 \mathcal{L}_4&=&-g^{\alpha \gamma} g_{\beta \lambda} g^{\mu \nu} \Gamma_{\mu \nu}^\lambda \Gamma_{\gamma \alpha}^\beta , \quad \mathcal{L}_5=-g^{\alpha \beta} \Gamma_{\lambda \alpha}^\lambda \Gamma_{\mu \beta}^\mu\;, 
\end{eqnarray}
and $\{\alpha_a \}$ are small coefficients. $ \{ \mathcal{L}_1, \cdots  \mathcal{L}_5 \}$ span all possible terms which contribute independently. See our Appendix \ref{depapp} for further explanation.

 Recall that only a specific combination of the diffeomorphism-violating terms in Eq.~(\ref{diffLs}) remains invariant under this transformation; namely, the combination $\alpha_1=-\alpha_2$ and $\alpha_3=\alpha_4=\alpha_5=0$, which yields the Einstein-Hilbert term up to a total derivative. As we will see below, the more general condition that yields the field equations of GR up to a redifiniton of Newton's constant is $\alpha_1+\alpha_2+\alpha_3+2\alpha_4=0$ together with $\alpha_4=\alpha_5=0$. Since the symmetry-breaking terms are in the same order as the Ricci scalar, they are not suppressed by extra factors of the inverse energy scale in the energy expansion of the effective theory, which makes the symmetry-breaking effects potentially more visible.  Ref~\cite{Anber:2009qp} considers one of the terms ($\mathcal{L}_3)$ in the post-Newtonian analysis. The most stringent bound comes from the absence of preferred
frame effects in pulsars, which requires $\alpha_3\lesssim 10^{-20}$. To the best of our knowledge, there is no other analysis for the terms given in Eq.~(\ref{diffLs}). 

Note that we do not focus on constraining the parameters of the theory in this paper. It is clear that the diffeomorphisms breaking effects are tightly constrained by the post-Newtonian effects. What we are interested in this paper is the possible inconsistencies between the nonsymmetric theory in the symmetric limit ($\alpha\rightarrow 0$) and the symmetric theory ($\alpha= 0$, i.e. GR) regarding the evolution of the cosmological scale factor. For instance, in the absence of diffeomorphism invariance, the scale factor solutions in the nonsymmetric theory may have a singular behaviour such as $1/\alpha$-type divergence (see footnote~\ref{example}), which would point to a discontinuity effect similar to the vDVZ discontinuity in the Pauli-Fierz massive gravity~\cite{Hinterbichler:2011tt}. Furthermore, if the solutions have a different time behaviour than GR. i.e, the deviation is not constant or does not asymptotically approach a constant value, even a small amount of symmetry-breaking will register large effects as time goes on. In this case, even if there is no discontinuity as mentioned above, a growing difference from GR with time would potentially cause inconsistencies with the observed evolution of the universe. Our investigations will show that the cosmological evolution is well-behaved within the diffeomorphism theory considered in this paper.

The Euler-Lagrange equations for the action above, obtained from $\delta \mathcal{S}/\delta g^{\alpha\beta}=0$, is given as
\begin{eqnarray}
\frac{\partial \sqrt{-g} \mathcal{L}}{\partial  g^{\mu \nu}}-\partial^\rho\left(\frac{\partial  \sqrt{-g} \mathcal{L}}{\partial  g^{\mu \nu, \rho}}\right)=0\;,
\end{eqnarray}

which yields
\begin{equation} 
\label{EE51}
G_{\mu\nu} + \sum_{a= 1}^{5} \alpha_a M_ {\mu\nu}^{(a)}   = 8 \pi G T_{\mu\nu}\;, 
\end{equation}
where $G_{\mu\nu} =R_{\mu\nu} - \frac{1}{2} g_{\mu\nu} R$ is the Einstein tensor and $T_{\mu\nu}=T_{\mu\nu}^m+T_{\mu\nu}^\Lambda$ is the total energy-momentum tensor including the contributions from the matter $T_{\mu\nu}^m=-2 \frac{1}{\sqrt{-g}} \frac{\delta S^m}{\delta g^{\mu \nu}}$ and the effective vacuum energy $T_{\mu\nu}^\Lambda=-\rho_{\Lambda} g_{\mu \nu}=-\frac{\Lambda}{8 \pi G} g_{\mu \nu}$ , which are the usual terms in diffeomorphism invariant theory.  $M_ {\mu\nu}^{(a)}$ are the diffeomorphism-breaking contributions, the explicit expressions of which are given in Appendix~\ref{expMapp}. 

Since Eq.~(\ref{EE51}) is not invariant under general coordinate transformations, a consistency condition is required. Recall that the Einstein tensor is identically divergenceless ($\nabla^\mu G_{\mu\nu}=0$). The condition that $\nabla^\mu T_{\mu\nu}=0$, which can be seen as the generalization of the flat-space matter energy-momentum conservation law to curved space-times, is obtained automatically from the diffeomorphism invariance of the matter action~\cite{Carroll:2004st}. Since the diffeomorphism breaking in our framework is employed only in the gravitational sector, as emphasized also in Ref.~\cite{Bello-Morales:2023btf}, the divergenceless condition is still valid in this case. Therefore, the consistency equation is obtained as
\begin{equation}
\label{cons5}
\alpha_a \nabla^\mu M_{\mu\nu}^{(a)} = 0\;.
\end{equation}
 Therefore, unlike the diffeomorphism invariant case, we do not have the freedom to choose the metric among the ones related by diffeomorphisms. Only the ones satisfying the constraint given in Eq.~(\ref{cons5}) must be considered. 

\subsection{Cosmological set-up\label{setup}}

We will now choose the form of the metric to look for solutions for the equations of motion of the diffemorphism breaking theory and the constraint equation given in Eqs.~(\ref{EE51}) and (\ref{cons5}), respectively. We will, of course, stay faithful to the observed cosmological behaviour, namely, a spatially flat expanding universe.

An immediate suggestion for cosmological metric (for our spatially flat universe) would be the regular FLRW metric $ds^2=-dt^2+a^2(t)(dx^2+dy^2+dz^2)$, which satisfies the constraint (\ref{cons5}). However, as it will be clear below (see Eq.~(\ref{fieldeqsgeneral})), the theory reduces to GR with this choice of metric up to a redefinition of the gravitational constant $G$. Therefore, we will work with a more general metric (modified FLRW)
\begin{eqnarray}
\label{metricwithf}
ds^2=-f^2(t)dt^2+a^2(t)(dx^2+dy^2+dz^2).
\end{eqnarray}
Having set the metric, below we focus on the effects of the diffeomorphism-breaking terms by discussing the related consistency and field equations in detail.

\subsubsection{The consistency equation}

Now, we are in a position to work out the constraint equation (\ref{cons5})
\begin{equation}
\label{constrainte}
\alpha_a\nabla^\mu M_{\mu\nu}^{(a)} = \alpha_ag^{\mu\lambda} \nabla_{\lambda}M_{\mu\nu}^{(a)} = 0,
\end{equation}
where we employ the modified FLRW metric (\ref{metricwithf}) explicitly. The are only a few non-vanishing Christoffel symbols
\begin{equation}
\Gamma^0_{ij} = \delta_{ij} \frac{a\dot{a}}{f^2},   \quad   \Gamma^0_{00} = \frac{\dot{f}}{f},  \quad  \Gamma^i_{0j} = \delta^i_j \frac{\dot{a}}{a}, 
\end{equation}
and in components LHS of (\ref{constrainte}) yields
\begin{subequations}
\begin{align}
\label{con1}
\nabla^\mu M_{\mu 0} &= \dot{M}_{00} + \left(  \frac{3\dot{a}}{a} -  \frac{2\dot{f}}{f}\right) M_{00} + \frac{\dot{a} f^2}{a^3} \delta^{ij} M_{ij} , \\
\nabla^\mu M_{\mu i} &=  \dot{M}_{0i} - \left( \frac{\dot{f}}{f} - \frac{3\dot{a}}{a}  \right) M_{0i},  
\label{con2}
\end{align}
\end{subequations} 
where $\{\dot{}\}$ is derivation with respect to the coordinate time $t$.
The second part (\ref{con2}) is trivially satisfied as $M_{0i}^{(a)} =0$. Thus, there is only one constraint equation, namely (\ref{con1}) to be worked out. 

An account of the individual terms of the constraint equation is presented in our Appendix~\ref{Appconstraint}. The contributions from the first three diffeomorphism breaking terms (\ref{diffLs}) are shown to be identical, see (\ref{conM1}), (\ref{conM2}) and (\ref{conM3}). However, the remaining $ \mathcal{L}_4$ and $ \mathcal{L}_5$ contribute differently both from them and from each other, see (\ref{conM4}) and (\ref{conM5}). Therefore, if we were to express the consistency equation (\ref{cons5}) in its full form, we would get 
\begin{eqnarray}
&&(\alpha_1 + \alpha_2 +\alpha_3 +\alpha_4 +\alpha_5)\left[ - \frac{\dddot{f} }{f} + \frac{6 \dot{f}\ddot{f}}{f^2} - \frac{6 \dot{f}^3}{f^3}\right] +(\alpha_4 -\alpha_5)\left[ \frac{3\dddot{a}}{a} + \frac{6\dot{a}^3}{a^3} + \frac{18 \ddot{a}\dot{a}}{a^2} \right] \nonumber\\
&&+ 3\big(  2(\alpha_1 +\alpha_2 +\alpha_3) + 3\alpha_4 +\alpha_5\big)\left[\frac{2 \dot{a} \dot{f}^2}{af^2} - \frac{\ddot{f}\dot{a}}{fa}   \right] \nonumber \\  
&&-3 (\alpha_1 +\alpha_2+\alpha_3 +3\alpha_4 -\alpha_5) \left[  \frac{\ddot{a}\dot{f}}{af} + \frac{2\dot{a}^2\dot{f}}{a^2 f}\right]  =0. 
\label{congenx}
\end{eqnarray}
Let us now discuss particular cases of this complicated result (\ref{congenx}).
Firstly, we remind the reader that in \cite{Anber:2009qp} authors considered third Lagrangian only with the contribution (\ref{M3}). Their consistency condition was $\nabla^\mu M_{\mu 0}^{(3)} =0$ (\ref{conM3}). Secondly, in \cite{Bello-Morales:2023btf}, a diffeomorphism-violating model based only on the fifth term $\mathcal{L}_5$ was worked out. Such a theory yields a modification of (\ref{M5}) to Einstein's equations with a different consistency condition $\nabla^\mu M_{\mu 0}^{(5)} =0$ (\ref{conM5}).

However, from a theoretical point of view, there is no reason for a single choice of those diffeomorphism-breaking contributions. In our study, we do not distinguish between those individual terms and include all of them in our framework. 

Note that in the linear analysis performed in Ref.~\cite{Anber:2009qp}, it was found that the most general case is obtained with the condition
\begin{equation}
\label{alphacondition}
\alpha_a\neq 0\;\quad\quad\mathrm{and} \quad\quad \alpha_4=\alpha_5\;.
\end{equation}
Hence, we will stick with this condition for the rest of this paper for our full (non-linear) cosmological analysis.
Adding (\ref{conM4}) and (\ref{conM5}) cancels the different terms and we get the same expression as in (\ref{conM1}), (\ref{conM2}) and (\ref{conM3})
\begin{equation}
\label{conM4M5}
\nabla^\mu M_{\mu 0}^{(4)}+ \nabla^\mu M_{\mu 0}^{(5)} = 2 \left[  - \frac{\dddot{f} }{f} + \frac{6 \ddot{f}}{f} \left( \frac{\dot{f}}{f} - \frac{\dot{a}}{a} \right) - \frac{3 \ddot{a} \dot{f}}{af} - \frac{6 \dot{f}^3}{f^3} - \frac{6 \dot{a}^2 \dot{f}}{a^2f} + \frac{12 \dot{a} \dot{f}^2}{af^2}  \right] .
\end{equation}

Thus, the consistency equation (\ref{cons5}) becomes
\begin{eqnarray}
\label{constraint2}
- \frac{\dddot{f}}{f} + \frac{6\ddot{f}\dot{f}}{f^2} - \frac{6\ddot{f}\dot{a}}{fa} - \frac{3\ddot{a}\dot{f}}{fa} - \frac{6 \dot{{f}}^3}{f^3} - \frac{6\dot{a}^2 \dot{f}}{a^2f} + \frac{12 \dot{a}\dot{f}^2}{af^2} =0\,,
\end{eqnarray}
with a prefactor\footnote{We shall note that without the condition $\alpha_4=\alpha_5$, we wouldn't have an overall $\alpha$ dependent factor; there are extra terms coming with the factors $(\alpha_4-\alpha_5)$ both in the field equations and the constraint equation, see Appendices \ref{Appconstraint} and \ref{EinsteinApp} for details. Our condition yields a much simpler equation by cancelling those extra terms.} $-K\equiv \alpha_1 + \alpha_2 +\alpha_3 + 2 \alpha_4$. The trivial solution for the consistency equation above would be $K=0$. However, this reproduces Einstein's equations, ala a redefinition of Newton's constant\footnote{If $\alpha_1+\alpha_2+\alpha_3+2\alpha_4=0\,$, the field equations become
\begin{eqnarray}
\nonumber
3(2\alpha_1+3\alpha_2+1)\frac{\dot a^2}{a^2}&=&8\pi G T_{00}\\
-(2\alpha_1+3\alpha_2+1)\left[\left(\dot a^2+2a\ddot a\right)-2a\dot a \frac{\dot f}{f} \right]&=&8\pi G f^2 T_{xx}\,,
\end{eqnarray} which reduces to GR via a redefinition of Newton's constant, $G_{eff}=G/(1+2\alpha_1+3\alpha_2)$. Notice that $f(t)$ can be eliminated in GR through a coordinate transformation to the cosmological time $d\tilde{t}=f(t)dt$, unlike in the theory with broken diffeomorphism, as we will see below.\label{footnote1}}, and hence, it is not in our interest. 

Throughout the paper, we keep the $\alpha$ coefficients as general as possible. In fact, the only conditions we will consider come from the linear analysis on the Minkowskian background~\cite{Anber:2009qp}; it is required that $\alpha_1+\alpha_2-3 \alpha_3-2 \alpha_4 \geqslant 0$   (condition-I)\footnote{ Notice the sign difference with Ref.~\cite{Anber:2009qp}, discussed in Appendix~\ref{expMapp} in this paper.} so that the extra massless mode in the linearized version of the action~(\ref{action}) is not a ghost (the saturation of this inequality decouples this extra mode). Moreover, as discussed in Ref.~\cite{Alvarez:2006uu}, the Minkowskian vacuum in the linear theory admits a classical instability associated with the vector modes, unless the condition $a_3+a_4=0$ (condition-II) is imposed.\footnote {The condition $a_3+a_4=0$, together with the saturation of the no-ghost inequality, $\alpha_1+\alpha_2-3 \alpha_3-2 \alpha_4 =0$, restores the full diffeomorphism invariance: satisfying these equations corresponds to the vanishing of $K \equiv -\alpha_1 - \alpha_2 -\alpha_3 - 2 \alpha_4$, which recovers GR, as can be seen from Eqs.~(\ref{EEc1}), (\ref{EEc1}), and (\ref{cec}) below.} If the latter condition is satisfied, the diffeomorphism symmetry in the linearized theory on Minkowski background ($g_{\mu\nu}=\eta_{\mu\nu}+h_{\mu\nu}$) is broken only down to transverse diffeomorphism~\cite{Anber:2009qp,Bello-Morales:2023btf,Alvarez:2006uu} (also called the restricted diffeomorphism); namely, instead of full diffeomorphism $h_{\mu \nu} \rightarrow h_{\mu \nu}+\partial_\mu \xi_\nu+\partial_\nu \xi_\mu$ (with arbitrary $\xi_\mu$), which can be seen from Eq.~(\ref{metrictransformation}), transverse diffeomorphisms comes with the condition $\partial_\mu \xi^\mu=0$~\cite{vanderBij:1981ym,Unruh:1988in,Buchmuller:1988wx,Alvarez:2006uu}. In our numerical analysis, to be discussed later in this paper, we choose our benchmark points (BPs), given in Table~\ref{BPs}, such that one, both, or neither of these conditions are satisfied; this will give us an idea of whether the instabilities in the linear theory carry over to the full, nonlinear theory.

\subsubsection{Einstein's field equations}

The energy-momentum tensor for a perfect fluid is
\begin{equation}
T_{\mu\nu} = (\rho + p) u_\mu u_\nu + p g_{\mu\nu},
\end{equation}
where the 4-velocity of the perfect fluid in its rest frame is given as $u^\mu = (1/f(t), 0, 0, 0)$. Then, the non-zero components of $T_{\mu\nu}$ are
\begin{equation}
T_{00} = f^2 \rho,  \quad  T_{ij} = \delta_{ij} a^2 p. 
\end{equation}
Above $\rho$ is the density of the fluid, and $p$ is its pressure in the rest frame.  Energy conservation $\nabla^\mu T_{\mu 0} = 0$ leads to
\begin{equation} 
\dot{\rho} + 3 \frac{\dot{a}}{a} (\rho + p) =0\;,
\end{equation}
which is independent of $f(t)$. Therefore, together with the equation of state $p = \omega \rho$ with $\omega$ being a constant, it yields the same relation as in GR, i.e., 
\begin{equation}
\rho = \rho_0 a^{-3(1+\omega)}.
\end{equation}
\\
For the metric~(\ref{metricwithf}), the components of Einstein's tensor can be found as 
\begin{equation}
G_{00} = 3 \frac{\dot{a}^2}{a^2},  \quad  G_{ij} = \frac{1}{2f^2} (  - \dot{a}^2 -2 a\ddot{a} + 2a \dot{a} \frac{\dot{f}}{f}).  
\end{equation}
We recall our Appendix \ref{EinsteinApp} where the LHS of Einstein's equations are given in full generality. Setting $\alpha_4 = \alpha_5$, they become
\begin{equation}
\label{00genalpha}
\begin{split}
 3 \frac{\dot{a}^2}{a^2} + \frac{3}{2}(\alpha_1 + 3 \alpha_2 - 3 \alpha_3 - 6 \alpha_4)   \frac{\dot{a}^2}{a^2}    + (\alpha_1 + \alpha_2 +\alpha_3 +2\alpha_4) \left( \frac{3}{2} \frac{\dot{f}^2}{f^2} - 3 \frac{\dot{a}\dot{f}}{af} - \frac{\ddot{f}}{f} \right) = 8 \pi G T_{00} 
\end{split}
\end{equation}
\begin{equation}
\label{ijgen}
\begin{split}
&\delta_{ij} \left(-1 - \frac{\alpha_1}{2} - \frac{3}{2}\alpha_2 + \frac{3}{2}\alpha_3 + 3\alpha_4  \right) \frac{\dot{a}^2}{f^2} + \delta_{ij} \left( 2 + \alpha_1 + 3\alpha_2 - 3\alpha_3 -6 \alpha_4  \right) \frac{a\dot{a} \dot{f}}{f^3} \\
& +\delta_{ij} \left(-2 -\alpha_1 -3\alpha_2 + 3\alpha_3 +6 \alpha_4  \right) \frac{a\ddot{a}}{f^2} + \delta_{ij} (-\alpha_1 - \alpha_2 -\alpha_3 -2\alpha_4) \frac{a^2 \dot{f}^2}{2 f^4}  = 8 \pi G T_{ij}
\end{split}
\end{equation}
Note that the condition $\alpha_4 =\alpha_5$ eliminates $\ddot{a}$ term in the first equation and $\ddot{f}$ term in the second one.
Augmented with multiple perfect fluid components, the field equations~(\ref{EE51}) become
\begin{subequations}
\label{fieldeqsgeneral}
\begin{align}
\frac{L}{2} \frac{\dot{a}^2}{a^2} + K \left( \frac{\ddot{f}}{3 f} + \frac{\dot{a}\dot{f}}{af} - \frac{1}{2} \frac{\dot{f}^2}{f^2} \right) &= \frac{8\pi G}{3} f^2 \sum_n \rho_{n 0} \;a^{-3(1+\omega_n)}\label{fieldeqsgeneral1}\\
 L\left(  \frac{a\dot{a} \dot{f}}{f^3} - \frac{1}{2} \frac{\dot{a}^2}{f^2} - \frac{a\ddot{a}}{f^2}  \right) +  K \frac{a^2 \dot{f}^2}{2 f^4} &= 8\pi G\; a^2 \sum_n \omega_n \;\rho_{n 0} \;a^{-3(1+\omega_n)}\label{fieldeqsgeneral2},
\end{align}
\end{subequations}
where we set 
\begin{equation}
\label{LKs}
L \equiv 2 + \alpha_1 + 3 \alpha_2 - 3 \alpha_3 - 6 \alpha_4,  \quad  \quad  K \equiv -\alpha_1 - \alpha_2 -\alpha_3 - 2 \alpha_4.
\end{equation}

The field equations Eqs.~(\ref{fieldeqsgeneral1}) and (\ref{fieldeqsgeneral2}) with the constraint equation (\ref{constraint2}), two of which are independent, constitute our system of equations. Considering Eq.~(\ref{constraint2}), the simplest choice would be $\dot{f}=0$, but just like the regular FLRW metric, as mentioned above, $\dot{f}=0$ provides GR up to rescaled Newton's constant and thus it is trivial. This will be clearer when we switch the cosmological time below. Therefore, the constraint must be satisfied for the non-trivial $f(t)$ and $a(t)$ that solve the field equations for the metric~(\ref{metricwithf}),  discussed below. 

\subsection{Switching to the cosmological time\label{sec:cosmotime}}

Since, unlike in GR, we don't have the full diffeomorphism invariance in our action, and since we would like to compare our results with those in GR, we must be in the same framework. The form of the spatial part of our metric is the same as the FLRW metric, but the definitions of time are different. For comparison purposes, we switch to the cosmological time ($\tilde{t}$)~\cite{Bello-Morales:2023btf}, defined via $dt=d\tilde{t}/f(\tilde{t})$. Note that this is not a coordinate transformation; this is merely a redefinition of our time coordinate in terms of the one commonly used in GR; therefore, tensorial objects do not transform under this redefinition. Performing this time definition before or after solving the field and consistency equations, of course, does not matter, as shown in a sample case in footnote~\ref{footnote2}, below. 

Note that applying this redefinition at the line element level does not yield the same result, as expected. As mentioned at the beginning of section \ref{setup}, which can also be seen from Eq.~(\ref{fieldeqsgeneral}), this redefinition at the line element level would effectively be equivalent to choosing $f=1$ and hence just leads to GR. Such dependence on the order of transformation is known to occur in diffeomorphism-breaking theories (e.g., see comments below Eq.~(67) in Ref.~\cite{Alvarez:2007nn}).  This is, in fact, not surprising. Using such a redefinition in this way fixes the geometry without knowing the consistency equation and hence causes information loss. In other words, applying this field redefinition in the line element would change the metric and hence would act as a coordinate transformation. As mentioned above, the time switch is implemented just to ensure the same definition of cosmic (FLRW) time, commonly used in GR, for comparing the scale factor solutions. As it is not a coordinate transformation, it does change the dynamical field of the theory,  the metric.\footnote{ Instead of performing this time switch in the diffeomorphism-violating theory, we can apply its inverse in GR. Then, the common form of scale factor in GR with various perfect fluid sources will be subjected to such a transformation, and a comparison with the solutions in the diffeomorphism-violating theory can be consistently made. This will be mentioned again with an example in footnote~\ref{footnote2}.}
 
Switching from $t\rightarrow\tilde{t}$ in Eqs.~(\ref{fieldeqsgeneral1}) and (\ref{fieldeqsgeneral2}) yields the field equations in cosmological time $\tilde{t}$ as
\begin{subequations}
\begin{align}
\label{EEc1}
\frac{ L}{2}\frac{(a')^2}{a^2} - K \left( \frac{1}{6} \frac{(f')^2}{f^2} - \frac{a' f'}{af} - \frac{1}{3}\frac{f''}{f} \right)   &= \frac{8\pi G}{3} \sum_n \rho_{n 0} \;a^{-3(1+\omega_n)}, \\
 \frac{K}{2} \frac{f'^2}{f^2} - \frac{L}{2} \left( \frac{a'^2}{a^2} + 2 \frac{a''}{a} \right)  &= 8\pi G\; \sum_n \omega_n \;\rho_{n 0} \;a^{-3(1+\omega_n)},
\label{EEc2}
\end{align}
\end{subequations}
where  $\{'\}$  denotes derivation with respect to $\tilde{t}$. Recall that $L$ and $K$ are defined in Eq.~(\ref{LKs}).  
In addition, we shall note that the constraint equation (\ref{constraint2}) in terms of the cosmological time $\tilde{t}$ becomes
\begin{eqnarray}
\label{cec}
-  \frac{f'''}{f} + 2 \frac{ f' f''}{f^2} - \frac{f'^3}{f^3} - 6 \frac{ a' f''}{a f} + 3 \frac{ f'^2 a'}{f^2 a} - 3 \frac{ f' a''}{f a} - 6 \frac{ f' a'^2}{f a^2} = 0\;,
\end{eqnarray}
again with the prefactor $K$. Again, only two of Eqs.~(\ref{EEc1}), (\ref{EEc2}), and (\ref{cec}) are independent.

It is more apparent in Eqs.~(\ref{EEc1}) and (\ref{EEc2}) how the $K=0$ case, which would automatically satisfy the constraint equation~(\ref{cec}), would take us directly to the diffeomorphism invariant theory (GR) up to a redefinition of Newton's constant $G_{eff}=G/(1+2\alpha_1+3\alpha_2)$, as mentioned in Footnote~\ref{footnote1}. Therefore, we are not interested in this condition, and instead, we look for solutions that satisfy the constraint equation~(\ref{cec}). As in the previous subsection, the simple choice $f'=0$ (for all parameter space) satisfies the constraint equation but eliminates all the symmetry-breaking effects and restores diffeomorphism invariance. This is clear from the Eqs.~(\ref{EEc1}) and (\ref{EEc2});  the derivatives of $f$ appear with the factor $K$; thus, $K=0$ and the $f=\mathrm{constant}$  have the same effect. Therefore, we will look for solutions that are not constant in $f$ in order to observe the effects of the broken symmetry.
\\


\section{Analytical Approach to Single-Component Universes\label{sec:analytical}}

In the previous sections, we introduced the basics of our formalism and worked out the modified version of Einstein's equations in the presence of diffeomorphism-breaking terms. If gravity is emergent as it is assumed in our framework, then we might expect small indications of this phenomenon at low energies  (otherwise, they should have been observed experimentally). 

Our aim is to investigate theoretically those effects of diffeomorphism breaking in the cosmological context. A natural starting point is the single-component universes, where the effect of diffeomorphism breaking can be tested analytically.  Our physical insight is that the solutions of our modified model should be smoothly linked to those of the standard theory in the limit $\alpha_a \to 0$. If physically acceptable solutions can be found for the simplest single-component universes, then similar solutions may be expected in more realistic multi-component scenarios, as we will discuss in Sections \ref{sec:radmatnumerical}, \ref{sec:matvacnumerical}, and \ref{sec:mixnumerical}. 

For all single-component universes, we provide solution ansatzes to begin with. They are modifications of the standard solutions, including diffeomorphism-breaking terms. In order to guess their form, we make use of perturbative results given in our Appendix \ref{EinsteinApp}. In the majority of our computations, we did not put any restrictions on the parameters $\alpha_a$. Only at the end, we discuss $\alpha_a \to 0$ limit to discuss their validity. 

We will investigate certain regimes specified with definite $\omega$-values; radiation ($\omega= 1/3$), matter ($\omega = 0$), and cosmological constant ($\omega= -1$). For each case, we will search for solutions of $a(\tilde{t})$ and $f(\tilde{t})$ to the field equations, given in Eqs.~(\ref{EEc1}) and (\ref{EEc2}), while satisfying the constraint equation given in Eq.~(\ref{cec}). We will work in dimensionless time defined as $ \tau\equiv \sqrt{\frac{8\pi G \rho_0}{3}}\; \tilde{t}$. 

In order to find non-trivial solutions, we will require $\dot{f}(t)\neq 0$, as explained above. Ideally, one should look for solutions where the scale factor $a(\tau)$ is close to the GR counterpart, as expected from slightly broken diffeomorphism invariance. This slight (inert) deviation from GR can be understood as the absence of relative time dependence in scale factor behavior between the GR solution and the solution in the broken theory; namely, $a_{\tiny{\mbox{GR}}}(\tau)/a(\tau)$ being time-independent. This is indeed the case for the radiation-only and the matter-only cases, as we will show analytically in the former case and numerically in the latter. In the vacuum-only case, we will find an analytic solution in the exponential form as in GR. Due to the characteristic of the exponential function, an inert behavior in $a_{\tiny{\mbox{GR}}}(\tau)/a(\tau)$ is not possible. Since the deviation changes rapidly with time, even for small $\alpha_a$, this could be a big problem regarding the possibility of diffeomorphism violation for such a (vacuum-only) universe. As we will show in the next section, this problem does not exist, and solutions are well-behaved in the multiple-fluid cases, as long as $\alpha_a < O(10^{-1})$.  

 Finally, the solutions $a(\tau)$ must reduce to the GR solutions in the limit of vanishing symmetry-breaking parameters (whereas such a requirement is not meaningful for $f(\tau)$ since it is removed from the theory in this limit, as can be seen in Eqs.~(\ref{EEc1}) and (\ref{EEc2})). 
 
\subsection{Radiation\label{sec:radiationexact}}
\label{exactrdu}

For the radiation case ($\omega = \frac{1}{3}$), to find an analytical solution in the required form, the following ansatz is used.
\begin{equation} 
\label{ansatzrdu}
f(\tau)=  \tau ^{n},  \quad a(\tau) = \left( \frac{2 A \tau}{A+ B(\alpha_a)} \right)^{\frac{1}{2}},
\end{equation}
where $A$, $B(\alpha_a)$, and $n$ are constants to be found. The ansatz is chosen motivated by the zeroth-order perturbative solution given in Eq.~(\ref{radiationperturbative}). Notice that the ratio $a_{\tiny{\mbox{GR}}}(\tau)/a(\tau)=\mathrm{constant}$, as desired. The Hubble parameter becomes the same as the GR case, $H=a'/a=1/{2\tau}$. 
Putting (\ref{ansatzrdu}) into (\ref{cec}) firstly, we obtain 2 roots for $n$: $n= 0$ and $n = 1/2$. Since the constant $f$ case yields trivial GR solutions and eliminates all the symmetry-breaking effects as discussed above, we admit $n = 1/2$ as the solution. 

Substituting either into (\ref{EEc1}) or (\ref{EEc2}) yields
\begin{equation}
B = A \left(-1+ \sqrt{1+ \alpha_2 - 2 \alpha_3 - 4 \alpha_4}\right)\;,
\end{equation}
and setting $A=1$ we obtain the final form of our solution as\footnote{\label{footnote2}Expressing in the coordinate time, the solutions given in Eq.~(\ref{radsolution}) become linear in $t$ as $f(t) = \frac{t}{2}$ and $a(t) = t /\left(  2\sqrt{1+ \alpha_2 - 2 \alpha_3 - 4 \alpha_4}\right)^{1/2} $.  The reader can verify that this set of solutions satisfies the field and constraint equations given in Eqs.~(\ref{fieldeqsgeneral}) and (\ref{constraint2}), respectively. As expected, performing the time switch to the cosmological time before or after finding the solutions to the equation of motion does not lead to any inconsistencies. On the other hand, performing this transformation at the line element level, before determining the equation of motion, results in the loss of information, as mentioned at the beginning of Section \ref{sec:cosmotime}. This is not an issue of concern.  For instance, instead of switching to the cosmological time in the diffeomorphism-violating theory (i.e., $dt\rightarrow d\tau/f(\tau)$), one can perform the inverse transformation in GR (i.e., $d\tau\rightarrow f(t) dt$), which is certainly allowed since GR is a generally covariant theory. So, we can make a consistent comparison between solutions in both theories. For the case of radiation, the GR solution ($a_{\mathrm{GR}}(\tau)=\sqrt{2\tau}$) becomes $a_{\mathrm{GR}}(t)=t/\sqrt{2}$, which is clearly consistent with the above solution for $\alpha_a\rightarrow 0$.}
\begin{eqnarray}
\label{radsolution}
f(\tau) =  \tau^{1/2},   \quad  a(\tau) = \left(  \frac{2 \tau}{ \sqrt{1+ \alpha_2 - 2 \alpha_3 - 4 \alpha_4}}\right)^{1/2}  \;.
\end{eqnarray}
As required, in the limit $\alpha_a$'s $\rightarrow 0$, the solution for the scale factor $a(\tau)$ reduces to that in GR.  In this limit, $f(\tau)$ has no well-defined meaning since it disappears from the field equations due to the full diffeomorphism invariance.  We also checked that the numerical solution agrees with the analytical solution, as expected.
  
 Note that the above result is nontrivial and not obtained automatically by construction. Although we began with an ansatz for $a(t)$, there is no a priori guarantee (as will be evident in the matter case below) that a corresponding $f(t)$ exists such that, together with $a(t)$, it solves both the equations of motion and the constraint equation. Remarkably, in our case, such an $f(t)$–$a(t)$ pair does exist. Moreover, the smooth transition of the scale factor solution to GR in the limit $\alpha_a \to 0$ is itself nontrivial, since no such condition was imposed by hand but instead arises directly from the solutions of the relevant equations. Such healthy behavior is necessary for the continuity of physical theories but is not mathematically guaranteed (as exemplified by Pauli–Fierz massive gravity), which motivates our analysis of cosmological evolution. While the existence of discontinuous or singular solutions to ordinary differential equations is well known, we include a simple illustrative example in a footnote for the reader’s clarity.\footnote{\label{example} Consider the differential equation $\beta \dfrac{dy(t)}{dt} =y(t)+t$ where $x\in \mathbb{R}, t\geqslant 0$ and $\beta$ is a constant. The initial condition is selected as $y(0)=0$. For $\beta \neq 0$, the solution is obtained as $y_{\beta}(t)=-t-\beta+\beta e^{t / \beta}$. In the limit $\beta\rightarrow 0^+$, the solution diverges as $ y_{\beta}(t)\rightarrow \infty^+$ for a finite $t$. However, if we set $\beta=0$ in the differential equation, the solution is, trivially, $y_{0}(t)=-t$. This is an example of discontinuous (or singular) behaviour at around $\beta=0$.} 


\subsection{Matter\label{sec:matterexact}}

 Motivated by our success in the radiation case and by the corresponding zeroth-order perturbative solution given in Eq.~(\ref{matterperturbative}), to find exact solutions for the matter-only universe, one could try the following ansatz
\begin{equation}
\label{ansatzmdu}
f(\tau) = \tau^n, \quad a(\tau) = \left( \frac{3\tau}{2(1+ B(\alpha_a))}\right)^{\frac{2}{3}}\;, 
\end{equation}
where, again, the form of $a(\tau)$ reflects the ideal case where the time dependence is the same as GR.  Plugging it into (\ref{cec}), we determine that $n=0$, yielding a constant $f(\tau)$, thus a trivial solution. Therefore, we discard (\ref{ansatzmdu}). We examined a wide range of alternative ansatzes; however, these attempts did not yield an analytical solution.

This leads us to numerical computations. 
The benchmark points (BPs) used throughout the paper are listed in Table~\ref{BPs}. Even though we expect some of the coefficients (at least $\alpha_3$~\cite{Anber:2009qp}) to be extremely small, we choose our benchmark points, for illustrative purposes, in larger values than anticipated. BP1 and BP2 are in the same order of magnitude, close to the nonperturbative limit. BP3 is in the $O(10^{-3})$ range.
 BP1 and BP2 are chosen considering the conditions for a good linearized theory, discussed below Eq.~(\ref{constraint2}). BP1 satisfies both condition I ($ \alpha_1+\alpha_2-3 \alpha_3-2 \alpha_4 \geqslant 0$) and condition II ($ \alpha_3 +\alpha_4= 0$),  BP2 satisfies neither. BP3 satisfies only the former condition. Note that BP3 is chosen only as an example set for much smaller values than the first two, and it will not yield extra information regarding the effects of non-satisfaction of the conditions for the regular linearized model.  This is because, as we will see, even cases with BP2 lead to no issues in cosmological evolution with respect to BP1; namely, the problems occur either in both BP1 and BP2 (as we will see in the cases that include vacuum), or neither of them (as in the cases containing no vacuum). Cases with BP3 (which satisfy only condition I) do not exhibit any problematic behaviors in any of the cases, indicating that the issues are about the magnitude of the benchmark points and are independent of whether or not the conditions above are satisfied. Similarly, even though not shown here, a BP that satisfies only condition II does not cause problems as long as they are small enough (as we have also checked the smaller versions of BP1 and BP2). Namely, the nonsatisfaction of these conditions does not cause instabilities in the behavior of $a(\tau)$. We have tried many other benchmark sets and have not encountered such a problem.

\begin{figure}[ht!]
\captionsetup[subfigure]{labelformat=empty}
\centering
\vspace{-1.5cm}
\hspace{-0.8cm}
\begin{tabular}{lll}
\subfloat[ a) ]{\label{MatterBP1af}\includegraphics[width=5.4cm]{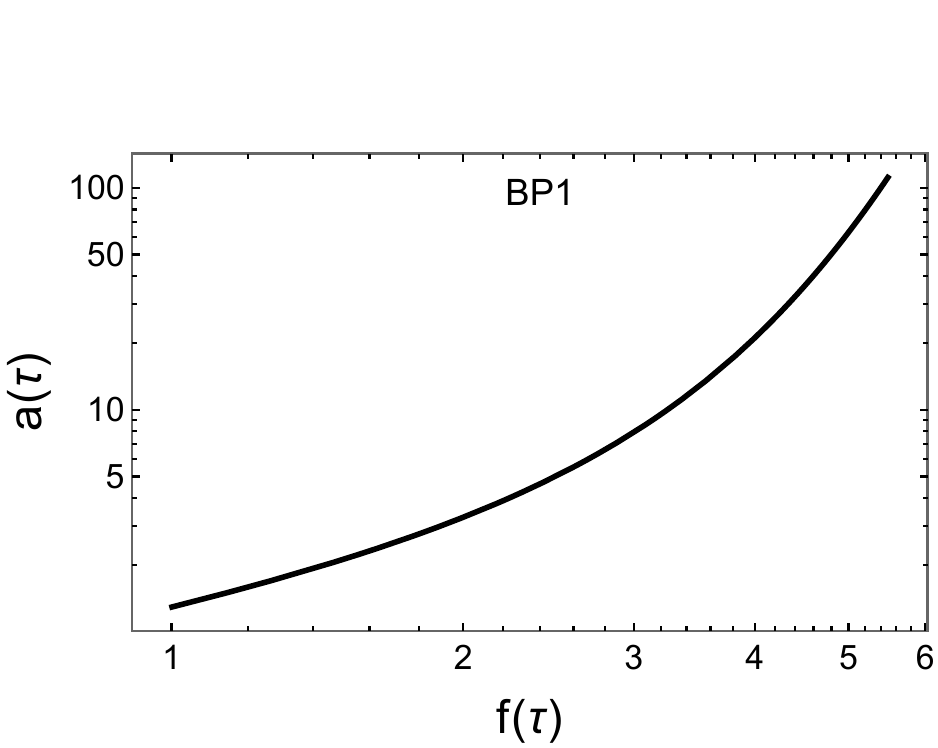}} \quad
\subfloat[  b) ]{\label{MatterBP2af}\includegraphics[width=5.4cm]{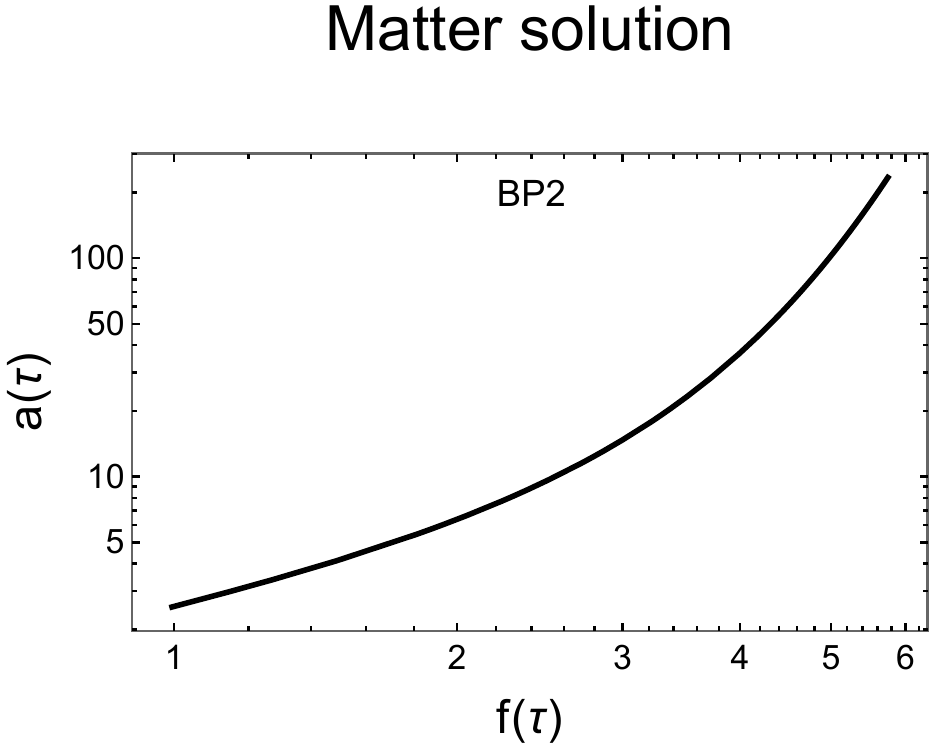}}\quad
\subfloat[  c)  ]{\label{MatterBP3af}\includegraphics[width=5.4cm]{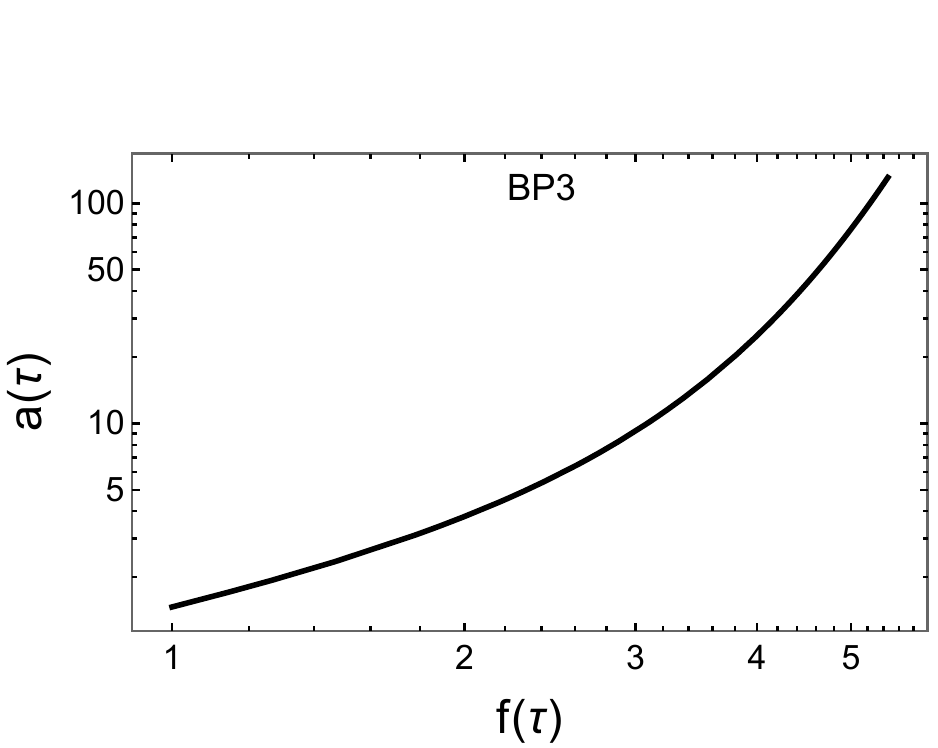}}
\end{tabular}
\caption{\label{Matterplotsaf} Logarithmic plots of the numerical solutions of $a(\tau)$ and $f(\tau)$ in the theory with broken diffeomorphism for the matter-only case for three benchmark points (BPs), given in Table~\ref{BPs}. The dimensionless time parameter $\tau$ is chosen to run from 1 to $1000$.
\vspace{0.6cm}
} 
\end{figure}

The numerical solutions for this system are given in Fig.~\ref{Matterplotsaf} in the form of $a(\tau)$-$f(\tau)$ parametric plot. We prefer to set the initial conditions for $a(\tau)$ near those of the GR solution. Since we don't have the analytic solution for this case, we select our initial conditions based on the analytic radiation-only solution, given in Eq.~(\ref{radsolution}).  We see that $f(\tau)$ is not an overall constant (even though it evolves into a slowly running function at later times), and hence, non-trivial: the solution for $a(\tau)$ results from non-trivial dynamics of the theory with broken diffeomorphism, as desired. 
 
\begin{figure}[h!]
\captionsetup[subfigure]{labelformat=empty}
\centering
\vspace{-0.3cm}
\hspace{-0.8cm}
\begin{tabular}{lll}
\subfloat[a)  ]{\label{MatterBP1GR}\includegraphics[width=5.4cm]{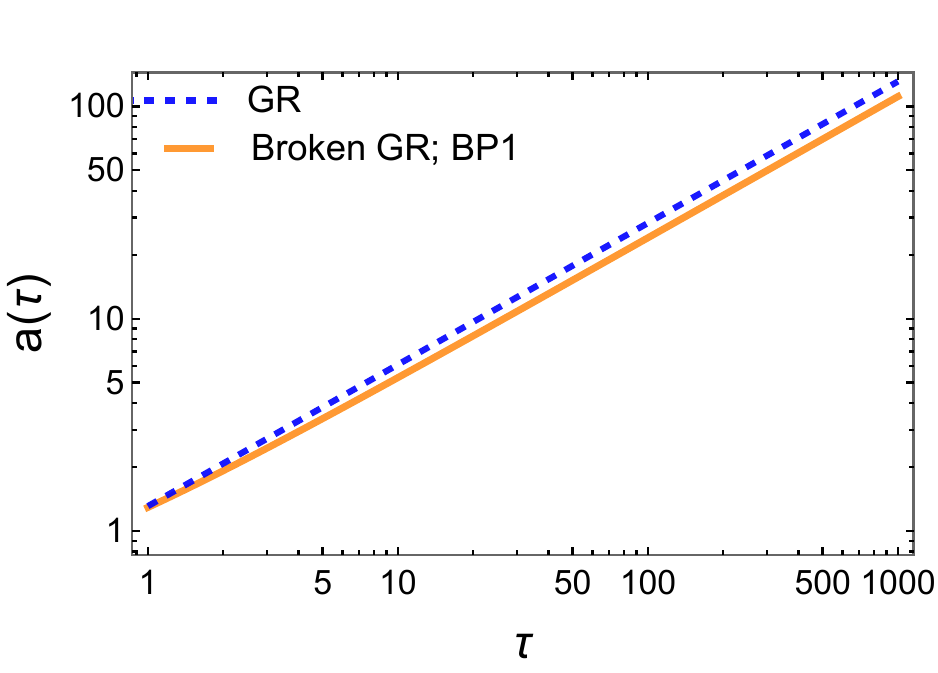}} \quad
\subfloat[b)]{\label{MatterBP2GR}\includegraphics[width=5.4cm]{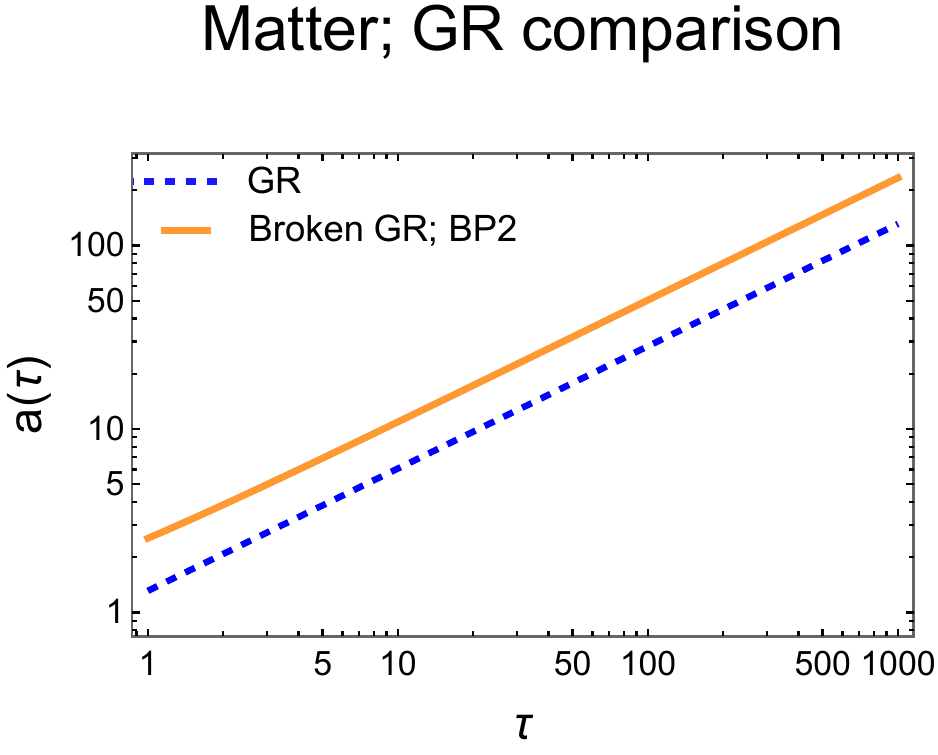}}\quad
\subfloat[c)]{\label{MatterBP3GR}\includegraphics[width=5.4cm]{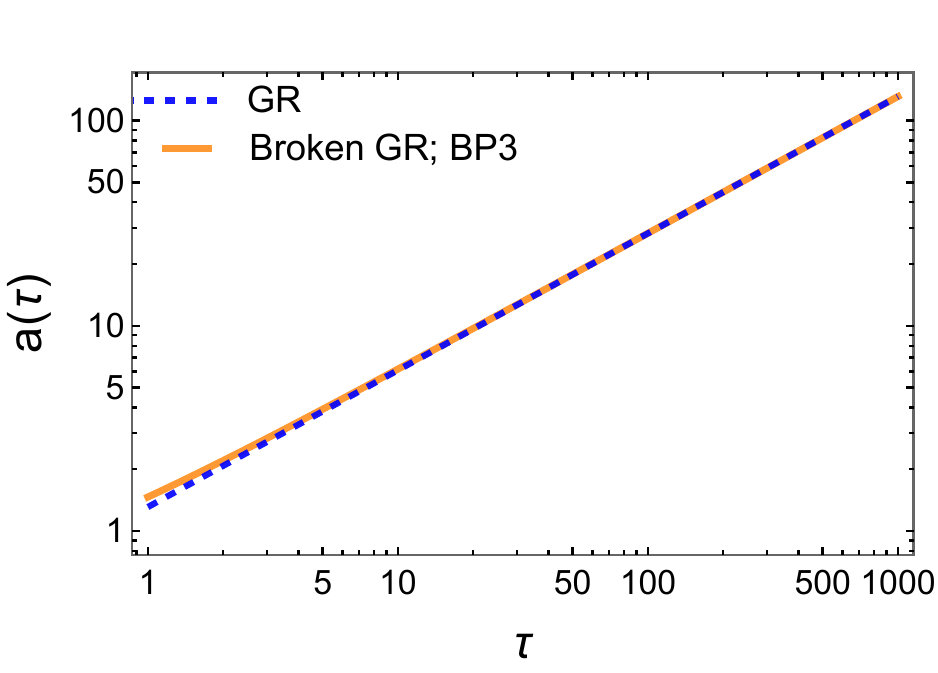}}\vspace{-0.2cm}\\
\subfloat[ d) ]{\label{MatterBP1GRratio}\includegraphics[width=5.4cm]{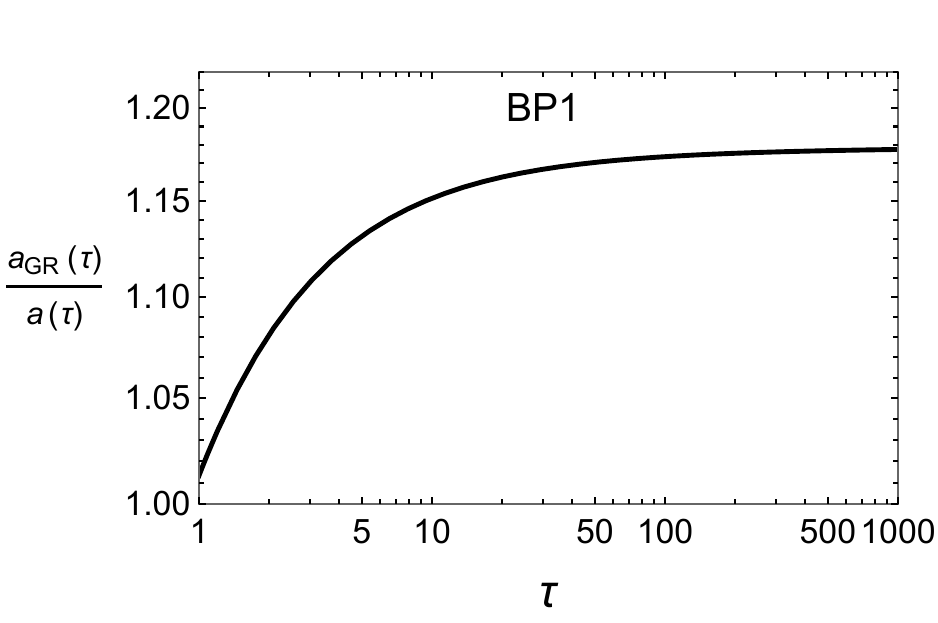}} \quad
\subfloat[  e) ]{\label{MatterBP2GRratio}\includegraphics[width=5.4cm]{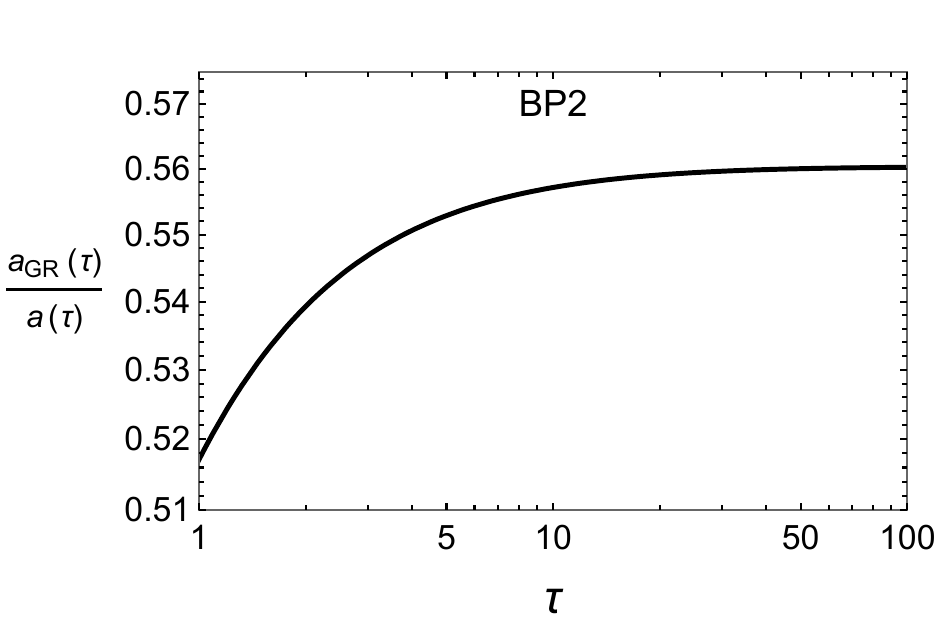}}\quad
\subfloat[  f)  ]{\label{MatterBP3GRratio}\includegraphics[width=5.4cm]{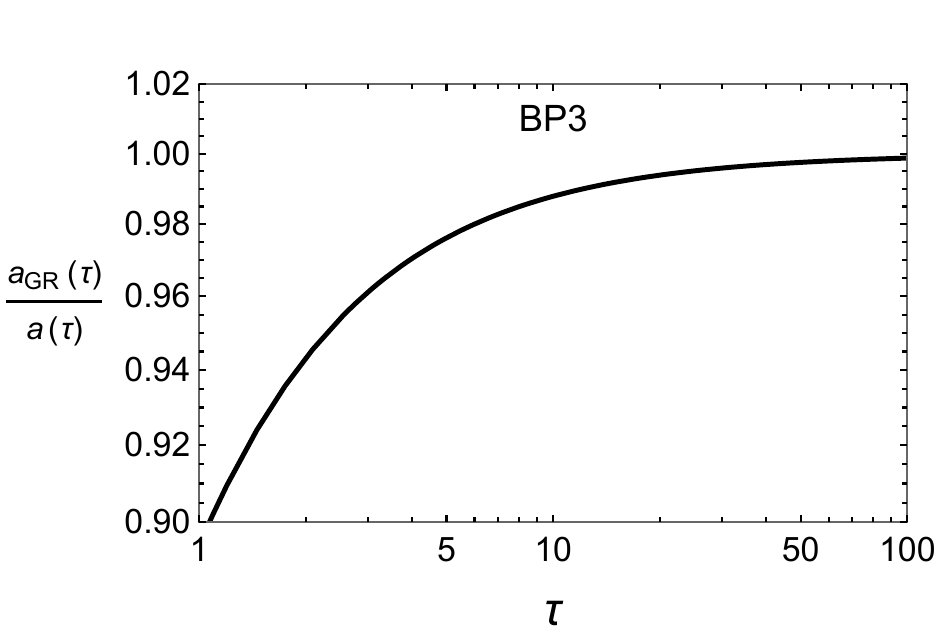}}\vspace{-0.4cm}\\
\quad\quad\quad\quad\quad\quad\quad\quad\quad\quad\quad\quad\quad\hspace{0.5cm}\subfloat[  i)  ]{\label{MatterBP3GRHubble}\includegraphics[width=5.4cm]{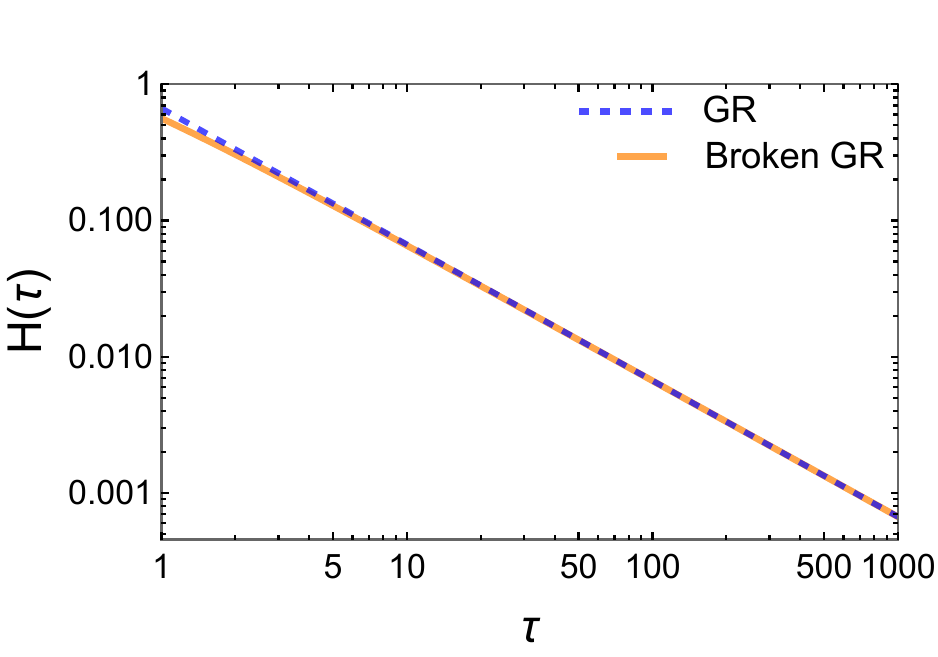}}
\end{tabular}
\caption{\label{MatterplotsGR}  In the top and mid panels, the comparison of the $a(\tau)$ solutions for the matter-only fluid in the theory with broken diffeomorphism (referred to as broken GR in the plots) to the GR case, $a_{\tiny{\mbox{GR}}}(\tau)=(\frac{3}{2}\tau)^{2/3}$, is displayed for each BP, given in Table~\ref{BPs}. In the bottom panel, we compare the Hubble parameters, which are the same for all three cases.}  
\end{figure}

In Fig.~\ref{MatterplotsGR}, we compare the diffeomorphism-violating theory and GR for the BPs, referred to above, as logarithmic plots. In the top and mid panels, the comparisons of the scale factors $a(\tau)$ are given in two different ways. We don't display the corresponding $f(\tau)$ solutions in Fig.~\ref{MatterplotsGR} since this figure is reserved for the GR comparison, and there is no GR counterpart for $f(\tau)$. (The $f(\tau)$ solutions can be seen in the mid panel of the next figure, where the orange lines correspond to the $f(\tau)$ solutions associated with the $a(\tau)$'s in Fig.~\ref{MatterplotsGR}.)  The initial conditions for $a(\tau)$ are chosen near those of the GR solution. The solutions begin with a slightly different slope (i.e., with a different time dependence) than the GR solution. However, they quickly settle into the required form of behavior: a solution slightly different from that of GR but with the same time dependence. For smaller $\alpha_a$, the solution gets closer and closer to the GR case. If a solution begins close enough to the GR solution, it eventually merges into the GR solution, as seen in the BP3 case. If not, it will still have (almost) the same behavior as the GR solution (the same time dependence) but differ from it with a constant factor due to the effects of the $\alpha_a$ dependence. Below, we will consider this point in relation to $f(\tau)$ settling down to a (nearly) constant behavior. Finally, in the bottom panel of Fig.~\ref{MatterplotsGR}, we display the Hubble parameter comparisons. All three cases have the same Hubble parameter for the same initial conditions. This can also be seen in the bottom panel of the next figure (orange lines). From all these plots, we observe that the $a(\tau)$ solutions are in the desired form given in Eq.~(\ref{ansatzmdu}).

\begin{figure}[h!]
\captionsetup[subfigure]{labelformat=empty}
\centering
\hspace{-0.8cm}
\begin{tabular}{lll}
\subfloat[ a) ]{\includegraphics[width=5.05cm]{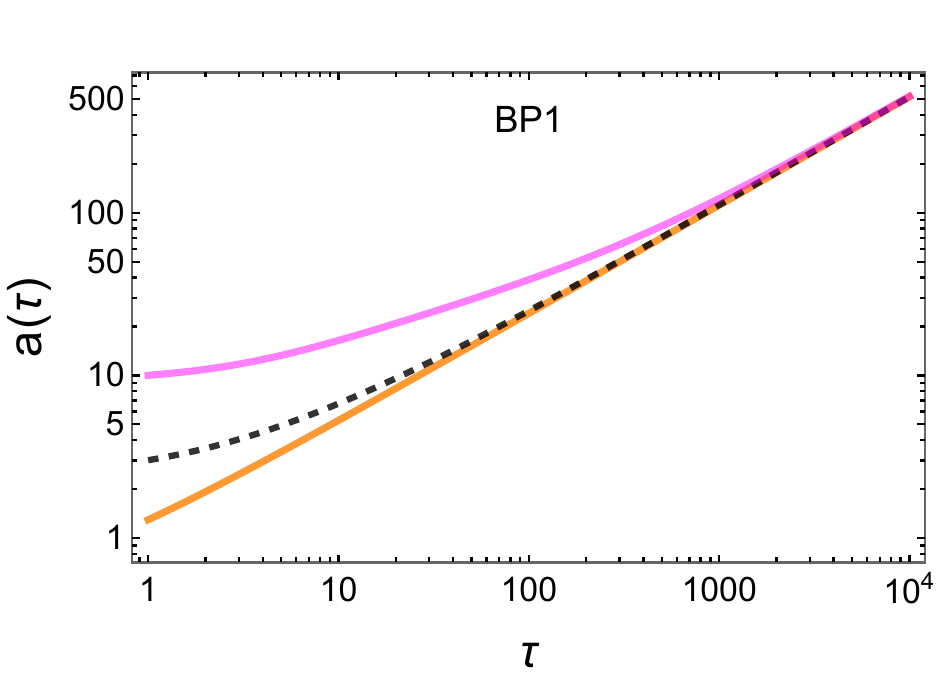}}\vspace{-0.2cm} \quad \hspace{0.03cm}
\subfloat[  b) ]{\includegraphics[width=5.05cm]{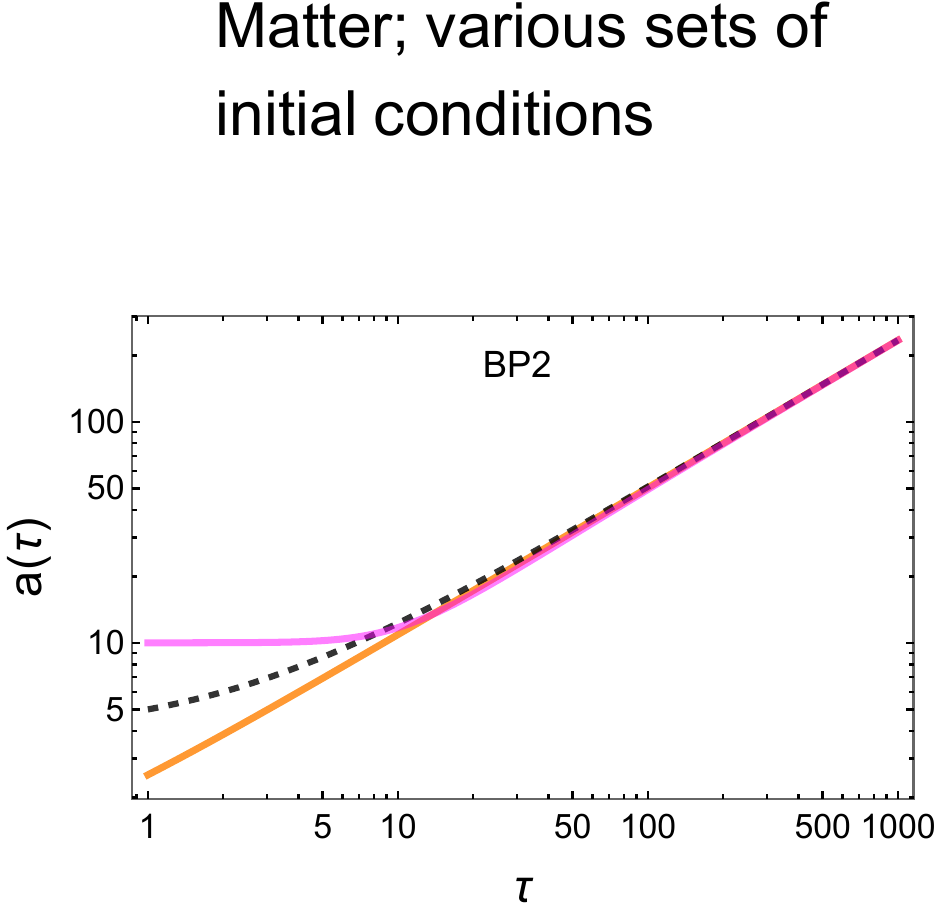}}\quad
\subfloat[  c)  ]{\includegraphics[width=5.05cm]{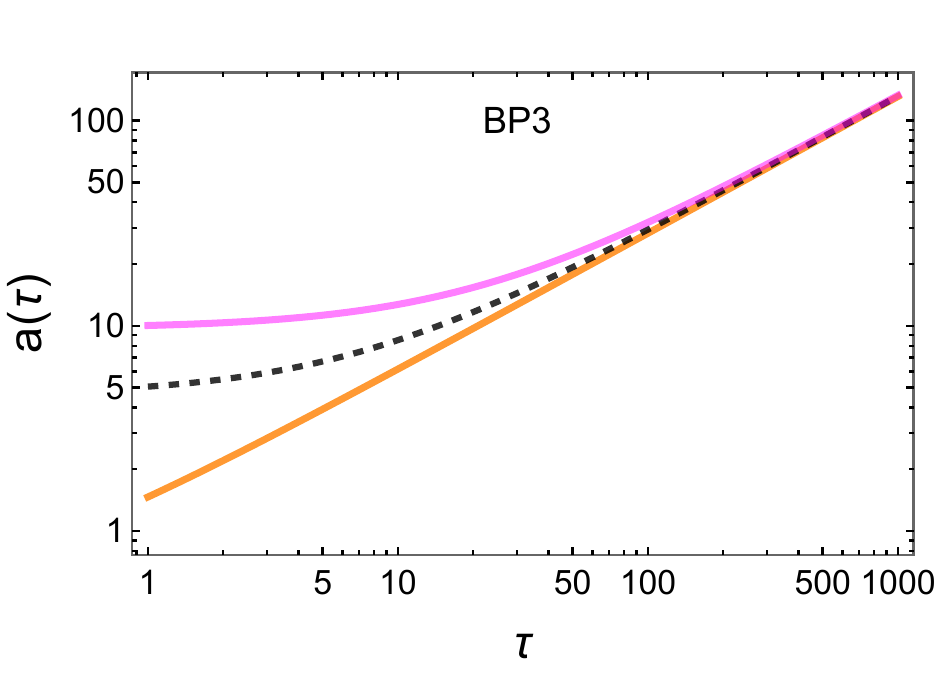}}\\
\subfloat[ d) ]{\includegraphics[width=5.1cm]{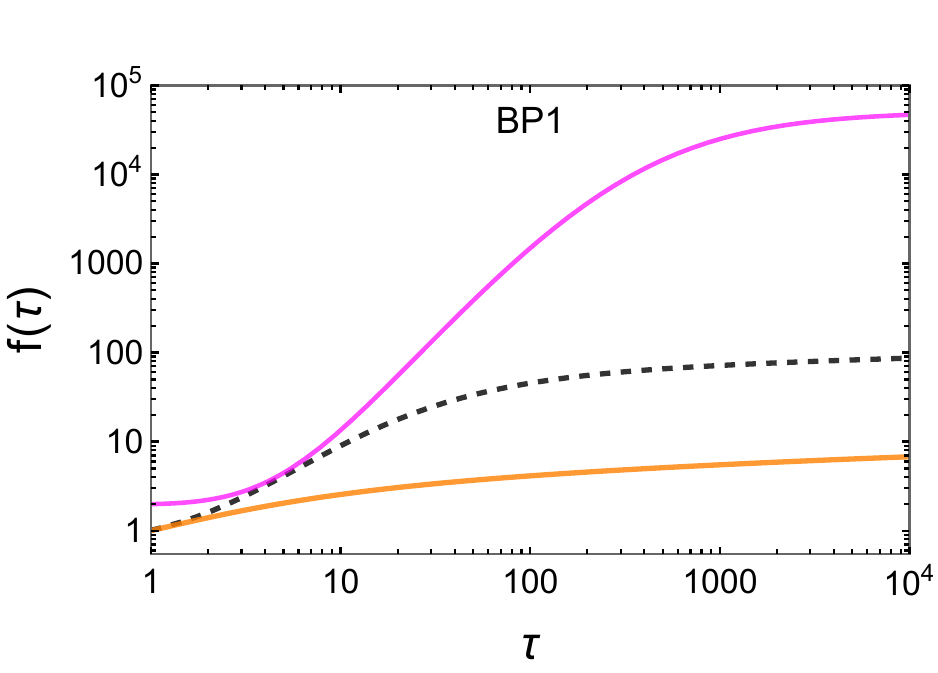}} \vspace{-0.3cm}\quad\hspace{0.03cm}
\subfloat[  e) ]{\includegraphics[width=5.2cm]{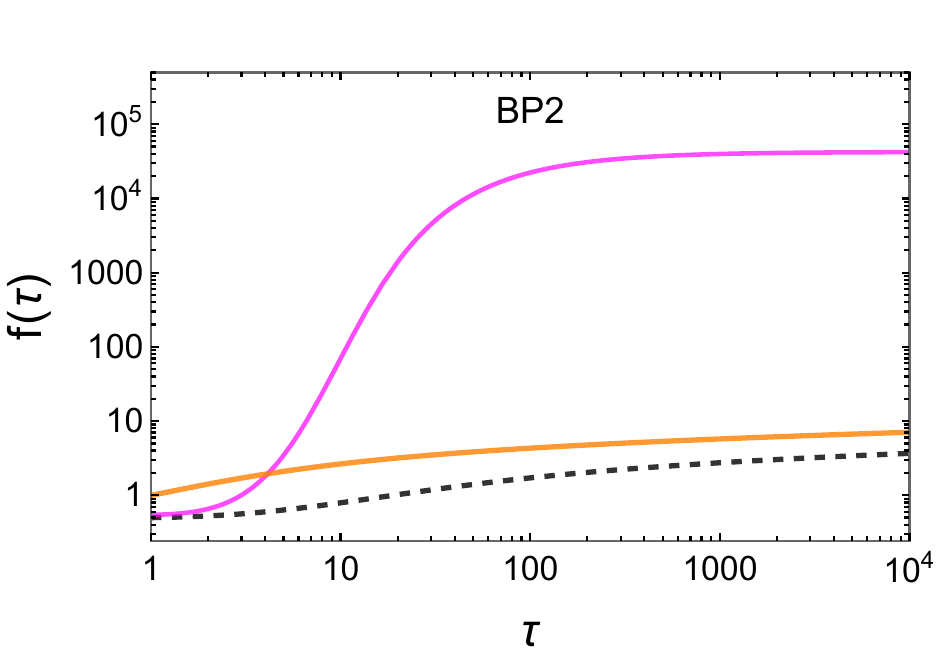}}\quad
\subfloat[  f)  ]{\includegraphics[width=5.0cm]{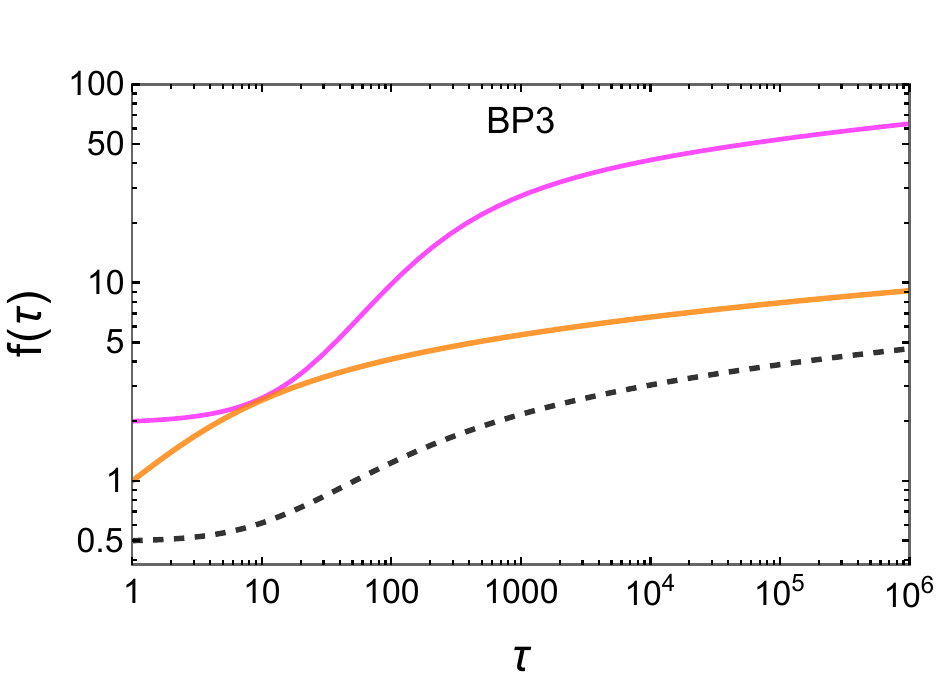}}\\
\hspace{-0.2cm}\subfloat[ g) ]{\includegraphics[width=5.4cm]{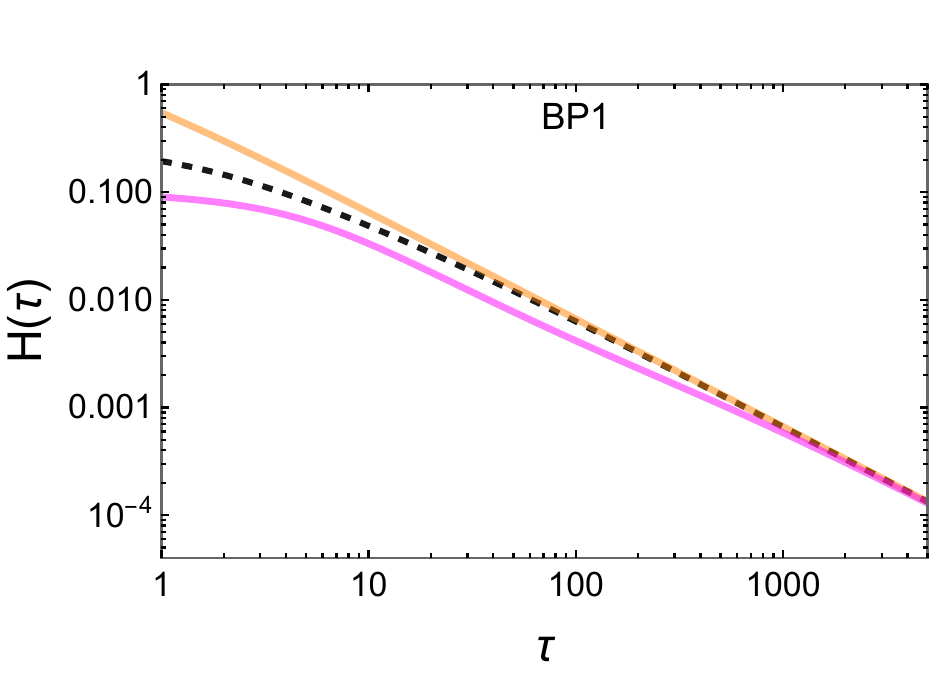}} \hspace{0.2cm}
\subfloat[  h) ]{\includegraphics[width=5.4cm]{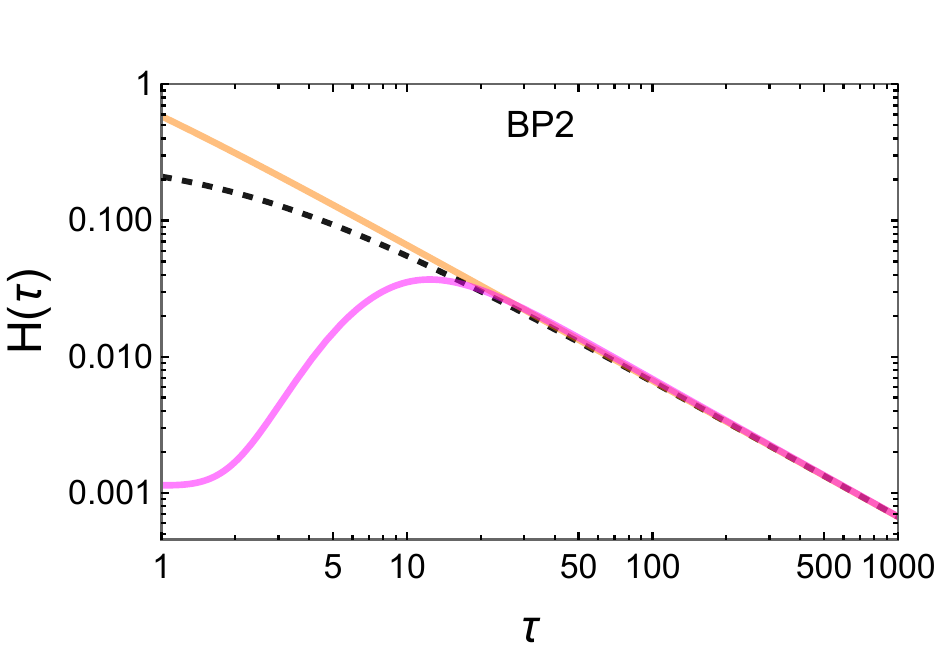}} \hspace{0.1cm}
\subfloat[  i)  ]{\includegraphics[width=5.3cm]{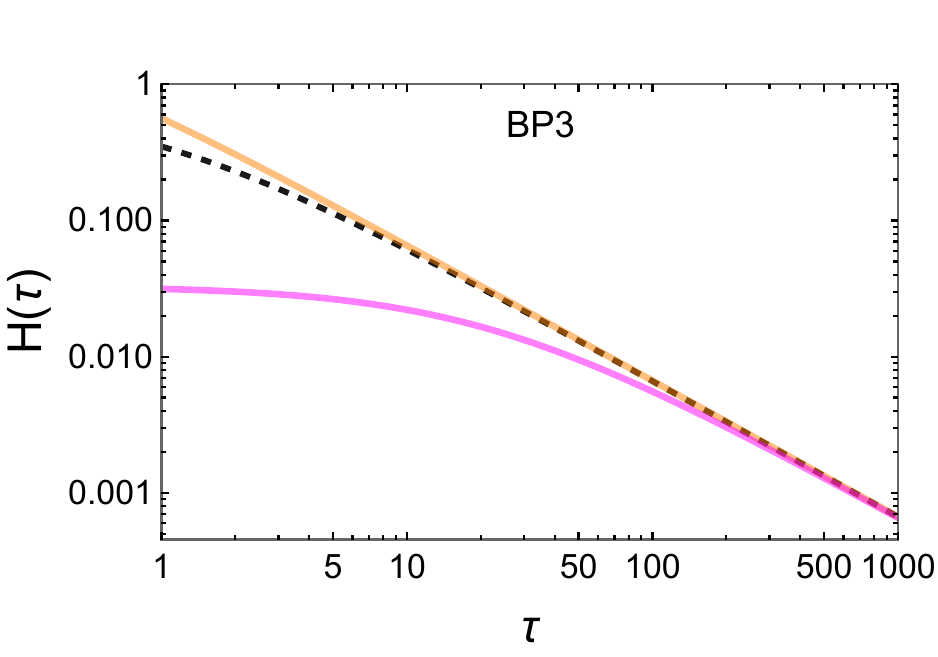}}
\end{tabular}
\caption{\label{Matterplotsinitialconditions} Comparison of $a(\tau)$, $f(\tau)$, and the Hubble parameter $H(\tau)$ for the matter-only case for different initial conditions, denoted by pink, orange, and dashed lines, for three benchmark points (BPs), given in Table~\ref{BPs}. Note that the orange lines correspond to the solutions in the previous figure.} 
\end{figure}

In Fig.~\ref{Matterplotsinitialconditions}, we display $a(\tau)$, $f(\tau)$, and the Hubble parameter $H(\tau)$ for substantially different conditions. The orange line in the top panel is the same as in the top panel of Fig.~\ref{MatterplotsGR}. As we see, the solutions quickly settle in the same behavior. As noted above, the numerical behavior of  $a(\tau)$ suggests that the solution is in the form given in Eq.~(\ref{ansatzmdu}).  The reason that an analytical solution (in which both $f(\tau)$ and $a(\tau)$ are in the form in Eq.~(\ref{ansatzmdu})) cannot be obtained is that $f(\tau)$, as displayed in the mid panel of Fig.~\ref{Matterplotsinitialconditions}, is not represented by a single function; it is a function with slowly varying behavior in different parameter ranges. As mentioned above, $f(\tau)$ is not an overall constant as required for non-trivial dynamics.  The location where $f(\tau)$ settles into a constant value signals the location where the $a(\tau)$ solution becomes GR-like (i.e. having the same time dependence), as mentioned above. Recall that $f(\tau)$ plateauing at a constant value does not cause a problem regarding the triviality; for a solution to be trivial in this sense, $f(\tau)$ has to be constant at all parameter ranges, in which case one can obtain GR by redefining an effective Newton's constant, as discussed below Eq.~(\ref{cec}). Finally, in the bottom panel, the corresponding Hubble parameters are shown, and as expected, they all eventually merge into the same behavior. \vspace{1.5cm}

\begin{table}[htbp]
\vspace{-1cm}
\caption{Benchmark Points}
\centering
\begin{tabular}{c c c c c }
\hline\hline
BP$\#$ & $\alpha_1$ & $\alpha_2$ & $\alpha_3$ & $\alpha_4 =\alpha_5$ \\ [0.5ex]
\hline 
1 & $0.25$ & $0.15$ & $0.20$ & $-0.20$ \\
2 & $-0.15$ & $-0.18$ & $\;\;\;0.12$ & $\;\;\;0.10$ \\
3 & $0.40\times 10^{-3}$ & $0.25\times 10^{-3}$ & $0.10\times 10^{-3}$ & $0.15\times 10^{-3}$  \\  [1ex]
\hline  
\end{tabular}
\vspace{2cm}
\label{BPs}
\end{table}

\newpage
\quad\
\newpage
\subsection{Cosmological Constant\label{sec:vacuumexact}}

\label{exactcc}
 In the case of the cosmological constant (or the vacuum-only), the only solution we have found is in the following form. 
\begin{equation}
\label{ansatzcc}
f(\tau) =  e^{- \frac{\tau}{C +  D(\alpha_a)}},   \quad  a(\tau) = e^{\frac{C\;\tau}{C+ D(\alpha_a)}}\;,
\end{equation}
where we have two unknowns $C$ and $D(\alpha_a)$. Again, this form of ansatz is motivated by the corresponding zeroth-order solution given in Eq.~(\ref{vacuumperturbative}). The dimensionless time is  $\tau= \sqrt{8\pi G \rho_\Lambda/3}\; \tilde{t}$. Substituting it first into the constraint equation (\ref{cec}) yields
\begin{equation}
-C+3 C\;^2=0\quad \longrightarrow \quad C=\frac{1}{3}\;,
\end{equation}
where the other solution, $C=0$, is ignored. Now it remains to find the other one $D$ with the help of field equations (\ref{EEc1}) and (\ref{EEc2}) for $\omega =-1$. Our ansatz (\ref{ansatzcc}), together with the solution $C=\frac{1}{3}$, puts them in such a form that their LHSs are also proportional with the same factor. Thus, we are left with one unknown and a single equation. Substituting (\ref{ansatzcc}) into either (\ref{EEc1}) or  (\ref{EEc2}) and imposing that $C+D>0$ determines $D$ in terms of $\alpha_a$ as 
\begin{equation}
\label{Dccalpha0}
D = \frac{1}{3}\left(-1 + \sqrt{1+2\alpha_1+3\alpha_2}\right) \;.
\end{equation}
Therefore, we have
\begin{eqnarray}
\label{ccsolution}
f(\tau) =  e^{- \frac{3\tau}{\sqrt{1+2\alpha_1+3\alpha_2}}},   \quad  a(\tau) = e^{\frac{\tau}{\sqrt{1+2\alpha_1+3\alpha_2}}}\;.
\end{eqnarray}
The Hubble constant is $H=a'/a=1/\sqrt{1+2\alpha_1+3\alpha_2}$ in dimensionless units. As expected, in the limit $\alpha_a$'s $\rightarrow 0$, the solution for the scale factor $a(\tau)$ reduces to that in GR. Since in this limit, we have the full diffeomorphism invariance, $f(\tau)$ disappears from the field equations, and thus, its exact form is not important as long as it does not have a singularity.  

Notice that the ratio $a_{\tiny{\mbox{GR}}}(\tau)/a(\tau)$ is time-dependent, unlike in the radiation-only and matter-only cases. Even for small $\alpha_a$, we have rapidly changing deviations from GR;  $a_{\tiny{\mbox{GR}}}(\tau)/a(\tau)\simeq e^{\tau(\alpha_1+3\alpha_2/2)}$.~Therefore, even small violations can introduce great deviations from GR, and this could cause problems for a diffeomorphism-violating theory for a universe composed of vacuum energy only. Fortunately, for more physical cases, namely multi-fluid universes, including the ones containing the vacuum as a component, we won't have such an issue.

\section{Radiation $+$ Matter} 
\label{sec:radmatnumerical}

In this section, we explore another commonly considered case, a small amount of matter on the top of a radiation-dominated universe, which is a good approximation for the early universe, where the cosmological constant can be ignored. This case is useful for understanding the transition from radiation to matter domination, which is sometimes difficult to observe when the vacuum is included since the vacuum takes over rather quickly, depending on the parameter choice, as we will see in the upcoming sections. 

In the radiation-matter system, the field equations (\ref{EEc1}) and (\ref{EEc2}) become
\begin{subequations}
\begin{align}
\label{EEcradmat1}
\frac{ L}{2}\frac{(a')^2}{a^2} - K \left( \frac{1}{6} \frac{(f')^2}{f^2} - \frac{a' f'}{af} - \frac{1}{3}\frac{f''}{f} \right)   &= \frac{8\pi G}{3}\rho_{r0}\left(\frac{1}{a^4}+\frac{r_1}{a^3} \right), \\
 \frac{K}{2} \frac{f'^2}{f^2} - \frac{L}{2} \left( \frac{a'^2}{a^2} + 2 \frac{a''}{a} \right)  &= \frac{8\pi G}{3} \;\rho_{r0} \left(\frac{1}{ a^4} \right)\;,
\label{EEcradmat2}
\end{align}
\end{subequations}
where $r_1\equiv\rho_{m0}/\rho_{r0}$. In addition, we have the constraint equation (\ref{cec}). Two of these three equations are independent, as usual. The derivative is with respect to the cosmological time. The dimensionless time we work with is defined as $ \tau\equiv \sqrt{\frac{8\pi G \rho_{r0}}{3}}\; \tilde{t}$. For concreteness, we set $\rho_{r0}=0.9 \;\rho_{\mathrm{tot},0}$ and $\rho_{m0}=0.1 \;\rho_{\mathrm{tot},0}$, with $\rho_{\mathrm{tot},0}=\rho_{r0}+\rho_{m0}$ being the total initial energy density of this  system.  
\vspace{1.0cm}

\begin{figure}[hbt!]
\captionsetup[subfigure]{labelformat=empty}
\centering
\hspace{-0.8cm}
\begin{tabular}{lll}
\subfloat[a)   ]{\label{PlotRadMatBP1a}\includegraphics[width=5.2cm]{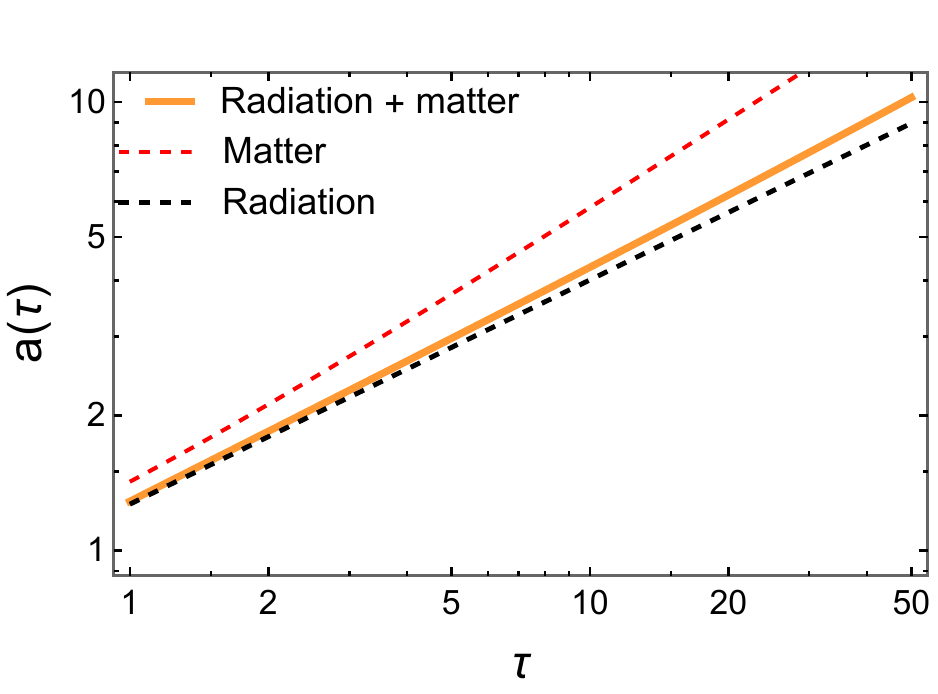}} \quad
\subfloat[b) ]{\label{PlotRadMatBP1b}\includegraphics[width=5.5cm]{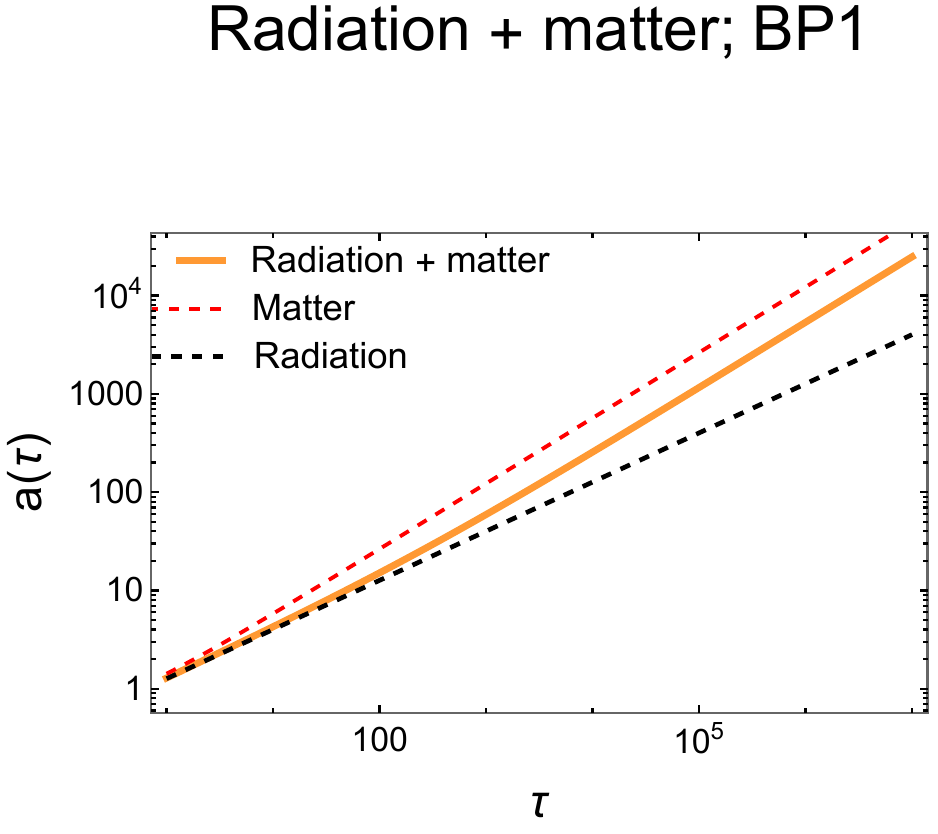}}\quad
\subfloat[c) ]{\label{PlotRadMatBP1c}\includegraphics[width=5.2cm]{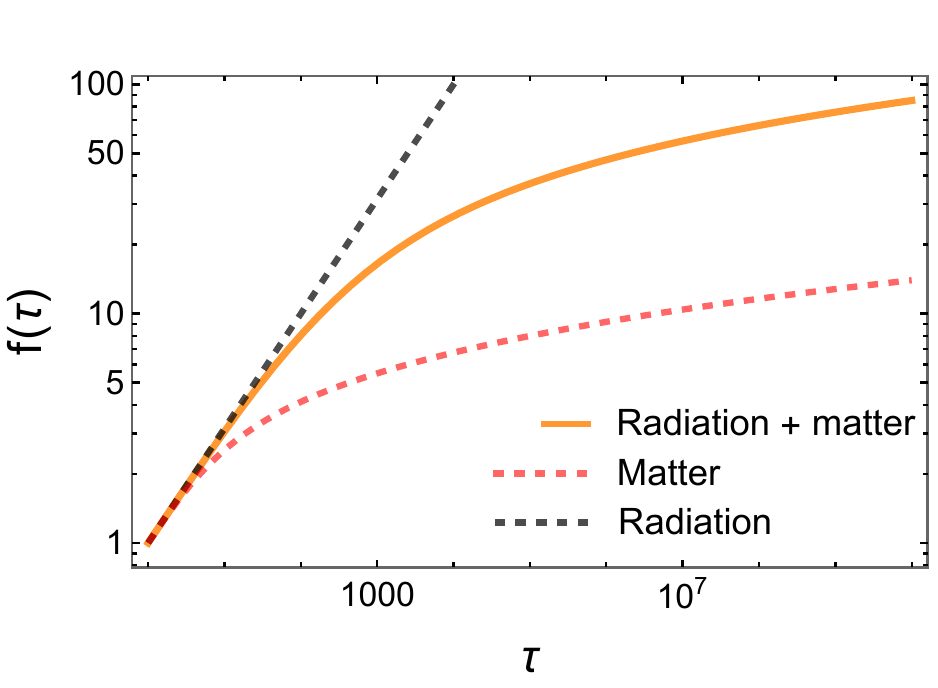}}\vspace{-0.9cm}\\
\quad\quad\quad\quad\quad\quad\quad\quad\quad\quad\quad\quad\quad\quad
\subfloat[d) ]{\label{PlotRadMatBP1d}\includegraphics[width=5.5cm]{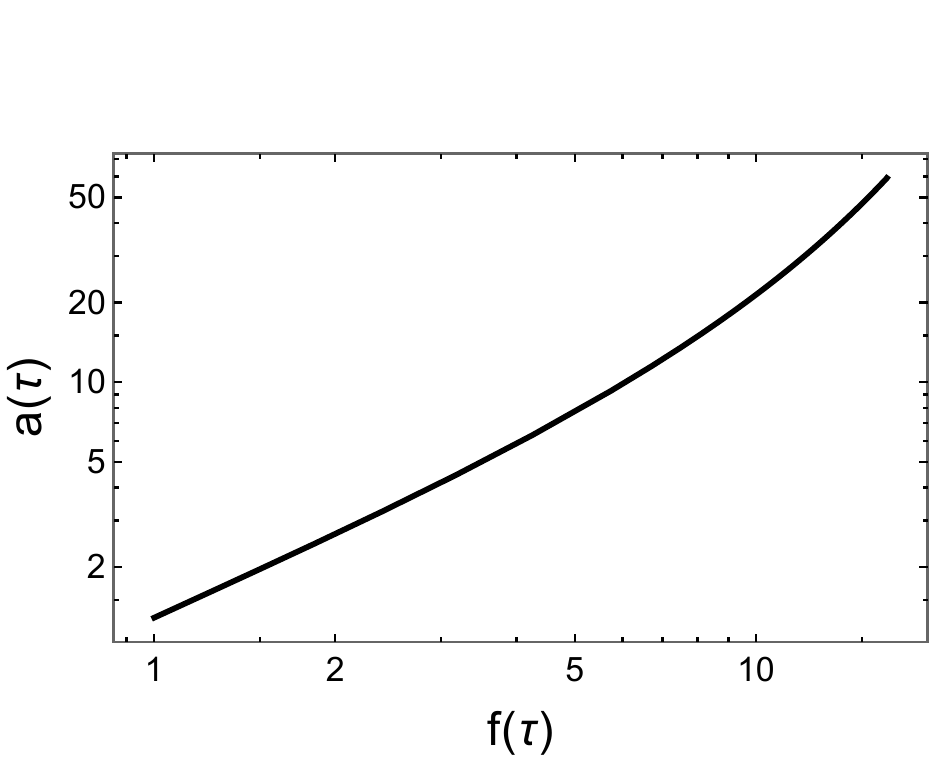}}
\end{tabular}
\caption{\label{PlotRadMatBP1} Logarithmic plots of the $a(\tau)$ and $f(\tau)$ solutions in theory with broken diffeomorphism (i.e. broken GR) for the case of a two-component fluid; a radiation-dominated system with small amount of matter. The plots are displayed for BP1, given in Table~\ref{BPs}. Figs.~\ref{PlotRadMatBP1a} and \ref{PlotRadMatBP1b} are the same plots in different ranges, given to demonstrate the evolution from the radiation solution through a matter-like realm. The $f(\tau)$ solution is displayed in Fig.~\ref{PlotRadMatBP1c}. In Fig.~\ref{PlotRadMatBP1d}, we display the $a(\tau)$-$f(\tau)$ parametric plot, for the reader's convenience.
}  
\end{figure}

\newpage
The numerical solutions for the scale factor $a(\tau)$ and $f(\tau)$ for the BP1 (given in Table~\ref{BPs})  for the radiation-matter system are shown in Fig.~\ref{PlotRadMatBP1}. The initial conditions for the system are chosen based on the radiation-only analytic solution, given in Eq.~(\ref{radsolution}), since the system in radiation-dominated to begin with.~In Figs.~\ref{PlotRadMatBP1a} and \ref{PlotRadMatBP1b}, we display the scale factor  $a(\tau)$ solution in two different plotting ranges to emphasize the transition from the radiation-like solution to the matter-like one. As expected, the solution at early times begins with a radiation-like behavior (with the same slope as the radiation solution in the logarithmic plot). And it gradually deviates toward the matter-like solution. In Fig.~\ref{PlotRadMatBP1c}, we show the solution of $f(\tau)$, where the transition from the radiation-domination to matter-domination is manifest. Finally, we display the parametric plot in Fig.~\ref{PlotRadMatBP1d}.

\begin{figure}[hbt!]
\captionsetup[subfigure]{labelformat=empty}
\centering
\hspace{-0.8cm}
\begin{tabular}{lll}
\subfloat[a) ]{\label{PlotsmatradGRBP1}\includegraphics[width=5.2cm]{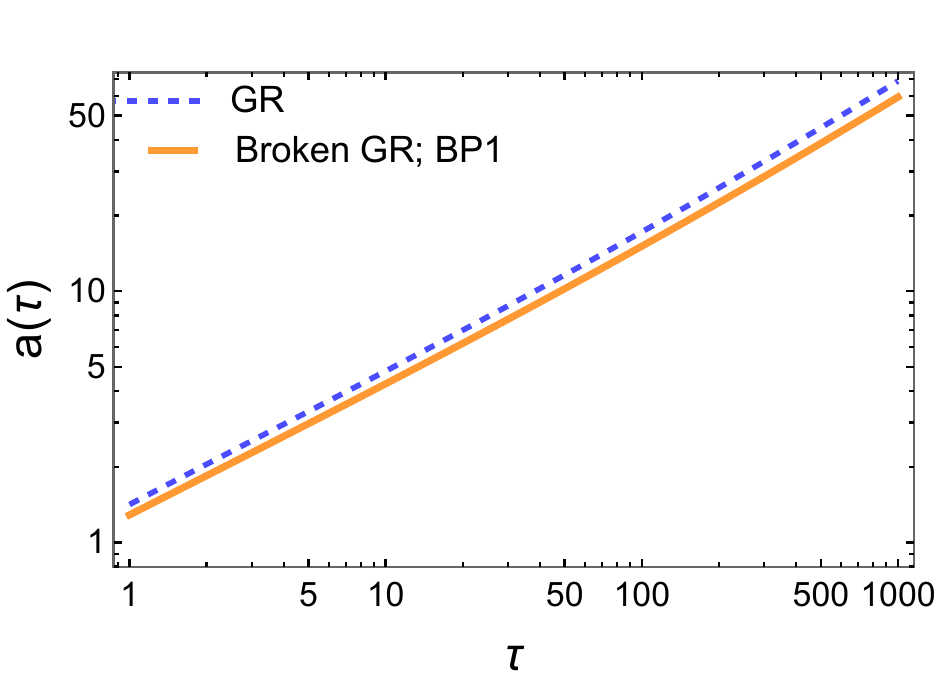}} \quad
\subfloat[b)]{\label{PlotsmatradGRBP2}\includegraphics[width=5.5cm]{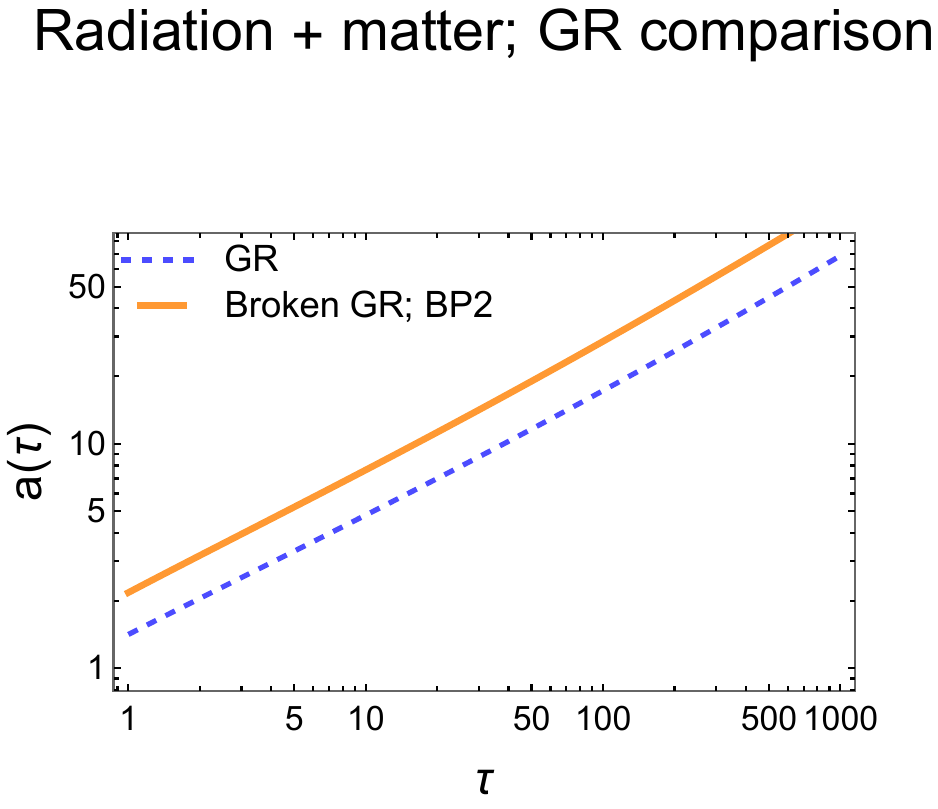}}
\subfloat[c)]{\label{PlotsmatradGRBP3}\includegraphics[width=5.1cm]{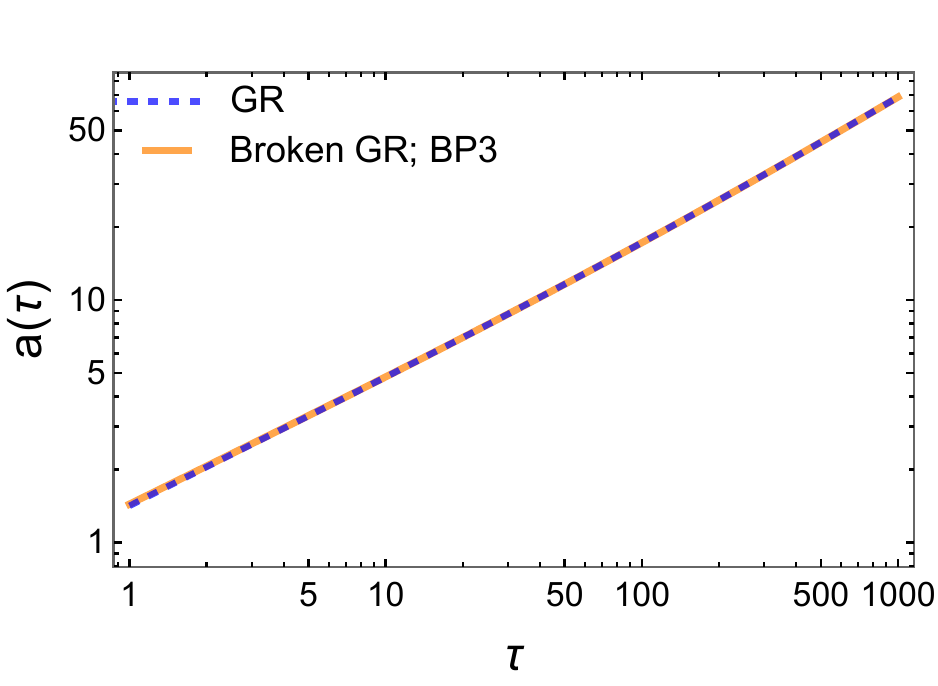}}\vspace{-0.3cm}\\
\hspace{-0.1cm}\subfloat[d)]{\label{PlotsmatradGRBP1ratio}\includegraphics[width=5.2cm]{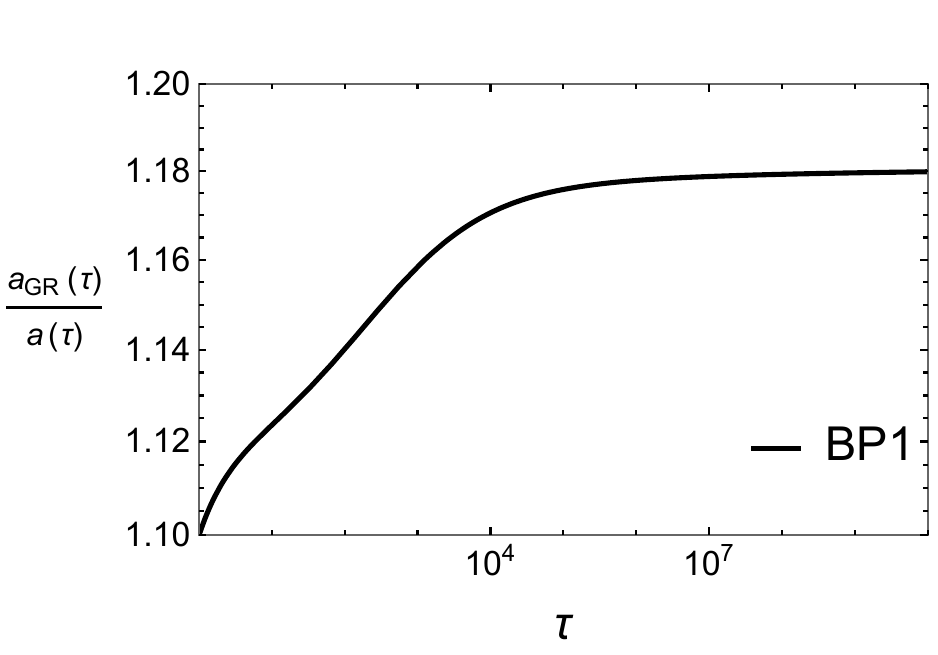}}\hspace{0.1cm}
\subfloat[e)]{\label{PlotsmatradGRBP2ratio}\includegraphics[width=5.3cm]{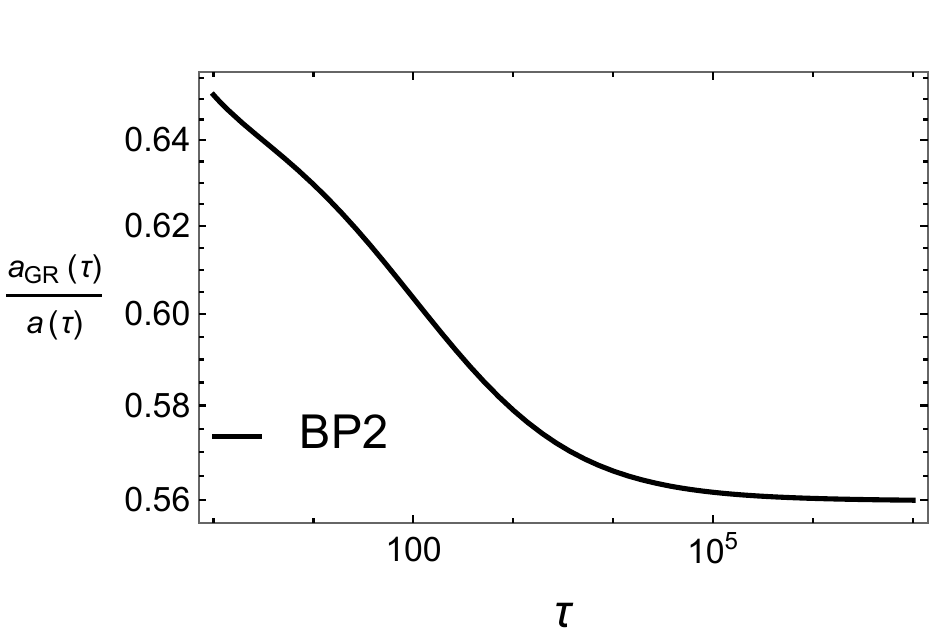}}\hspace{0.1cm}
\subfloat[f)]{\label{PlotsmatradGRBP3ratio}\includegraphics[width=5.35cm]{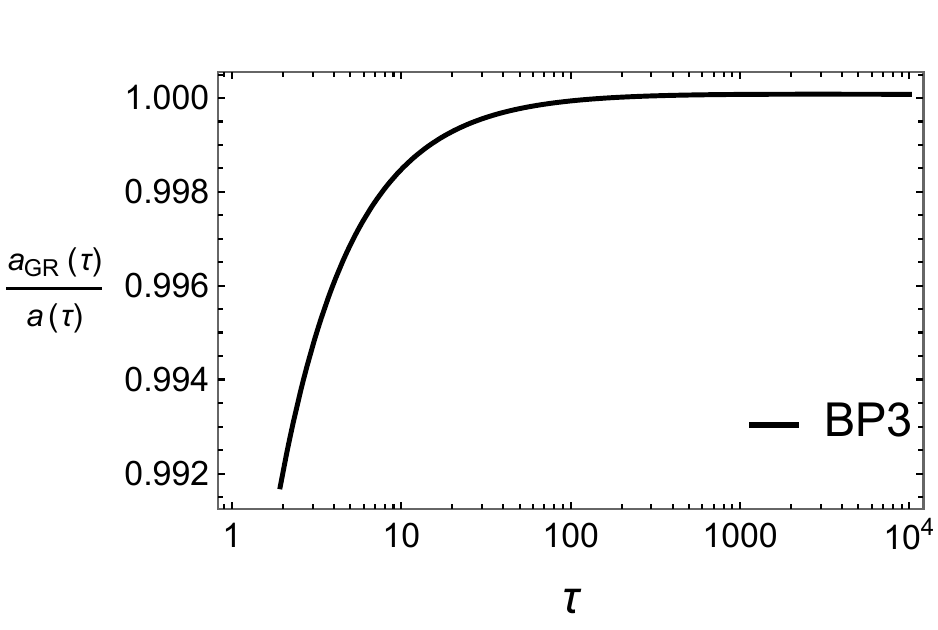}}\vspace{-0.3cm}\\
\quad\quad\quad\quad\quad\quad\quad\quad\quad\quad\quad\quad\quad\hspace{0.1cm}\subfloat[g)]{\label{PlotsmatradGRHubble}\includegraphics[width=5.3cm]{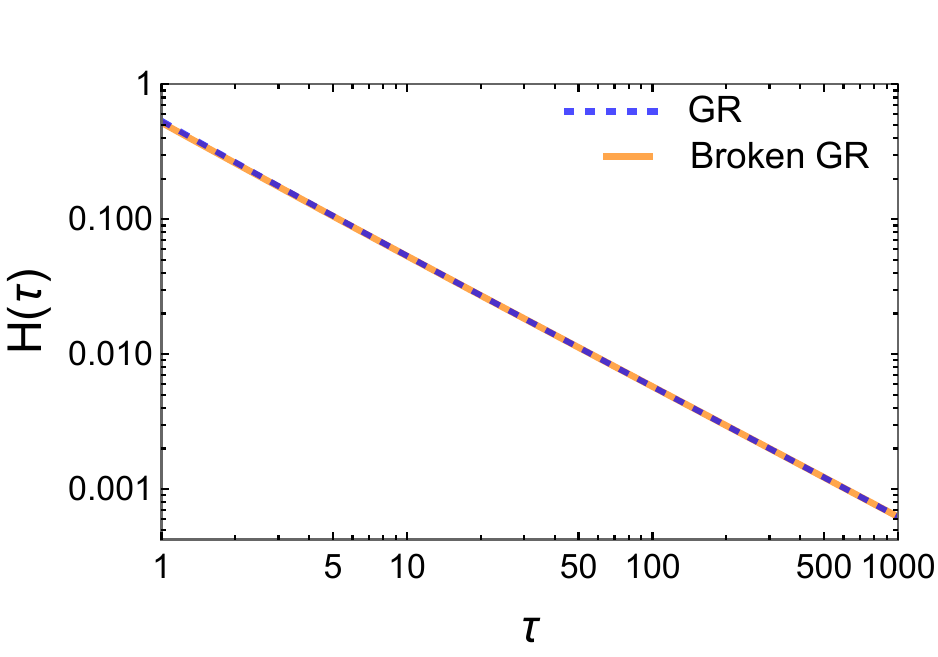}}
\end{tabular}
\caption{\label{PlotscomparisonmatradGR} Comparison of $a(\tau)$ solution in the broken theory to that of GR for different BPs (given in Table~\ref{BPs}) for the radiation-matter case in the top and mid panels. The comparison of the Hubble parameters is displayed at the bottom, which is the same for all three cases.
}   
\end{figure}

In Fig.~\ref{PlotscomparisonmatradGR}, we display the comparison of the scale factor $a(\tau)$ in the radiation-matter system to the GR counterpart, with the initial conditions chosen based on the radiation-only analytic solutions, given in Eq.~(\ref{radsolution}) for the broken case and $a_{\tiny{\mbox{GR}}}(\tau)=\sqrt{2\tau}$ in the regular GR solution. The top and mid panels are different illustrations of the $a(\tau)$ comparison. We observe that the deviation from GR is well-behaved and goes to a constant value at later times.  As in the case of matter-only, for small enough values of $\alpha_a$, the $a(\tau)$ merges with the GR solution, as seen in the BP3 case given in Fig.~\ref{PlotsmatradGRBP3ratio}; if not, it sill becomes GR-like (the same time dependence) but differs from it with an $\alpha_a$ dependent factor, as can be seen for BP1 and BP2 cases, given in Figs.~\ref{PlotsmatradGRBP1ratio} and \ref{PlotsmatradGRBP2ratio}, respectively. $a(\tau)$ becoming GR-like is due to $f(\tau)$ evolving into a constant value (as can be seen from the orange lines in the mid panel of the next figure, Fig.~\ref{Plotscomparisonmatrad}). Finally, in the bottom panel of Fig.~\ref{PlotscomparisonmatradGR}, we give the Hubble parameter, which is the same for all three BPs, indicating that it is independent of $\alpha_a$ coefficients, as in the case of radiation-only and matter-only cases.  This can be seen in the bottom panel of the next figure (orange lines).

Now, let's look at the dependence on the initial conditions, which is demonstrated in Fig.~\ref{Plotscomparisonmatrad}. The orange lines represent the same situation given in  Fig.~\ref{PlotscomparisonmatradGR}. We provide the $a(\tau)$ solutions in the top panel and the Hubble parameters in the bottom.  We see that, for a wide range of initial conditions, the solutions quickly settle into similar behavior. We display the corresponding $f(\tau)$ solutions in the mid-panel. As in the case of matter-only, $f(\tau)$ approaches a constant value as $a(\tau)$ becomes GR-like.

\begin{figure}[hbt!]
\captionsetup[subfigure]{labelformat=empty}
\centering
\hspace{-0.8cm}
\begin{tabular}{lll}
\subfloat[ a) ]{\label{PlotsmatradBP1X}\includegraphics[width=5.1cm]{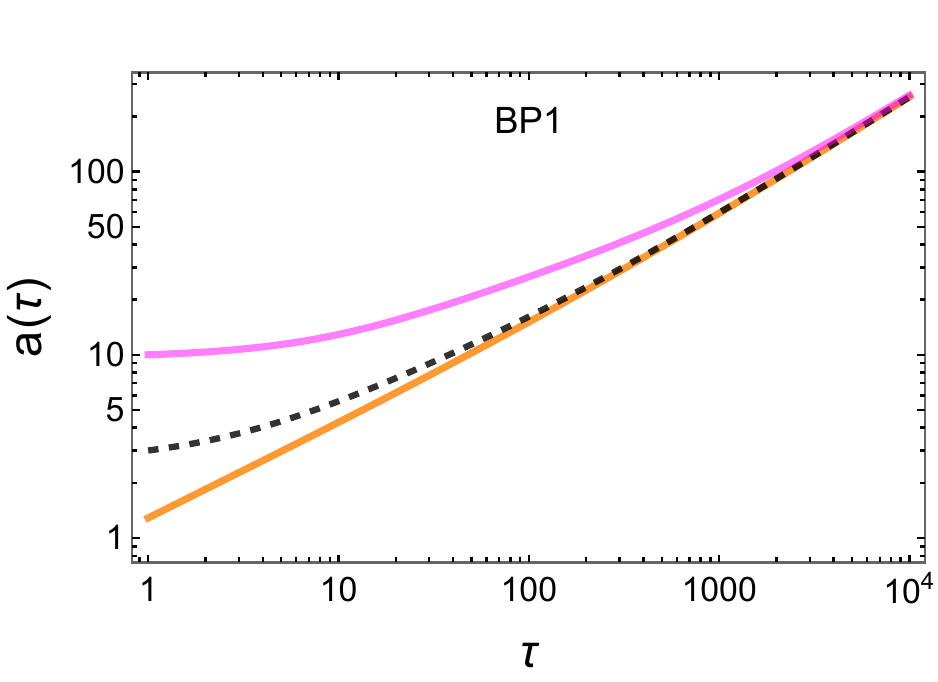}} \quad \vspace{-0.3cm}
\subfloat[b)]{\label{PlotsmatradBP2X}\includegraphics[width=5.3cm]{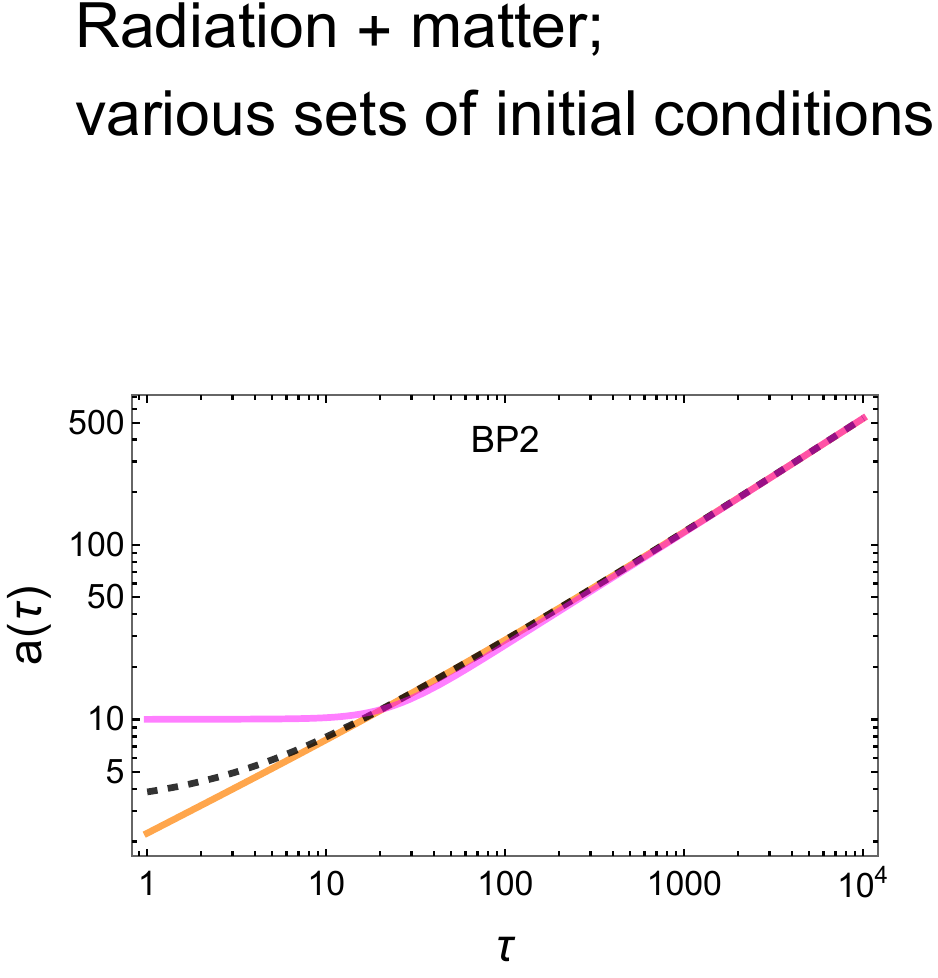}}
\subfloat[c)]{\label{PlotsmatradBP3X}\includegraphics[width=5.0cm]{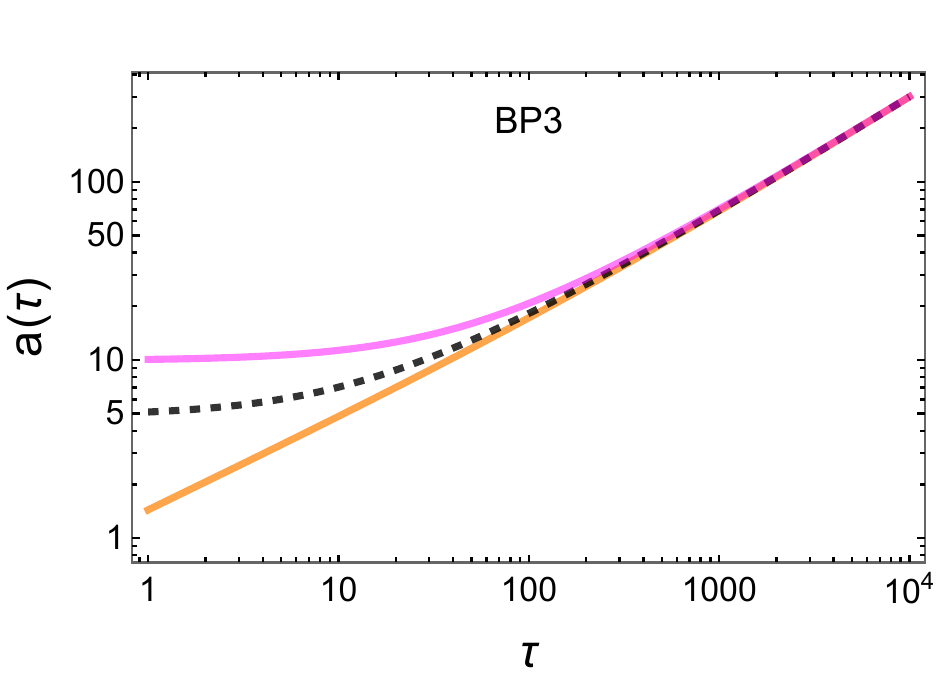}}\\
 \hspace{-0.2cm} \subfloat[d) ]{\label{PlotsmatradBP1Xf}\includegraphics[width=5.05cm]{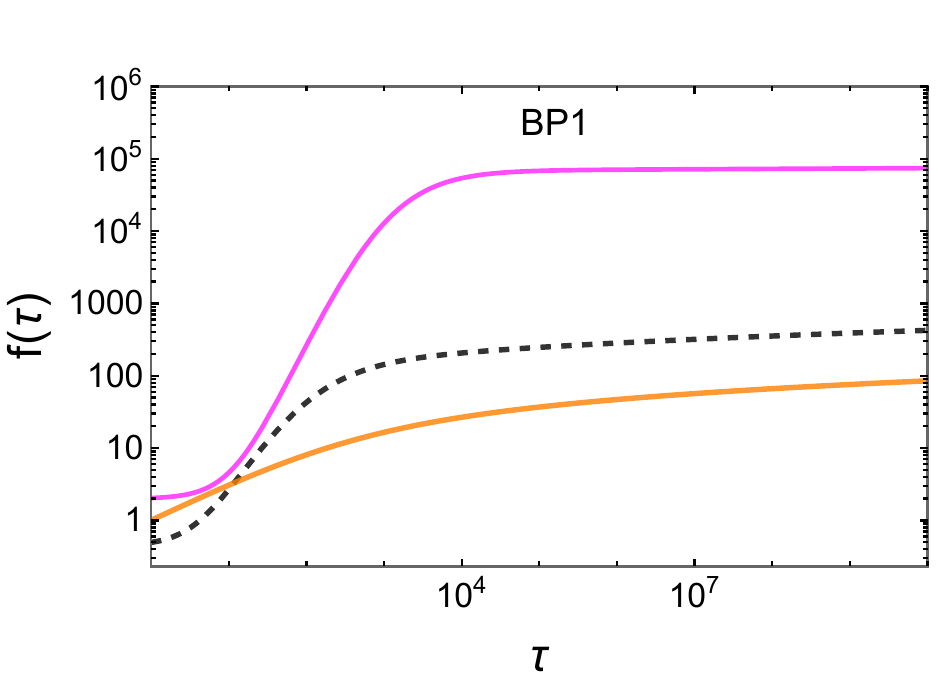}} \hspace{0.1cm} \vspace{-0.3cm}
\hspace{0.2cm}\subfloat[e)]{\label{PlotsmatradBP2Xf}\includegraphics[width=5.05cm]{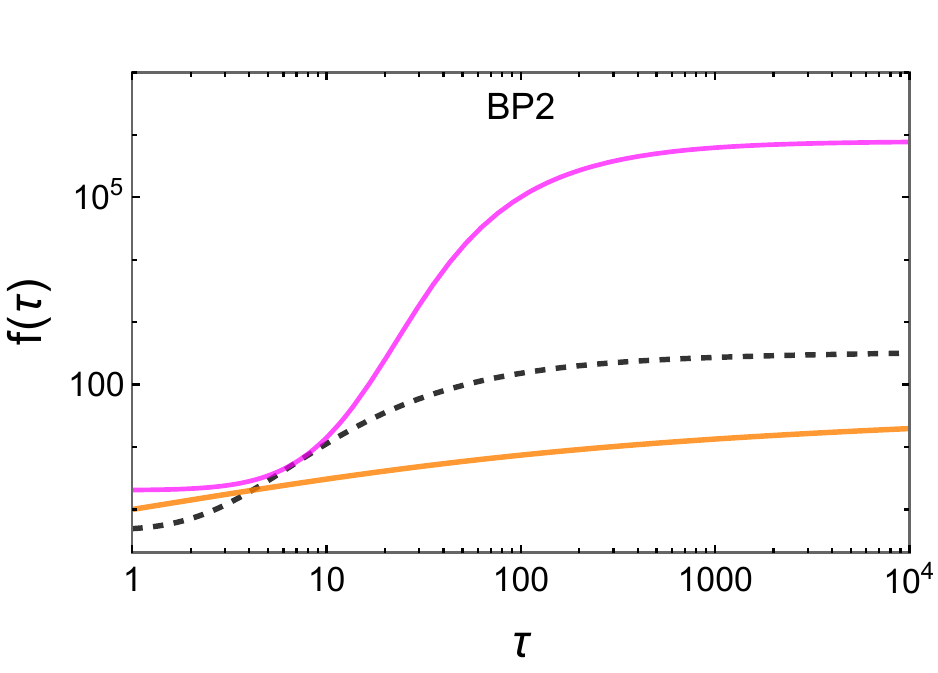}} \hspace{0.1cm} 
\subfloat[f)]{\label{PlotsmatradBP3Xf}\includegraphics[width=5.10cm]{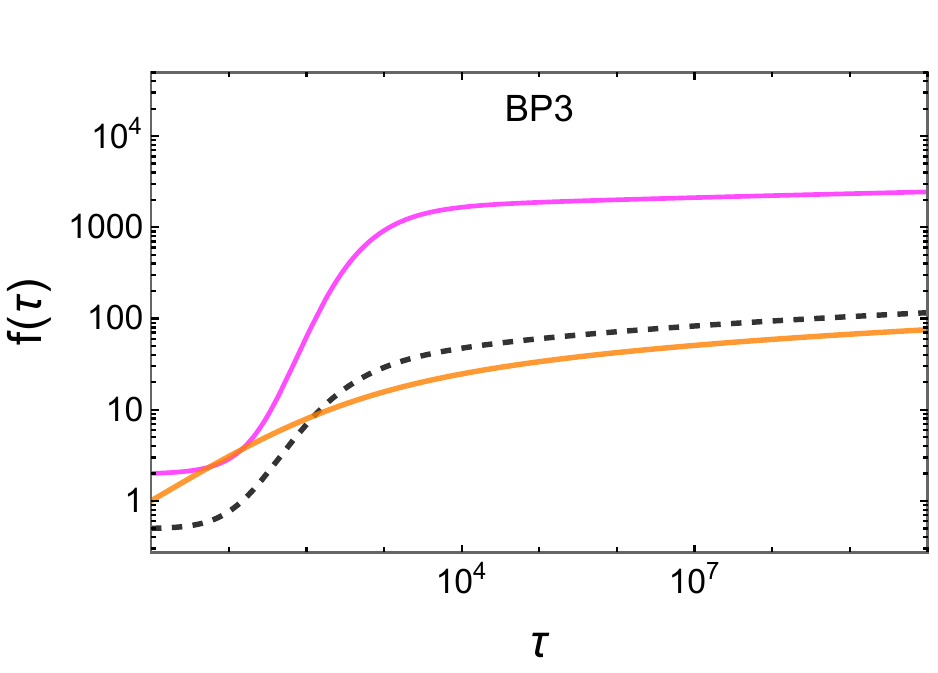}}\\
 \hspace{-0.3cm}
\subfloat[g) ]{\label{PlotsmatradBP1XHubble}\includegraphics[width=5.1cm]{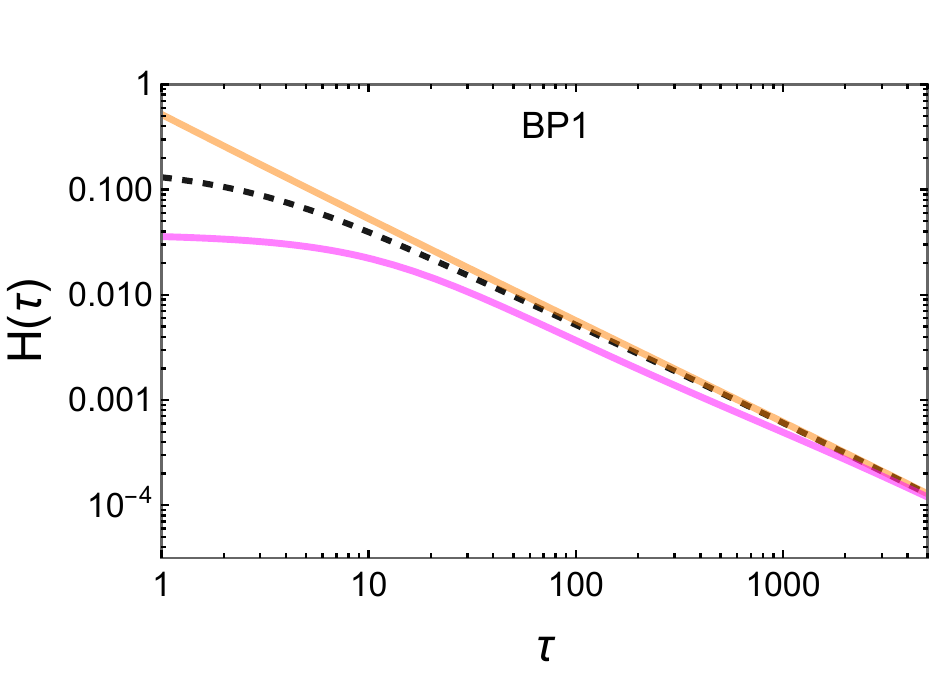}}  \hspace{0.2cm} 
\subfloat[h)]{\label{PlotsmatradBP2XHubble}\includegraphics[width=5.1cm]{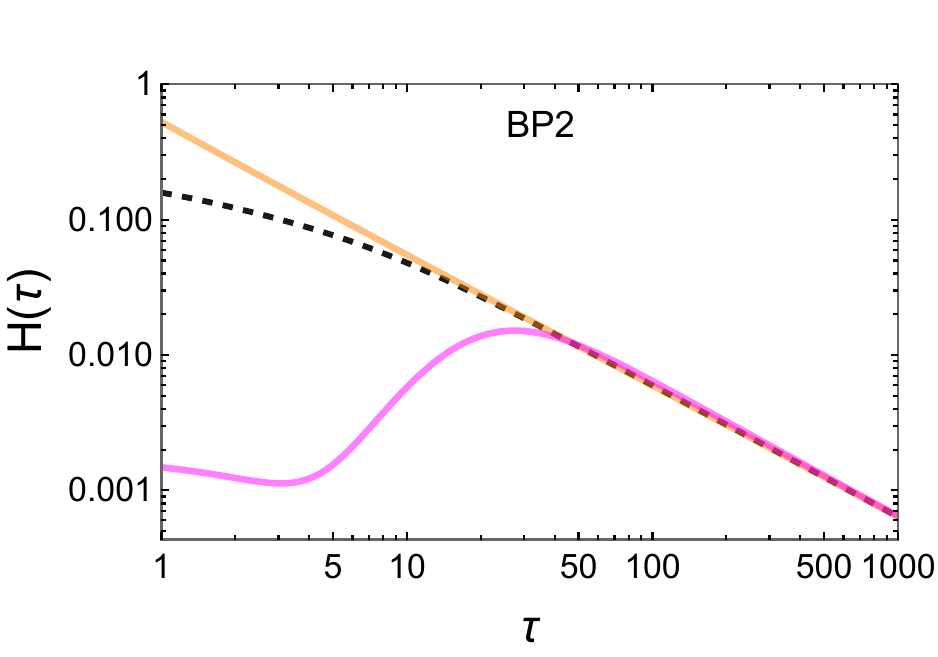}} \hspace{0.2cm} 
\subfloat[i)]{\label{PlotsmatradBP3XHubble}\includegraphics[width=5.1cm]{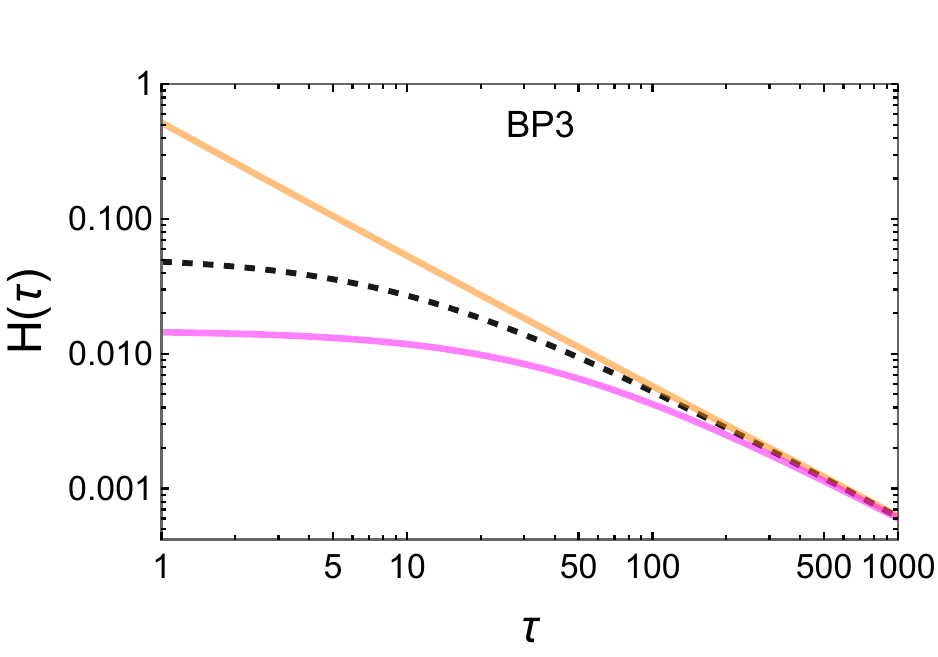}}
\end{tabular}
\caption{\label{Plotscomparisonmatrad} Comparison of $a(\tau)$,  $f(\tau)$, and  Hubble parameters for different initial conditions, denoted by pink, orange, and dashed lines, for the matter-radiation system. The orange lines correspond to the solutions in the previous figure.}   
\end{figure}

\newpage
\quad
\newpage
\section{Matter $+$ Vacuum} 
\label{sec:matvacnumerical}
Here, we will look at another two-fluid system: a small vacuum energy on the top of matter. In this system, the field equations (\ref{EEc1}) and (\ref{EEc2}) become
\begin{subequations}
\begin{align}
\label{EEcmatvac1}
\frac{ L}{2}\frac{(a')^2}{a^2} - K \left( \frac{1}{6} \frac{(f')^2}{f^2} - \frac{a' f'}{af} - \frac{1}{3}\frac{f''}{f} \right)   &= \frac{8\pi G}{3}\rho_{m0}\left(\frac{1}{a^3} +r\right), \\
 \frac{K}{2} \frac{f'^2}{f^2} - \frac{L}{2} \left( \frac{a'^2}{a^2} + 2 \frac{a''}{a} \right)  &= \frac{8\pi G}{3} \;\rho_{m0} \left(-3 r \right)\;,
\label{EEcmatvac2}
\end{align}
\end{subequations}
which, again, will be considered with the constraint equation given in Eq.~(\ref{cec}). Here, $r\equiv\rho_{\Lambda}/\rho_{m0}$, with  $\rho_{m0}$ and $\rho_{\Lambda}$ being the initial energy density for matter and vacuum, respectively. For concreteness, we will set $\rho_{\Lambda}=0.0001$ and $\rho_{m0}=0.9999$. We work with the dimensionless time $ \tau\equiv \sqrt{\frac{8\pi G \rho_{m0}}{3}}\; \tilde{t}$, where $\tilde{t}$ is the usual cosmological time.

\begin{figure}[hbt!]
\vspace{-0.5cm}
\captionsetup[subfigure]{labelformat=empty}
\centering
\hspace{-0.9cm}
\begin{tabular}{lll}
\quad\quad\quad\quad \quad \quad \quad \quad\quad \quad\quad\quad\quad\quad\quad\subfloat[ ]{ \label{plotlabel}\includegraphics[width=4.2cm]{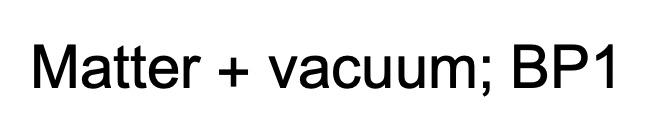}}\vspace{-0.9cm}\\ 
\subfloat[a)   ]{\label{PlotMatVacBP1a}\includegraphics[width=5.3cm]{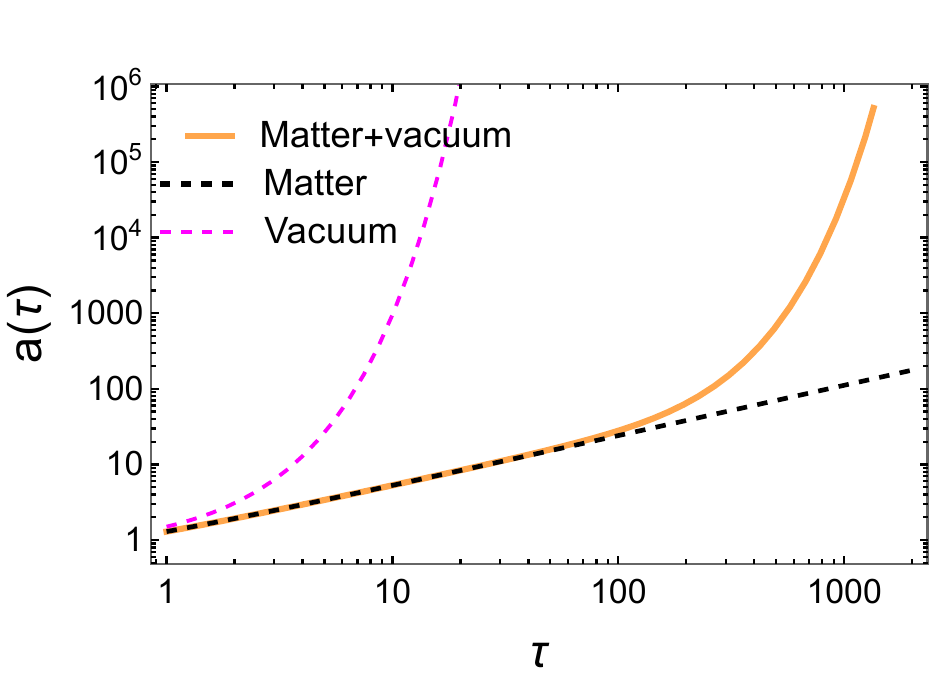}}  \quad
\subfloat[b) ]{\label{PlotMatVacBP1b}\includegraphics[width=5.0cm]{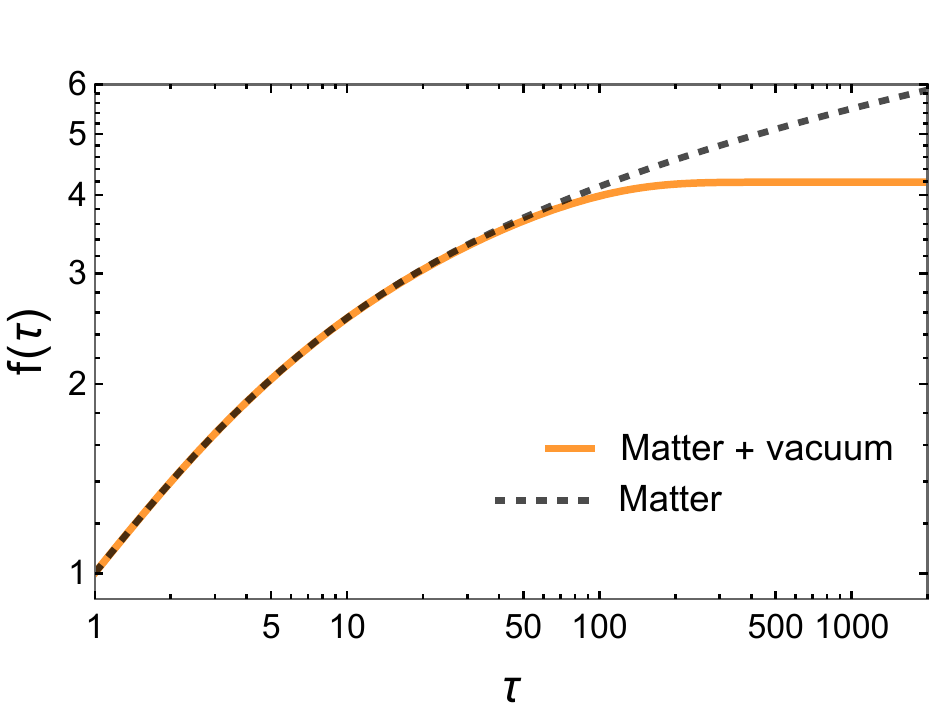}} \quad
\subfloat[c) ]{\label{PlotMatVacBP1c}\includegraphics[width=5.5cm]{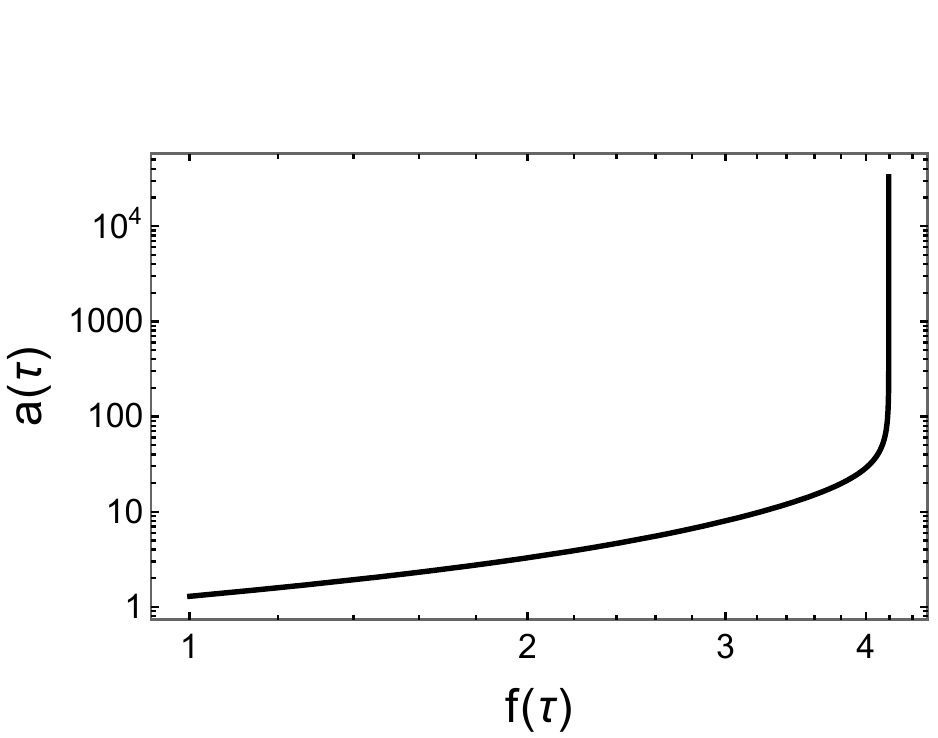}}
\end{tabular}
\caption{\label{PlotMatVacBP1} Logarithmic plots of the $a(\tau)$ and $f(\tau)$ solutions for a matter-dominated system with a small amount of vacuum energy density. The plots are displayed for BP1, given in Table~\ref{BPs}.}.   
\end{figure}

The solutions for $a(\tau)$ and $f(\tau)$, with initial conditions near those in the GR case, are given in Fig.~\ref{PlotMatVacBP1} for BP1. Since we do not have an analytical solution for the matter-only case, we choose the initial conditions based on the radiation-only solution as we did in the numerical solution matter-only case above; this does not have a big effect since the matter-vacuum solutions are valid for a wide range of initial conditions, as we will see below. We see that $a(\tau)$ begins on the matter-only track and gradually deviates as the vacuum energy dominates, as expected. 
 As for $f(\tau)$; it similarly begins on the matter-only track but then settles to a constant value as usual. 

\begin{figure}[hbt!]
\captionsetup[subfigure]{labelformat=empty}
\centering
\hspace{-0.8cm}
\begin{tabular}{lll}
\subfloat[a) ]{\label{PlotsmatvacGRBP1}\includegraphics[width=5.2cm]{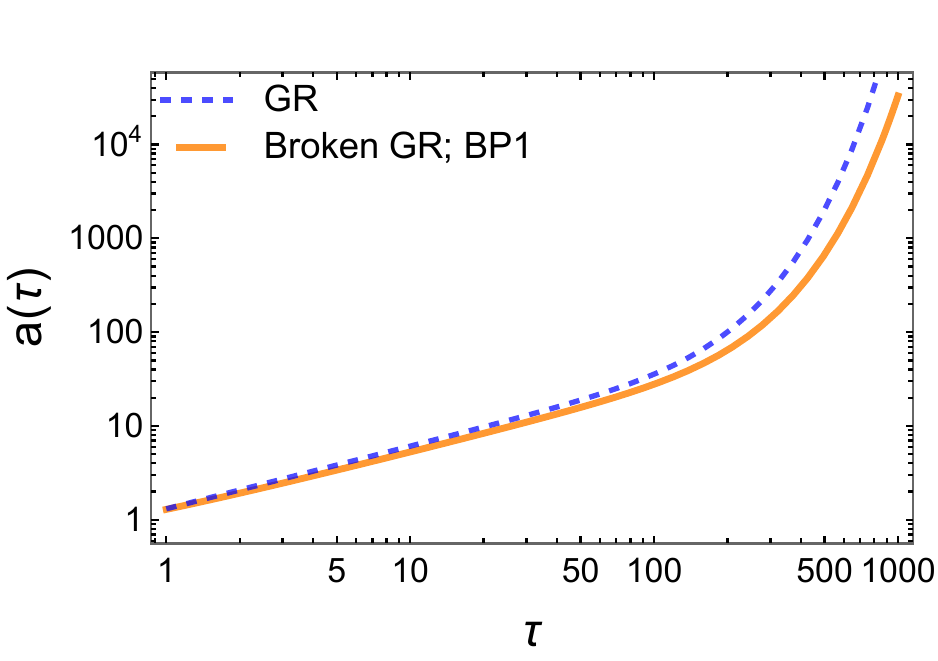}} \hspace{0.3cm}
\subfloat[b)]{\label{PlotsmatvacGRBP2}\includegraphics[width=5.5cm]{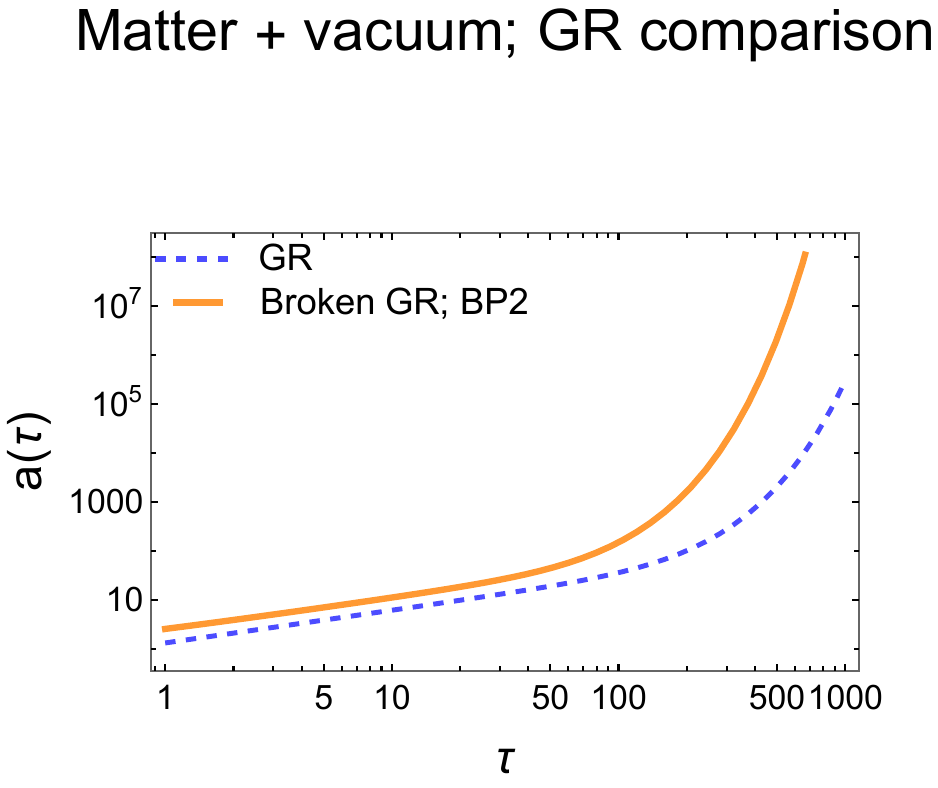}}
\subfloat[c)]{\label{PlotsmatvacGRBP3}\includegraphics[width=5.1cm]{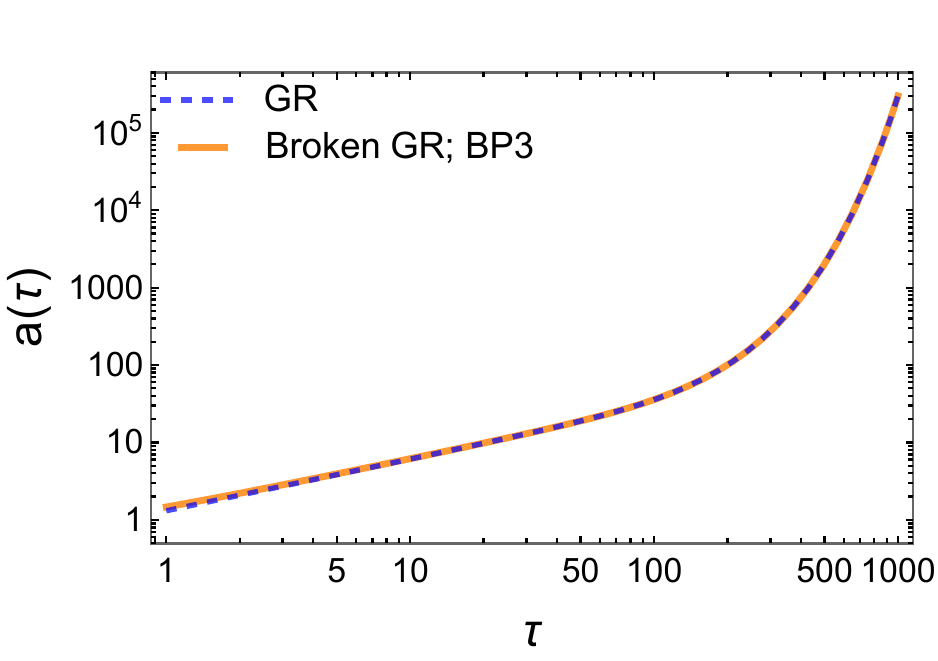}}\\
\subfloat[d)]{\label{PlotsmatvacGRBP1ratio}\includegraphics[width=5.2cm]{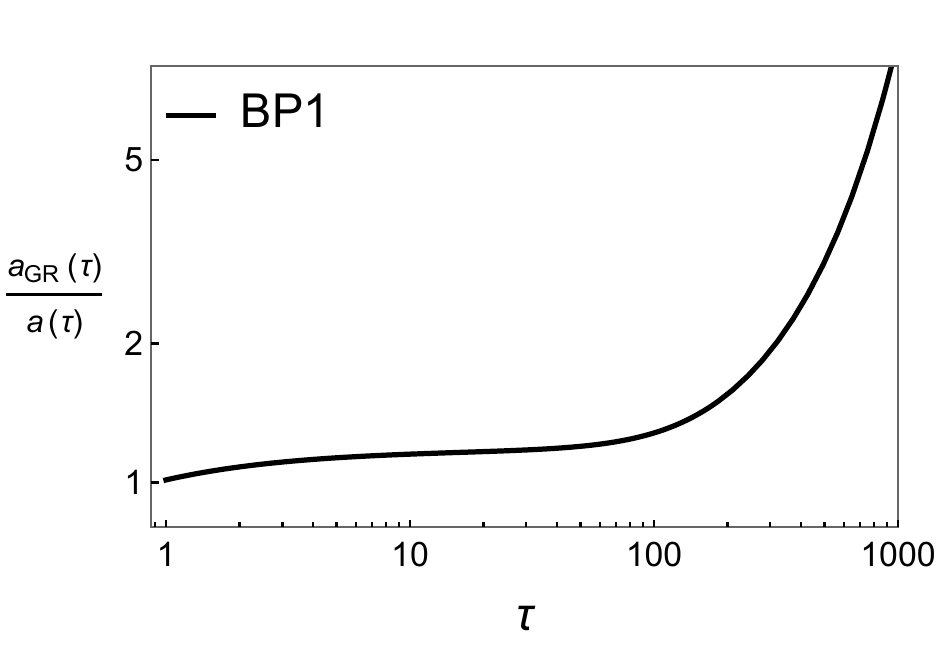}}\hspace{0.1cm}
\subfloat[e)]{\label{PlotsmatvacGRBP2ratio}\includegraphics[width=5.4cm]{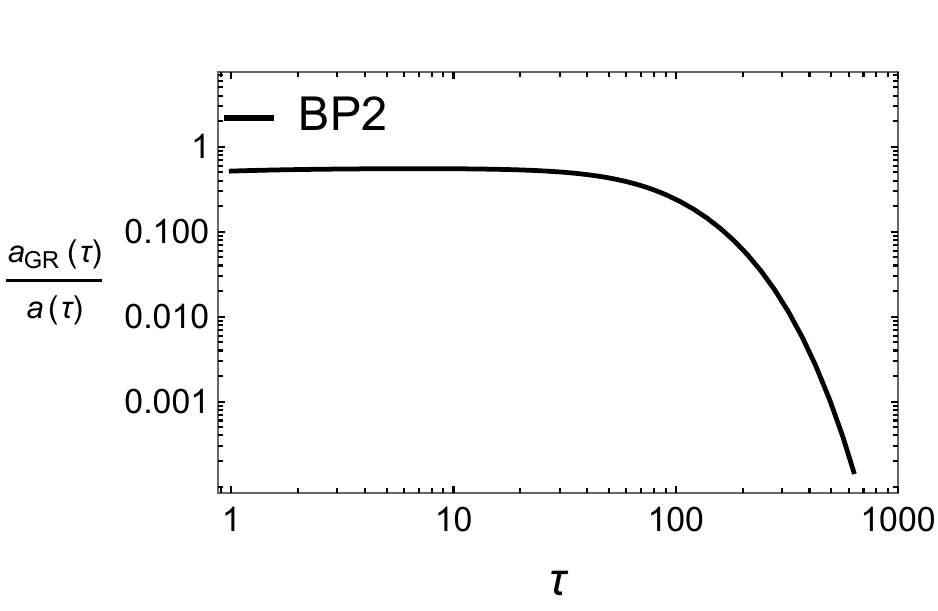}}\hspace{0.1cm}
\subfloat[f)]{\label{PlotsmatvacGRBP3ratio}\includegraphics[width=5.3cm]{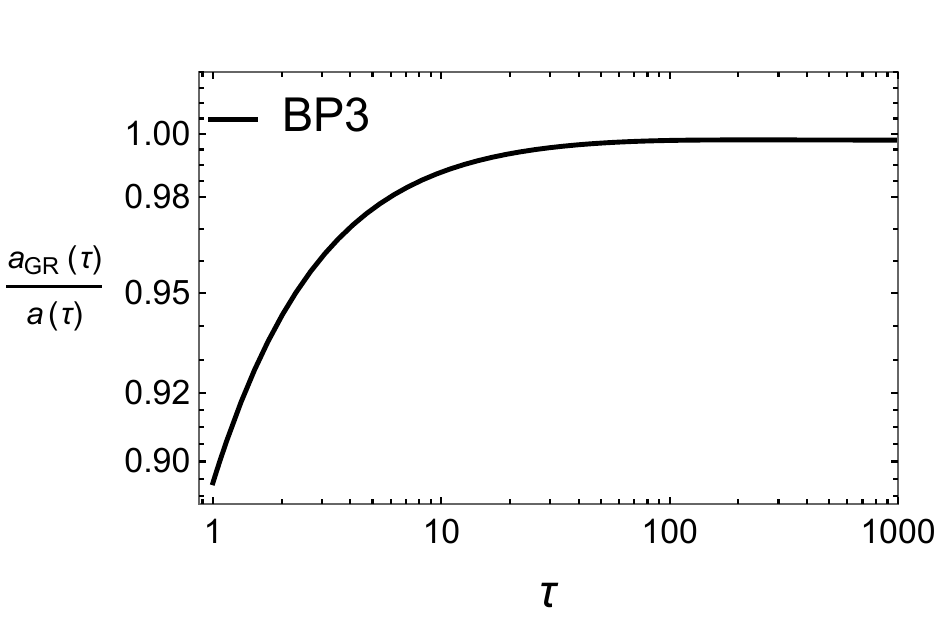}}\\
\subfloat[g)]{\label{PlotsmatvacGRBP1Hubble}\includegraphics[width=5.2cm]{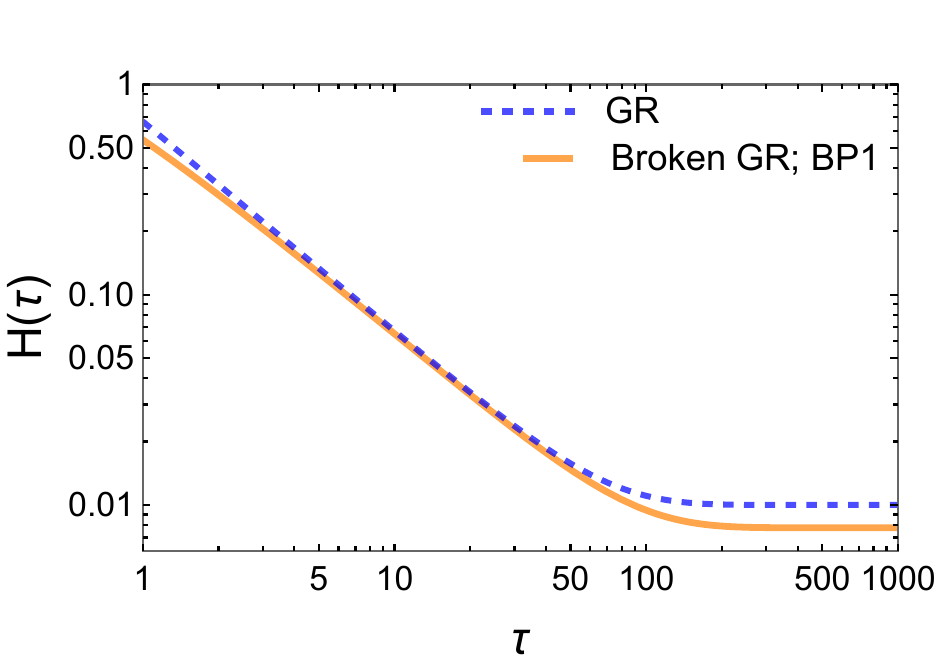}} \hspace{0.4cm}
\subfloat[h)]{\label{PlotsmatvacGRBP2Hubble}\includegraphics[width=5.1cm]{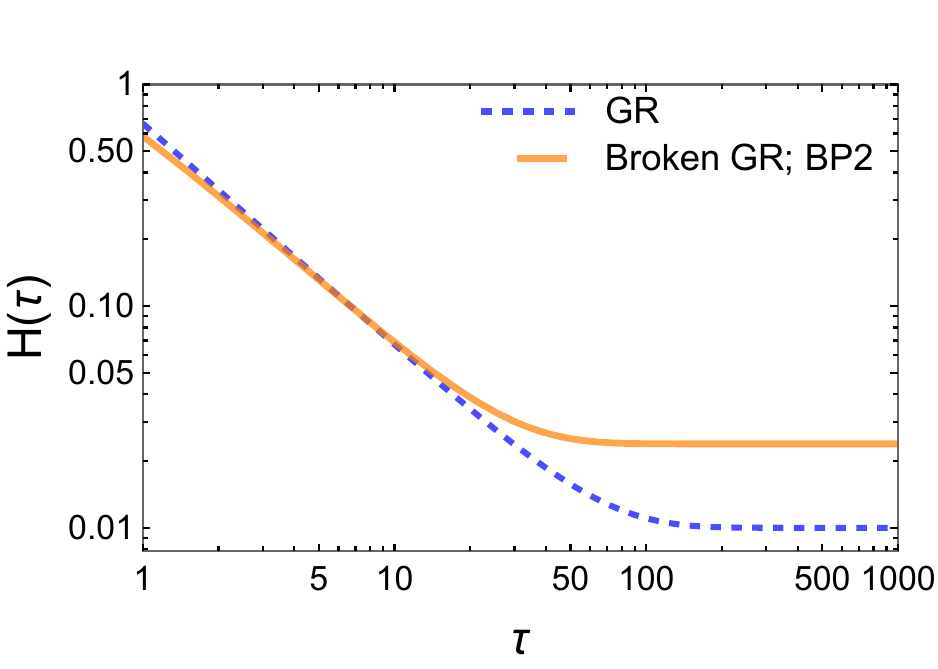}}\hspace{0.3cm}
\subfloat[i)]{\label{PlotsmatvacGRBP3Hubble}\includegraphics[width=5.1cm]{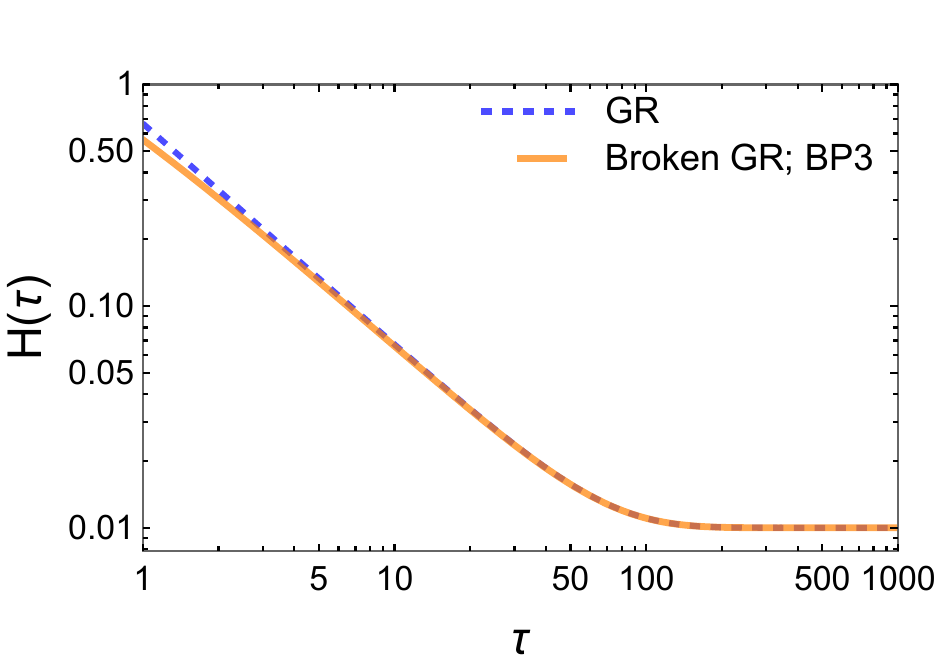}}
\end{tabular}
\caption{\label{PlotscomparisonmatvacGR} Comparison of $a(\tau)$ solutions to GR for different BPs (given in Table~\ref{BPs}) for the matter-vacuum case. 
}   
\end{figure}

The comparison to the GR case is given in Fig.~\ref{PlotscomparisonmatvacGR}. In the top and mid panel, we display the $a(\tau)$ comparison. Unlike in the matter-only and radiation-only cases, the desired, eventual time independence of $a_{\tiny{\mbox{GR}}}(\tau)/a(\tau)$ is not generic here due to the vacuum energy domination stage, as expected from the vacuum-only case. For BP1 (BP2), shown in Fig.~\ref{PlotsmatvacGRBP1ratio} (Fig.~\ref{PlotsmatvacGRBP2ratio}) the ratio increases (decreases) rapidly with time. For small enough values, however, we do have the eventual constancy of the ratio, as in the case of BP3, shown in Fig.~\ref{PlotsmatvacGRBP3ratio}. The Hubble parameter, displayed in the bottom panel of Fig.~\ref{PlotscomparisonmatvacGR}, is generically well-behaved and settles into a constant value; the smaller the $\alpha_a$ coefficients, the closer the Hubble parameter gets to the GR value, as desired. 

\begin{figure}[hbt!]
\captionsetup[subfigure]{labelformat=empty}
\centering
\hspace{-0.8cm}
\begin{tabular}{lll}
\subfloat[ a) ]{\label{PlotsmatvacBP1X}\includegraphics[width=5.1cm]{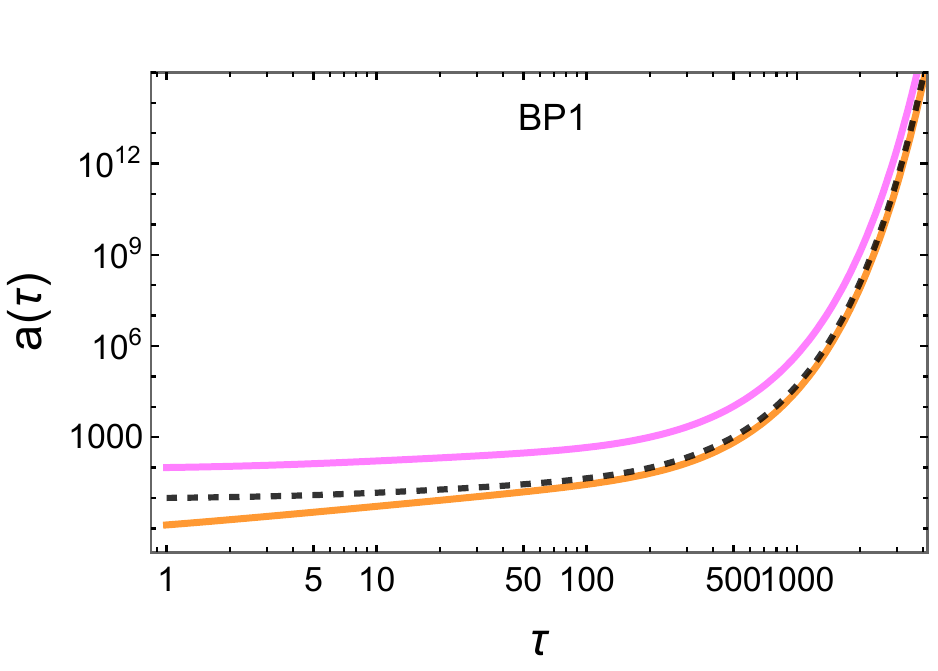}} \quad 
\subfloat[b)]{\label{PlotsmatvacBP2X}\includegraphics[width=5.4cm]{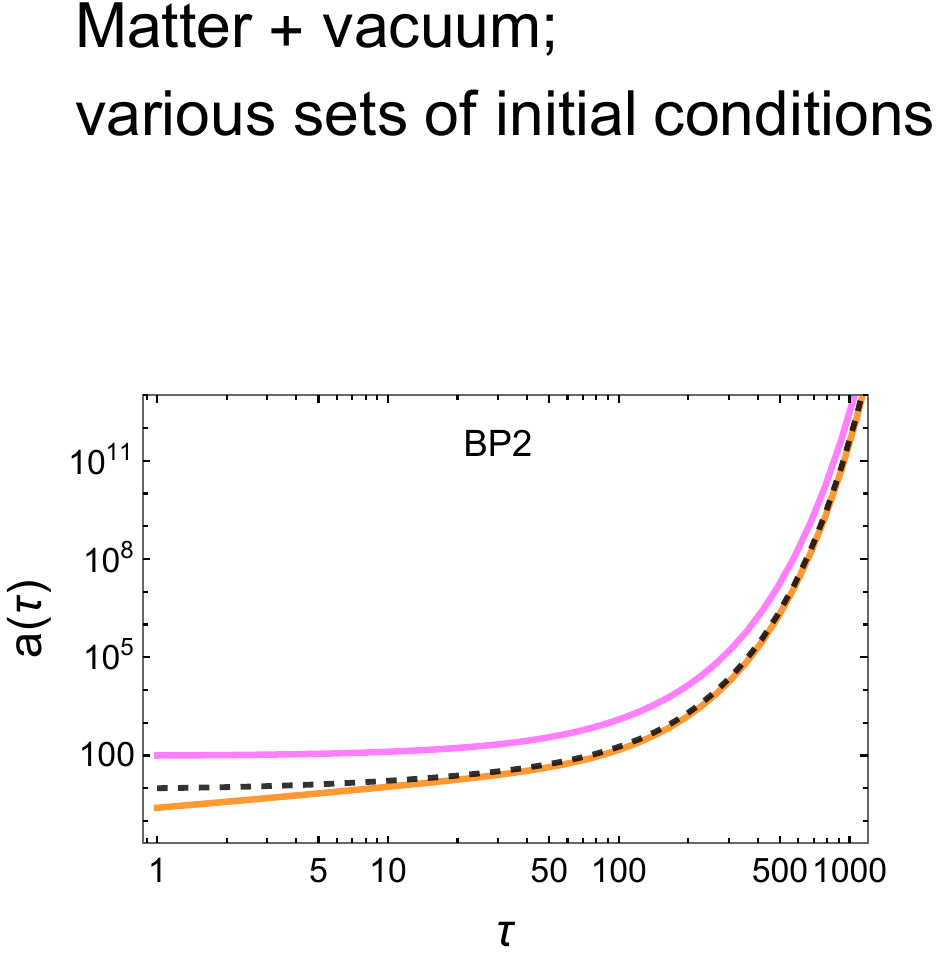}}
\subfloat[c)]{\label{PlotsmatvacBP3X}\includegraphics[width=5.0cm]{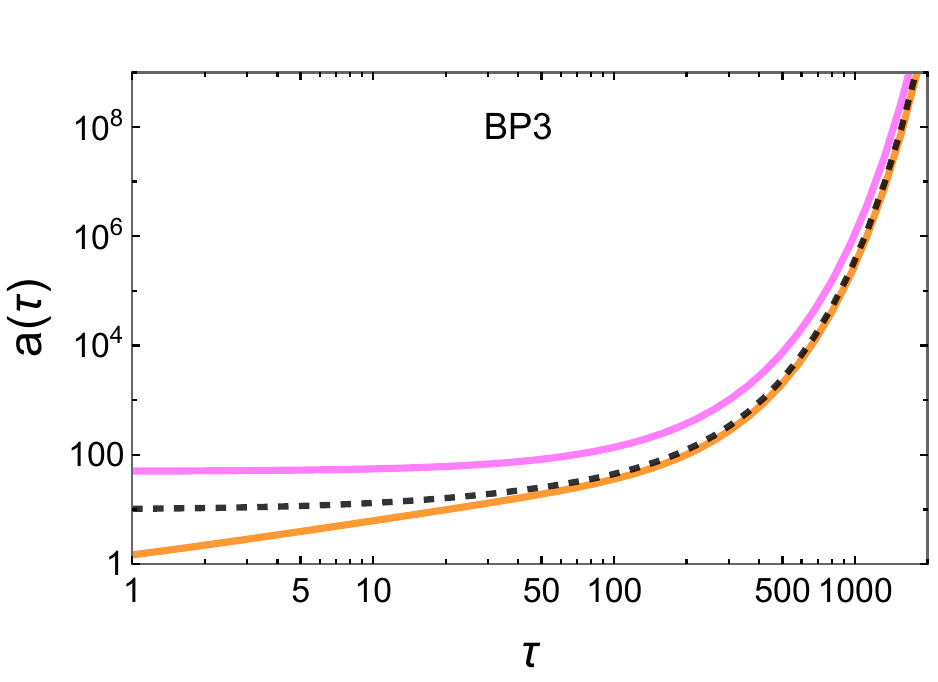}}\\
 \hspace{-0.2cm} \subfloat[d) ]{\label{PlotsmatvacBP1Xf}\includegraphics[width=5.1cm]{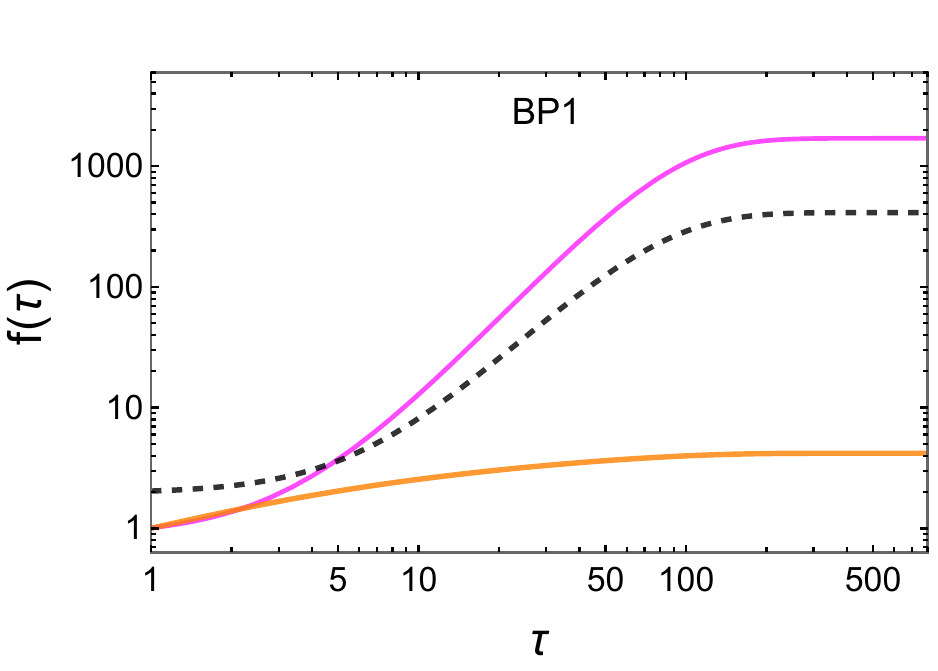}} \hspace{0.1cm} \vspace{-0.3cm} \hspace{0.1cm} 
\subfloat[e)]{\label{PlotsmatvacBP2Xf}\includegraphics[width=5.1cm]{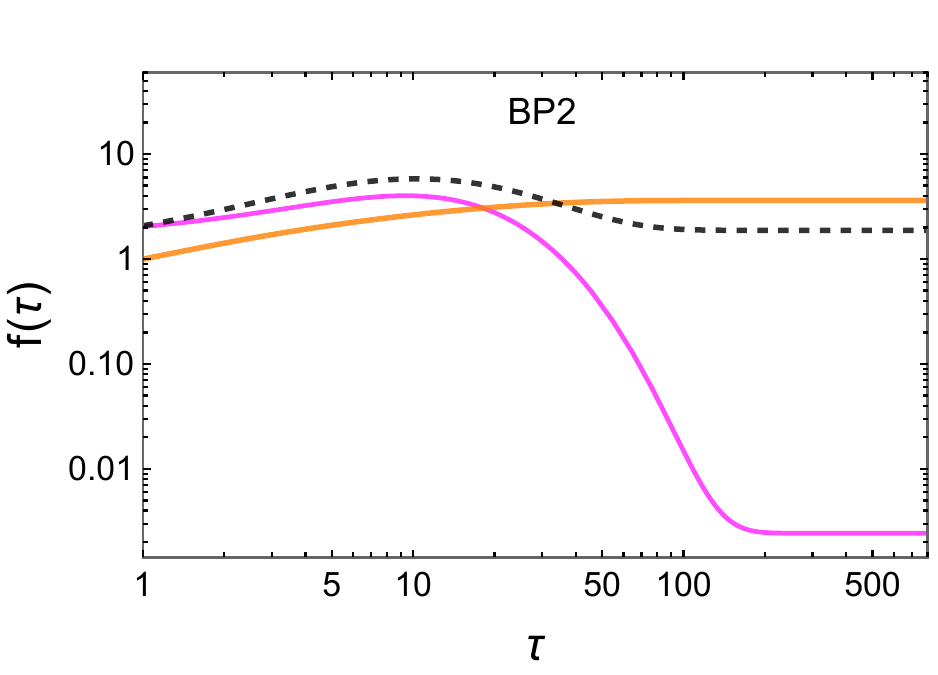}} \hspace{0.1cm} 
\subfloat[f)]{\label{PlotsmatvacBP3Xf}\includegraphics[width=5.1cm]{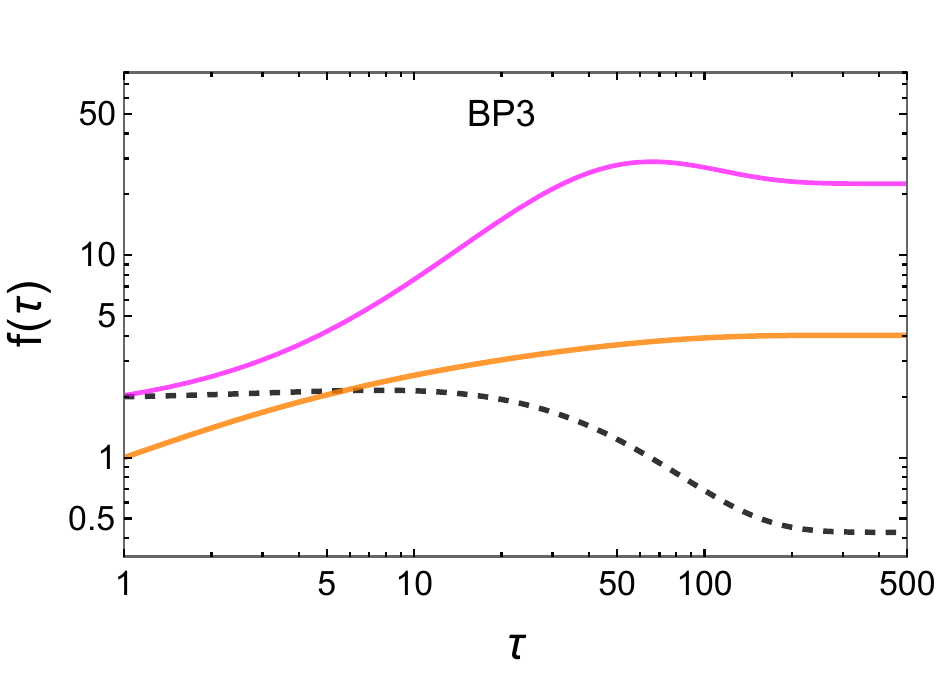}}\\
 \hspace{-0.3cm}
\subfloat[g) ]{\label{PlotsmatvacBP1XHubble}\includegraphics[width=5.4cm]{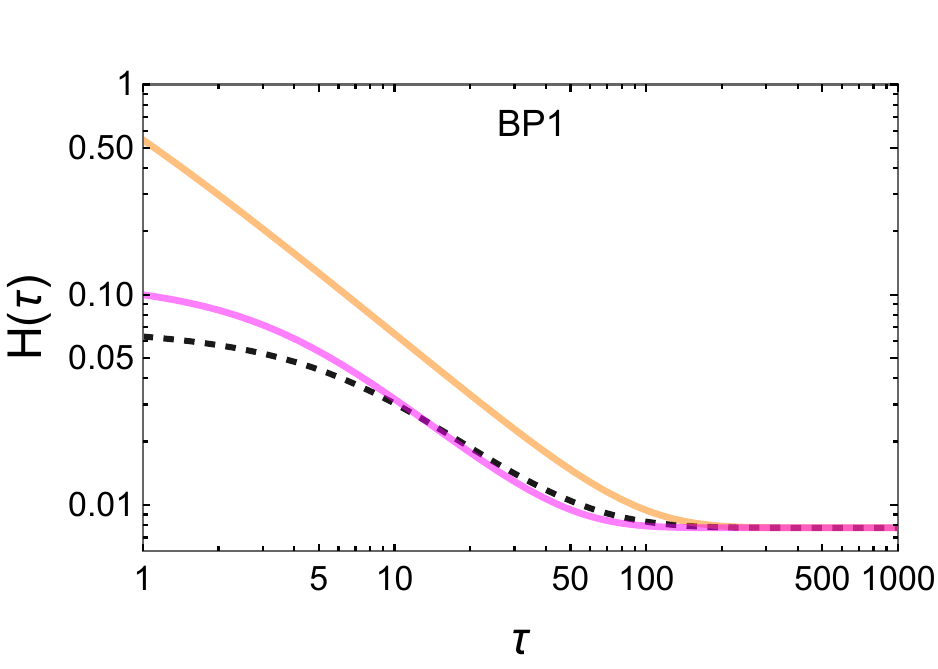}}  \hspace{0.1cm} 
\subfloat[h)]{\label{PlotsmatvacBP2XHubble}\includegraphics[width=5.3cm]{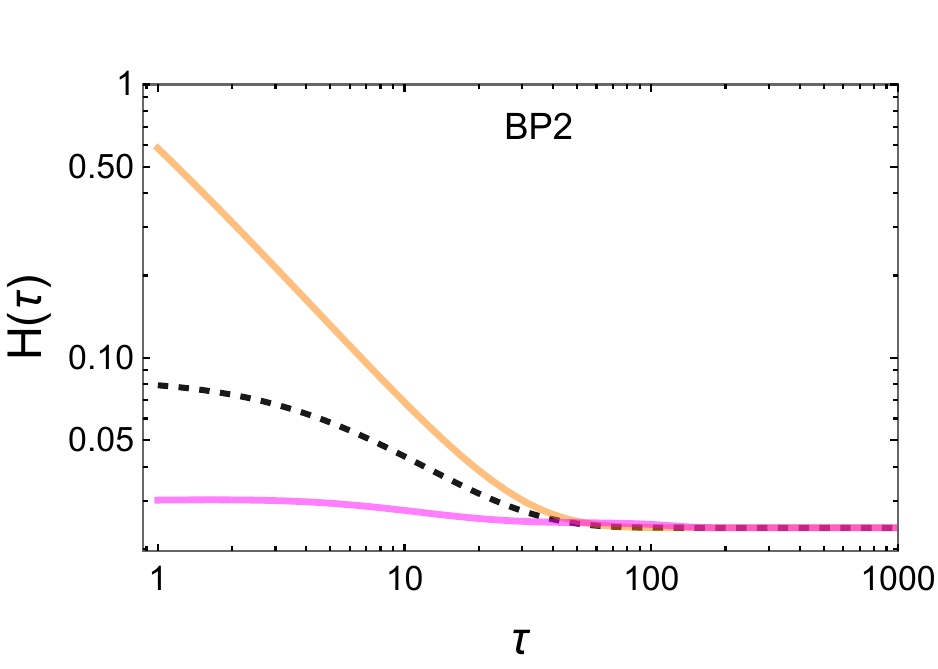}} \hspace{0.1cm} 
\subfloat[i)]{\label{PlotsmatvacBP3XHubble}\includegraphics[width=5.3cm]{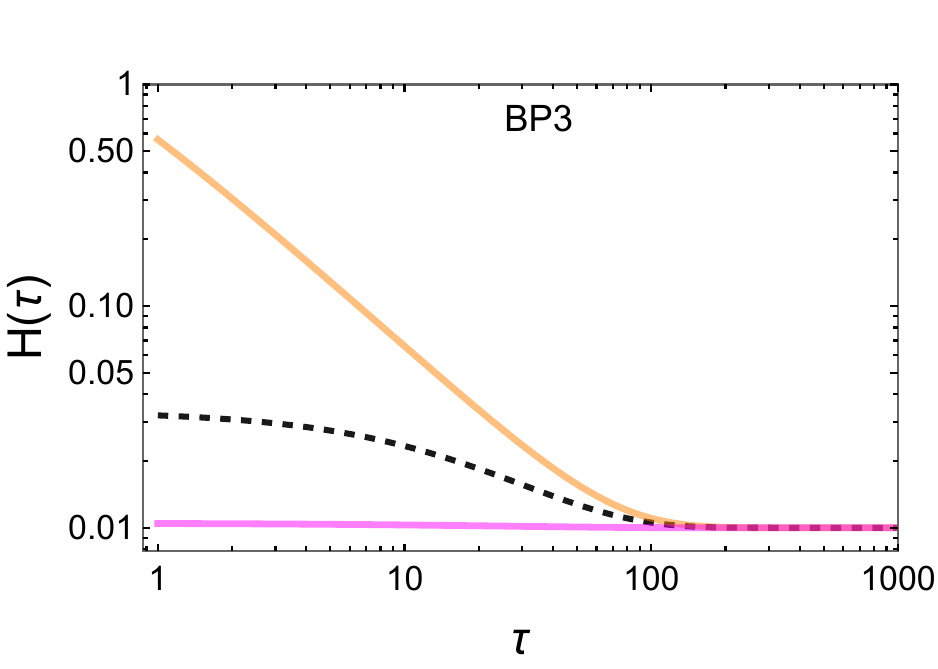}}
\end{tabular}
\caption{\label{Plotscomparisonmatvac} Comparison of $a(\tau)$,  $f(\tau)$, and  Hubble parameters for different initial conditions, denoted by pink, orange, and dashed lines, for the matter plus vacuum case. The orange lines correspond to the  solutions in the previous figure.}   
\end{figure}

The comparison with the various sets of initial conditions is given in Fig.~\ref{Plotscomparisonmatvac}. The $a(\tau)$ solutions and the Hubble parameters gradually approach similar behaviors. The $f(\tau)$ solutions generically settle into a constant value, as in the other cases.  
\\

\newpage
\quad
\newpage

\section{Radiation $+$ Matter $+$ Vacuum \label{sec:mixnumerical}}

Here, we will consider the most realistic scenario faced in cosmology: a universe starting with radiation domination with a small amount of matter density and even lesser vacuum energy density. The field equations~(\ref{EEc1}) and (\ref{EEc2}) become 
\begin{subequations}
\begin{align}
\label{EEcthreecomponent1}
\frac{ L}{2}\frac{(a')^2}{a^2} - K \left( \frac{1}{6} \frac{(f')^2}{f^2} - \frac{a' f'}{af} - \frac{1}{3}\frac{f''}{f} \right)   &= \frac{8\pi G}{3}\rho_{r0}\left(\frac{1}{a^4}+\frac{r_1}{a^3}+r_2 \right), \\
 \frac{K}{2} \frac{f'^2}{f^2} - \frac{L}{2} \left( \frac{a'^2}{a^2} + 2 \frac{a''}{a} \right)  &= \frac{8\pi G}{3} \;\rho_{r0} \left(\frac{1}{ a^4}-3 r_2 \right)\;,
\label{EEcthreecomponent2}
\end{align}
\end{subequations}
where $r_1\equiv\rho_{m0}/\rho_{r0}$ and $r_2\equiv \rho_{\Lambda}/\rho_{r0}$ with $\rho_{r0}$, $\rho_{m0}$, and $\rho_{\Lambda}$ being the initial energy density for radiation, matter, and vacuum, respectively. As before, the derivatives are with respect to the cosmological time $\tilde{t}$, and here we switch to the dimensionless time $ \tau\equiv \sqrt{\frac{8\pi G \rho_{r0}}{3}}\; \tilde{t}$. For concreteness, we set $\rho_{r0}=0.9 \;\rho_{\mathrm{tot,0}}$, $\rho_{m0}=0.09 \;\rho_{\mathrm{tot,0}}$, and $\rho_{\Lambda}=0.01 \;\rho_{\mathrm{tot,0}}$, where $\rho_{\mathrm{tot,0}}$ is the initial total energy density.

\begin{figure}[hbt!]
\captionsetup[subfigure]{labelformat=empty}
\centering
\hspace{-0.8cm}
\begin{tabular}{lll}
\subfloat[a) ]{\label{PlotMixBP1a}\includegraphics[width=5.3cm]{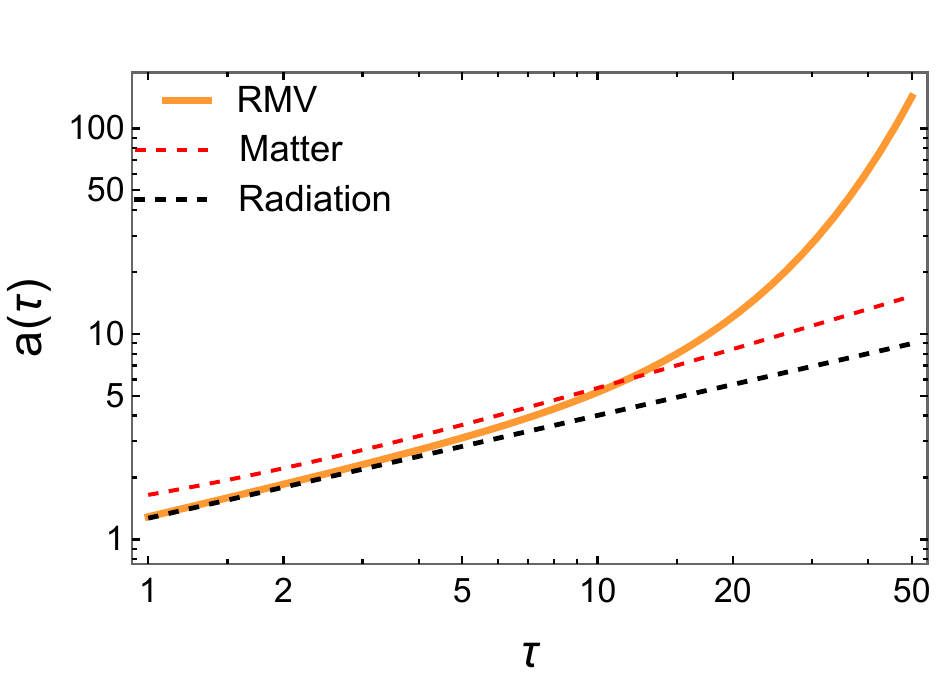}} \quad
\subfloat[b) ]{\label{PlotMixBP1b}\includegraphics[width=5.5cm]{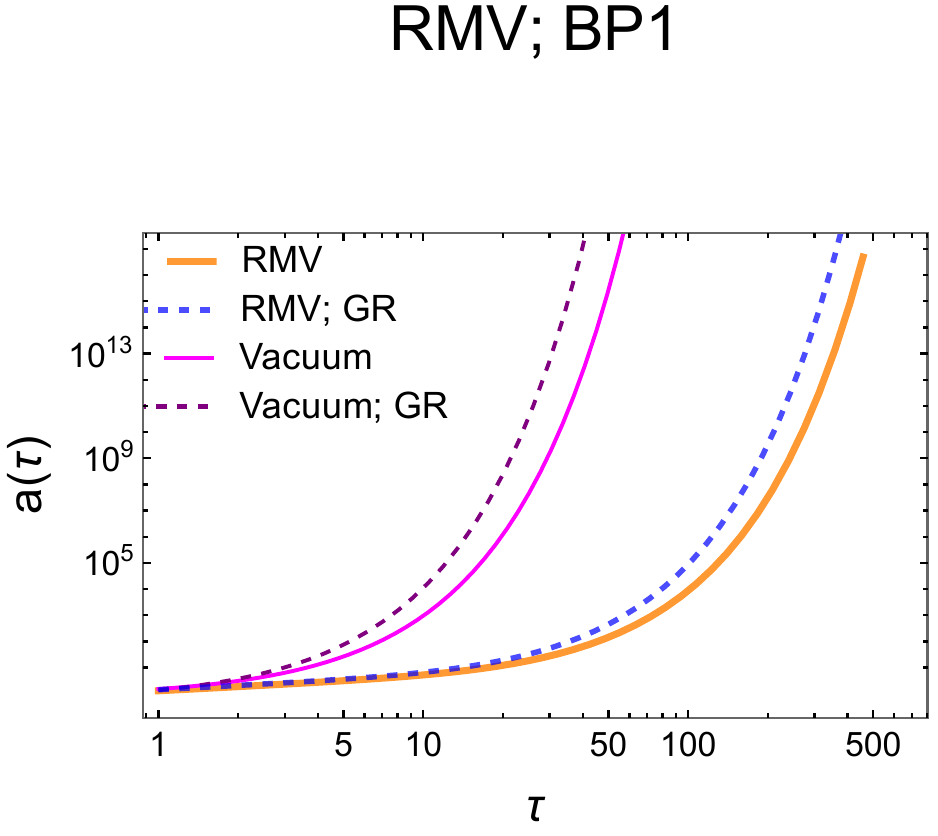}}\quad
\subfloat[c) ]{\label{PlotMixBP1c}\includegraphics[width=5.2cm]{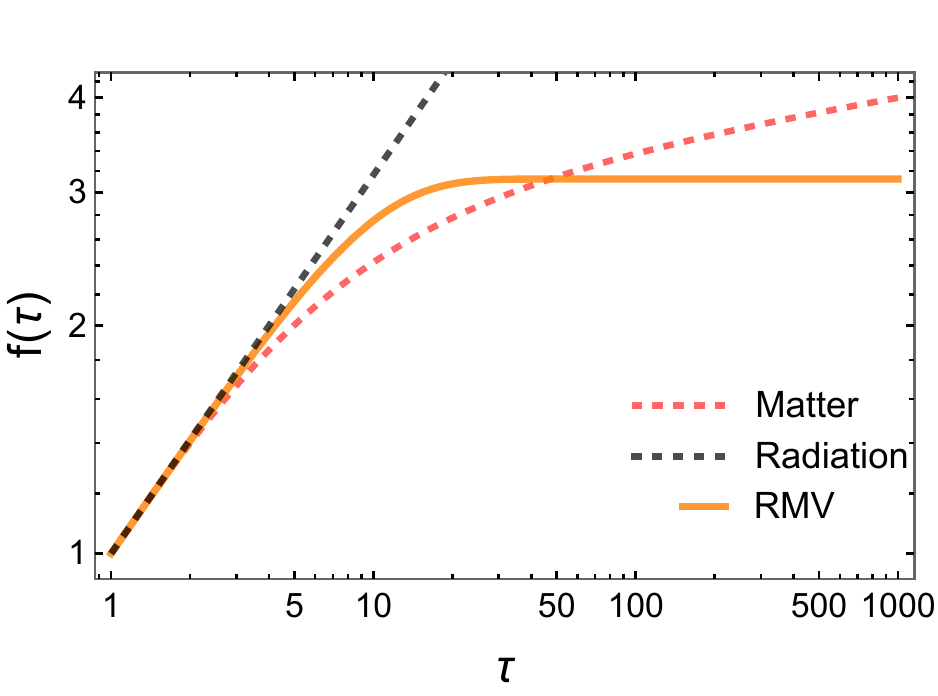}}\vspace{-0.5cm}\\
\quad\quad\quad\quad\quad\quad\quad\quad\quad\quad\quad\quad\quad\quad\subfloat[d) ]{\label{PlotMixBP1d}\includegraphics[width=5.5cm]{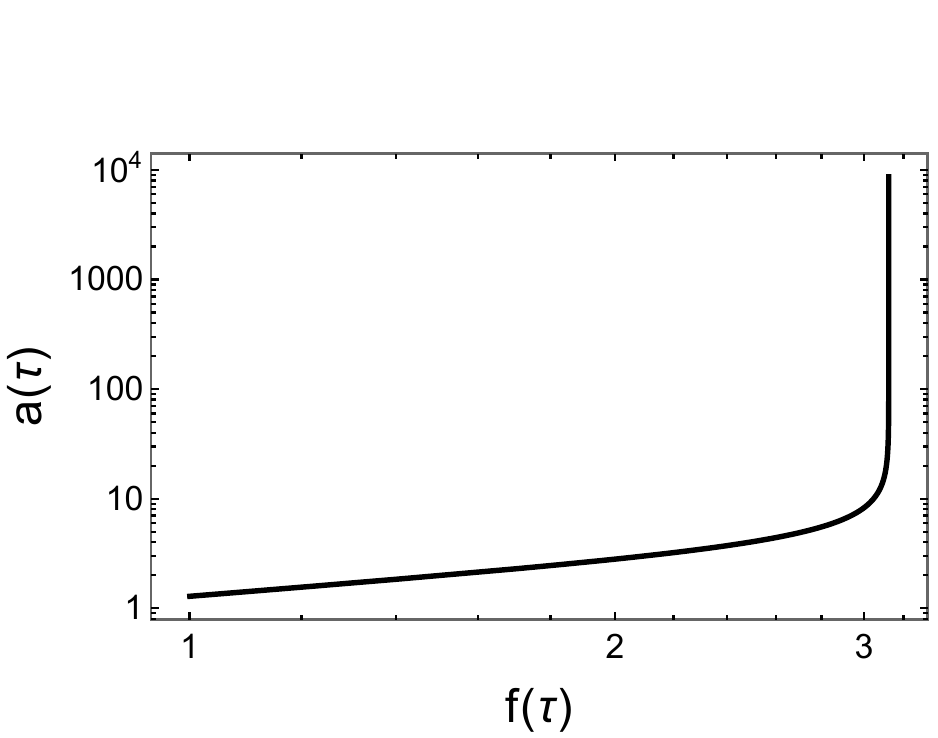}}
\end{tabular}
\caption{\label{PlotmixBP1} Logarithmic plots of the $a(\tau)$ and $f(\tau)$ for the case of three-component fluid: radiation plus matter plus vacuum (RMV). Namely, we have a radiation-dominated system with a small amount of matter density and lesser vacuum energy density. The plots are displayed for BP1, given in Table~\ref{BPs}.  Figs.~\ref{PlotMixBP1a} and \ref{PlotMixBP1b} are the same plots in different plotting ranges to demonstrate the evolution from the radiation solution through a matter-like realm to a vacuum-like solution. The vacuum-only solution in the broken theory is also included in Fig.~\ref{PlotMixBP1b}, as well as the RMV and vacuum-only GR solutions for comparison. In Fig.~\ref{PlotMixBP1c}, the $f(\tau)$ solution is displayed. Finally, in Fig.~\ref{PlotMixBP1d}, the $a(\tau)$-$f(\tau)$ parametric plot is displayed for convenience.}.   
\end{figure}

\begin{figure}[ht!]
\captionsetup[subfigure]{labelformat=empty}
\centering
\hspace{-0.8cm}
\vspace{-0.1cm}
\begin{tabular}{lll}
\subfloat[a) ]{\label{PlotMixGRBP1}\includegraphics[width=5.3cm]{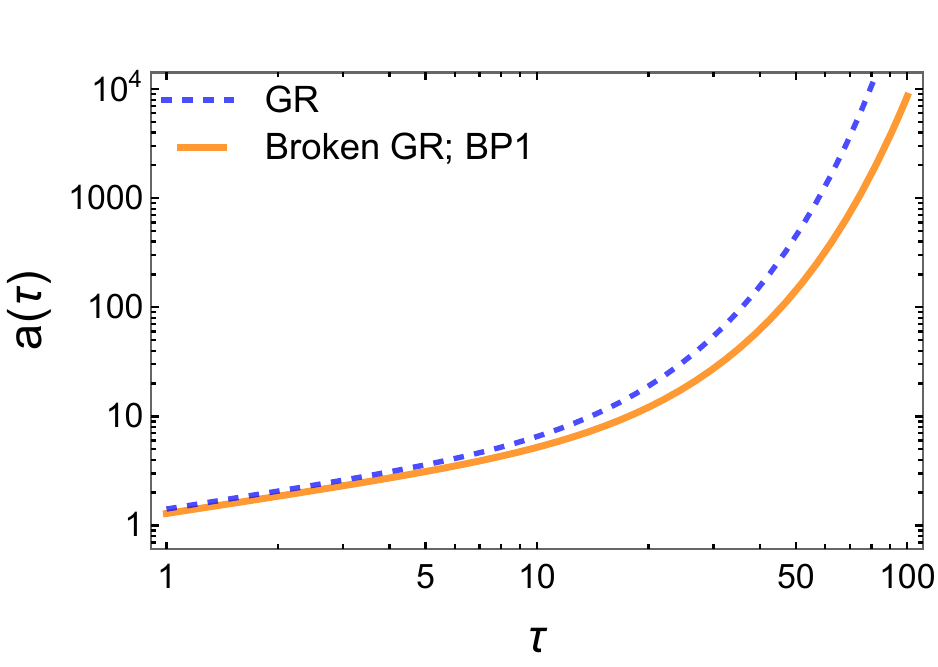}} \quad \vspace{-0.1cm}
\subfloat[b)]{\label{PlotMixGRBP2}\includegraphics[width=5.3cm]{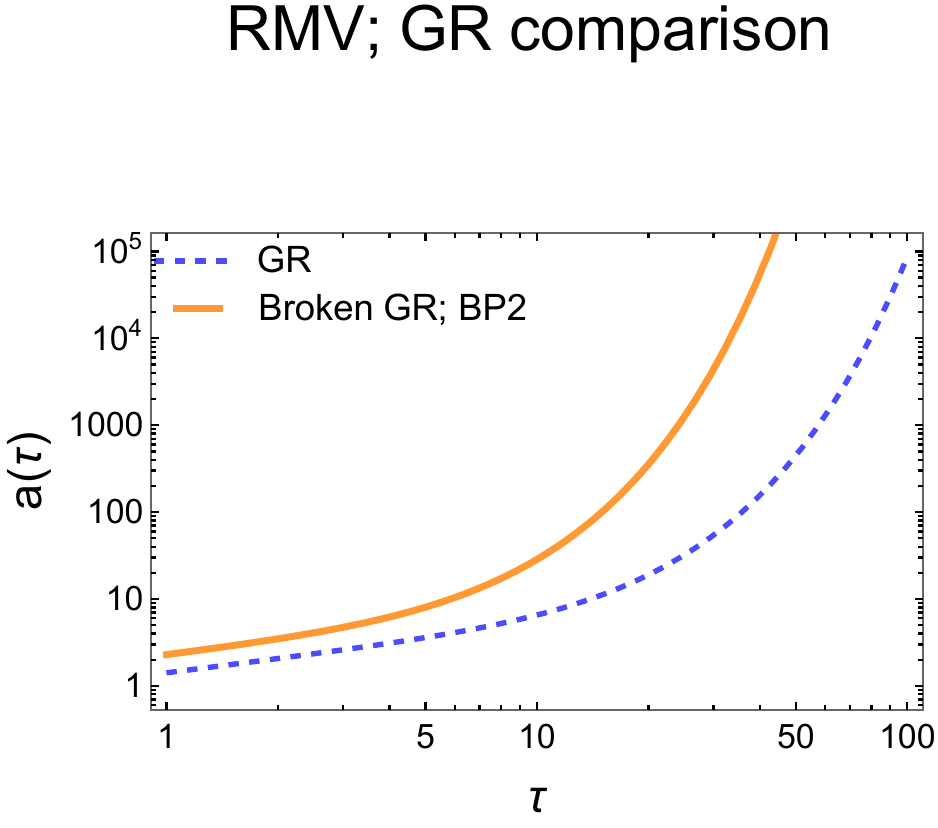}}\quad
\subfloat[c)]{\label{PlotMixGRBP3}\includegraphics[width=5.3cm]{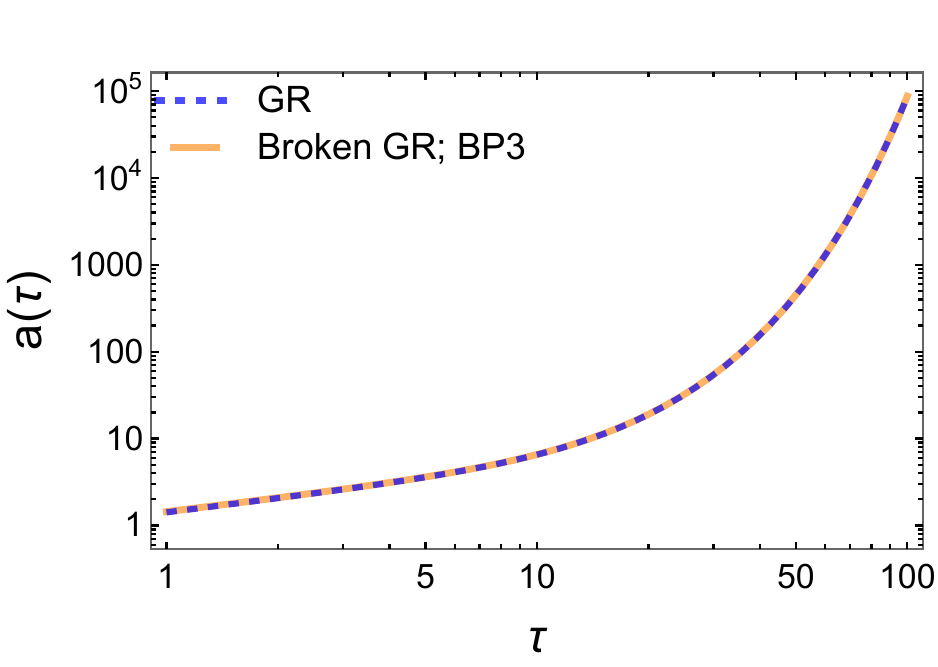}}\\
\subfloat[d) ]{\label{PlotsmixGRBP1ratio}\includegraphics[width=5.3cm]{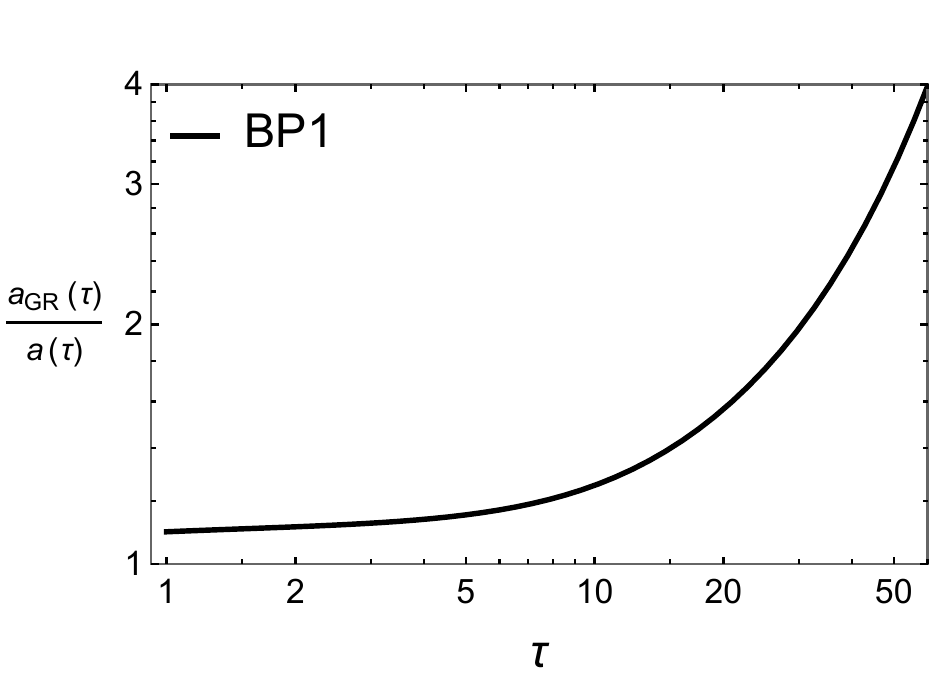}} \hspace{0.1cm} \vspace{-0.1cm}
\subfloat[e)]{\label{PlotsmixGRBP2ratio}\includegraphics[width=5.6cm]{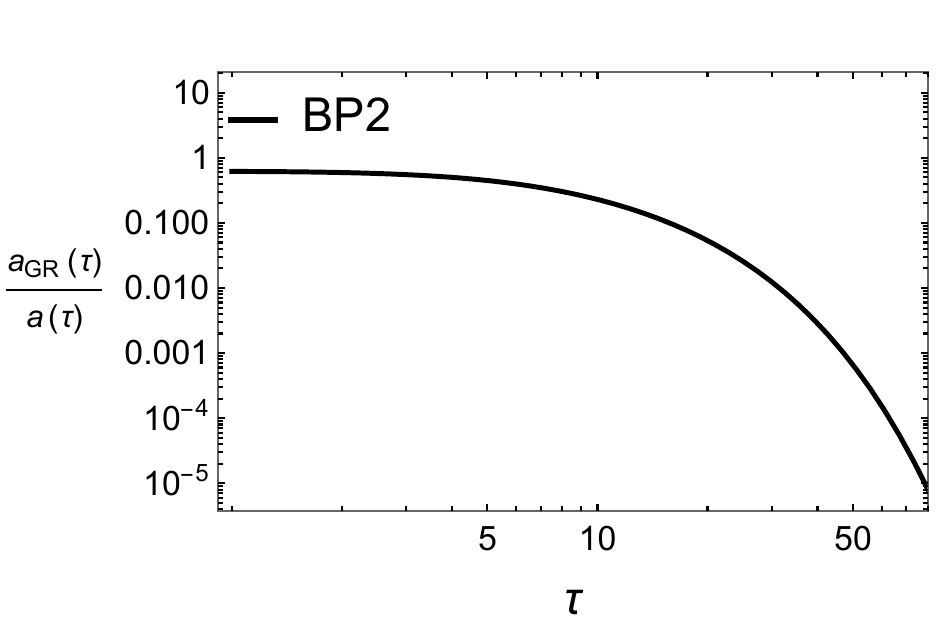}}\hspace{0.1cm}
\subfloat[f)]{\label{PlotsmixGRBP3ratio}\includegraphics[width=5.5cm]{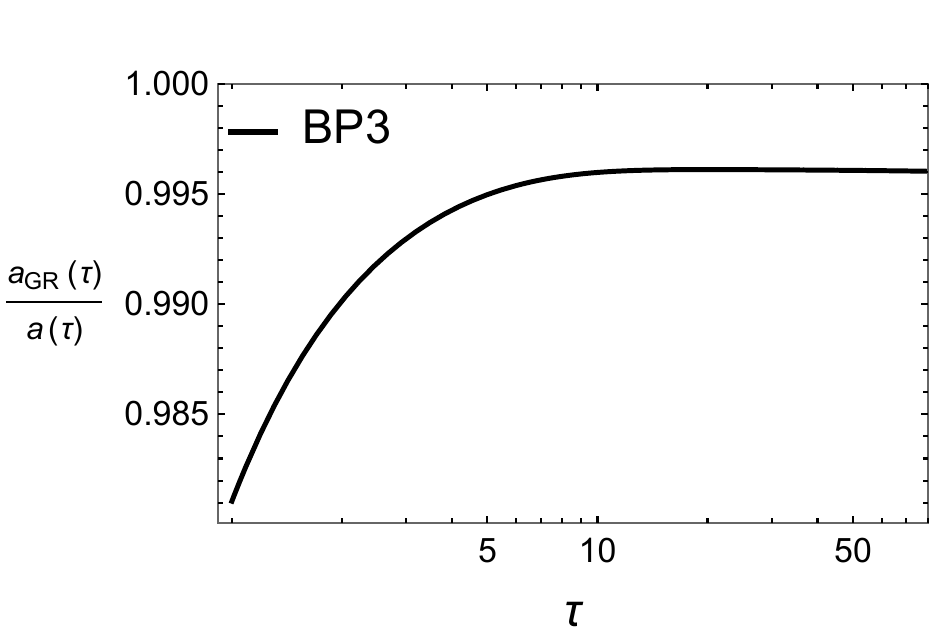}}\\
\hspace{0.2cm}\subfloat[g) ]{\label{PlotsmixBP1Hubble}\includegraphics[width=5.3cm]{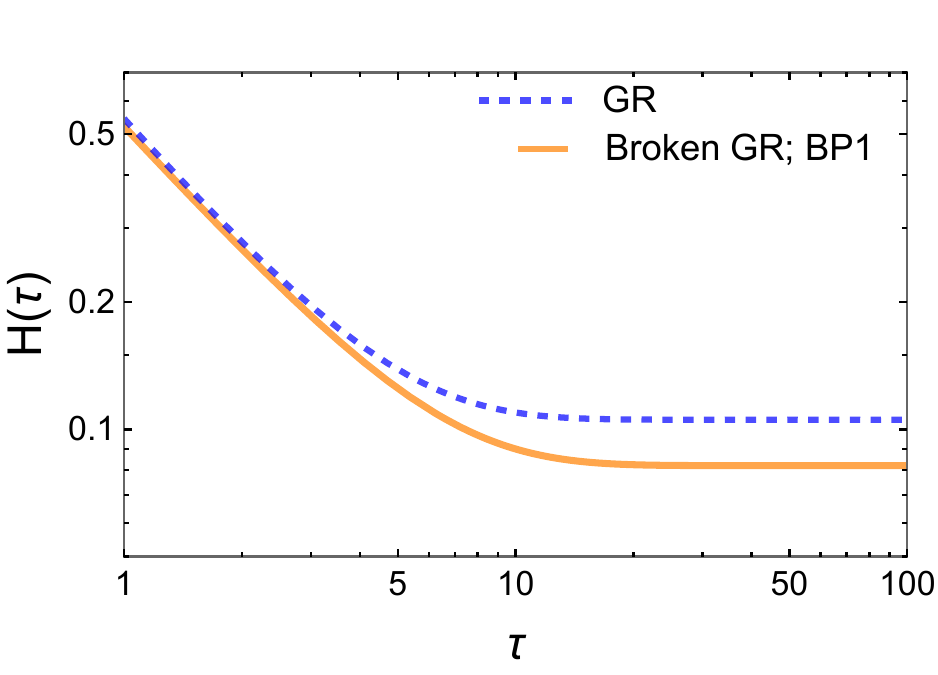}} \quad \hspace{0.05cm}
\subfloat[h)]{\label{PlotsmixBP2Hubble2}\includegraphics[width=5.3cm]{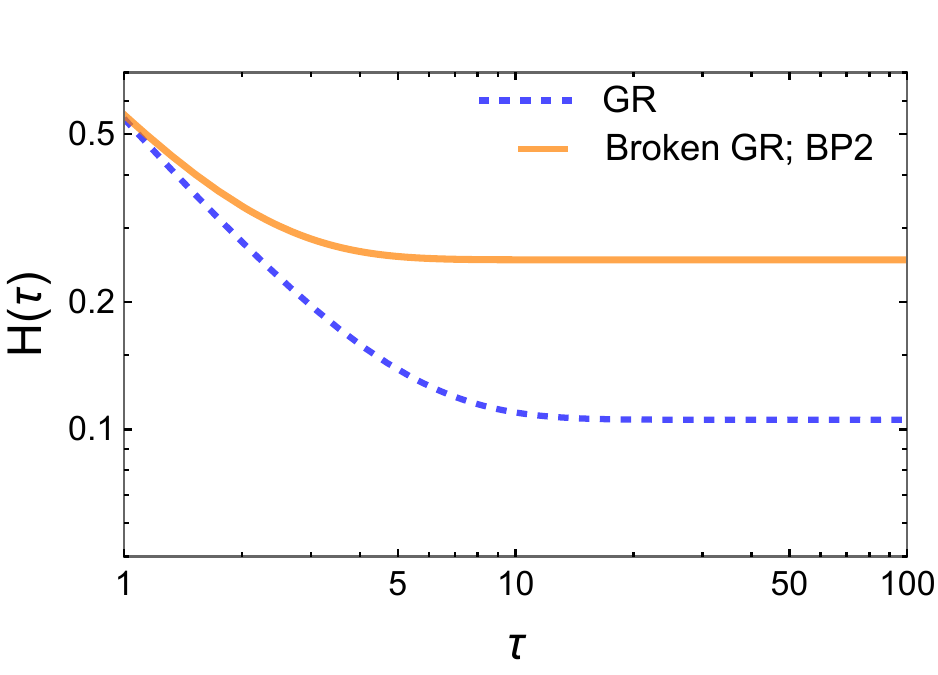}}\quad
\subfloat[i)]{\label{PlotsmixBP3Hubble}\includegraphics[width=5.3cm]{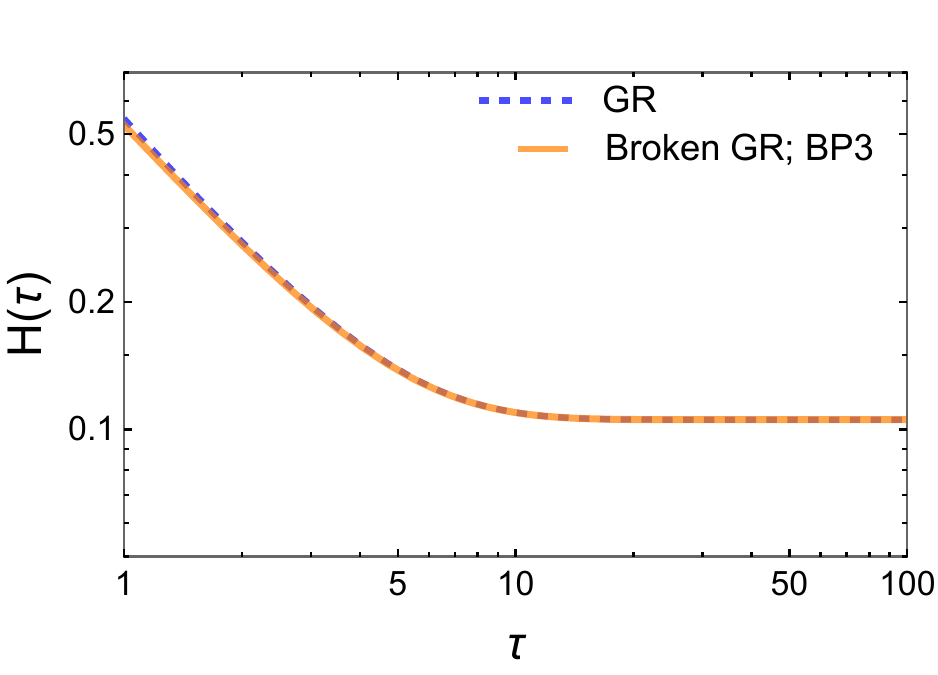}}
\end{tabular}
\caption{\label{PlotsmixcomparisonGR} Comparison of $a(\tau)$ solutions for different BPs (given in Table~\ref{BPs}) in the three-component (RMV) case.}   
\end{figure}

\begin{figure}[hbt!]
\captionsetup[subfigure]{labelformat=empty}
\centering
\hspace{-0.8cm}
\begin{tabular}{lll}
\subfloat[a) ]{\label{PlotMixBP1X}\includegraphics[width=5.2cm]{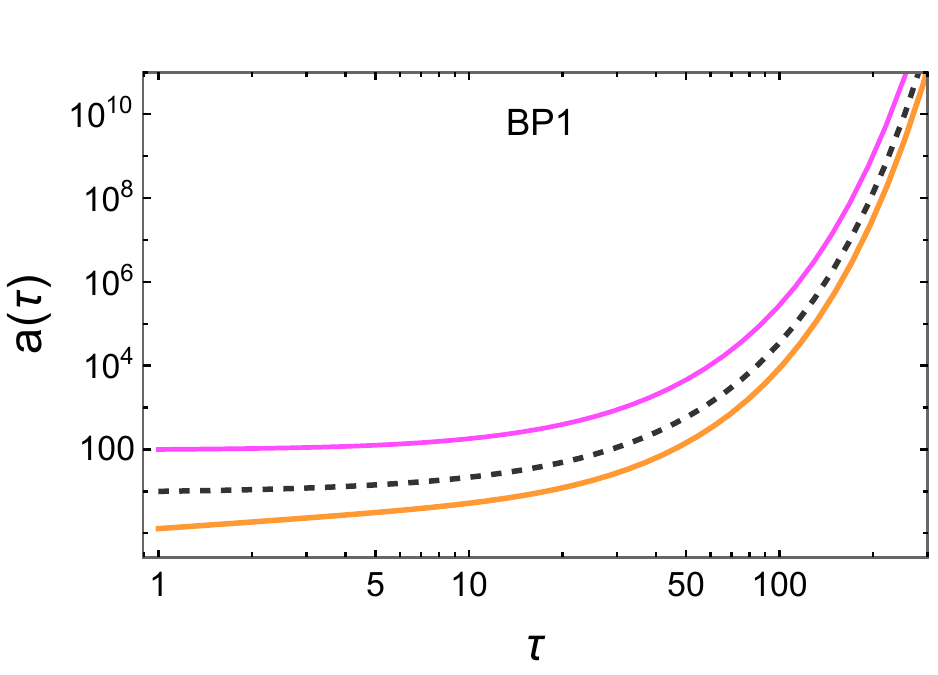}} \quad
\subfloat[b)]{\label{PlotMixBP2X}\includegraphics[width=5.25cm]{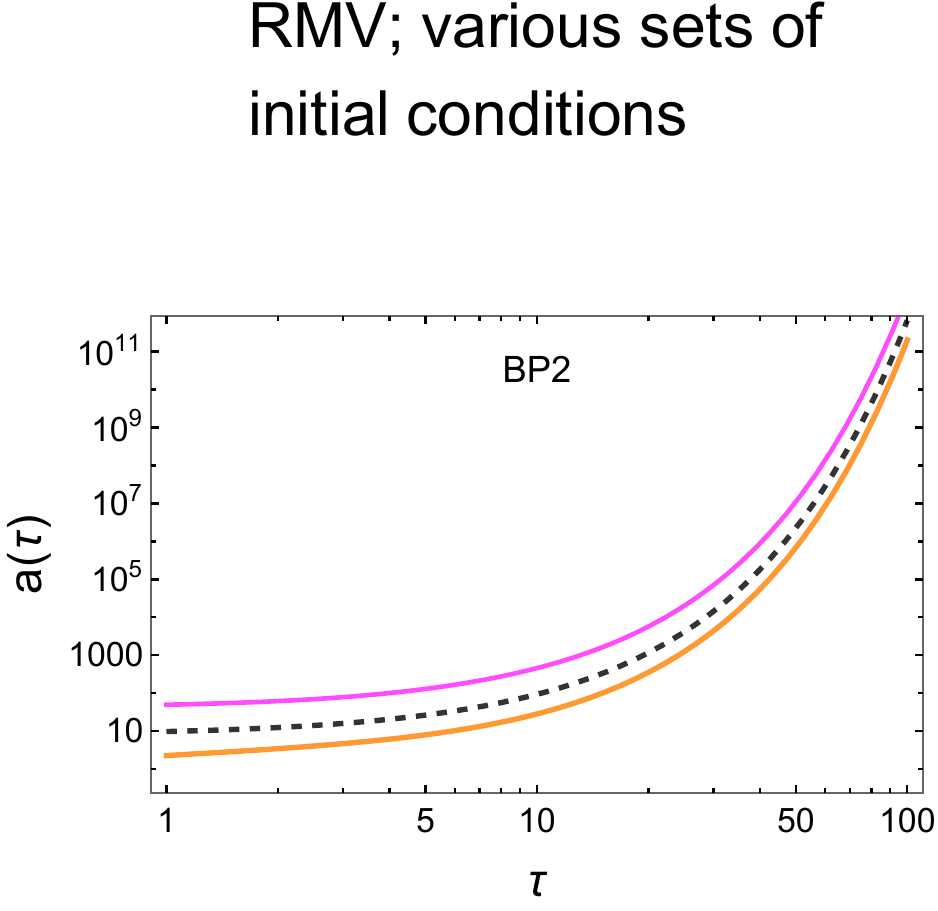}}\quad
\subfloat[c)]{\label{PlotMixBP3X}\includegraphics[width=5.2cm]{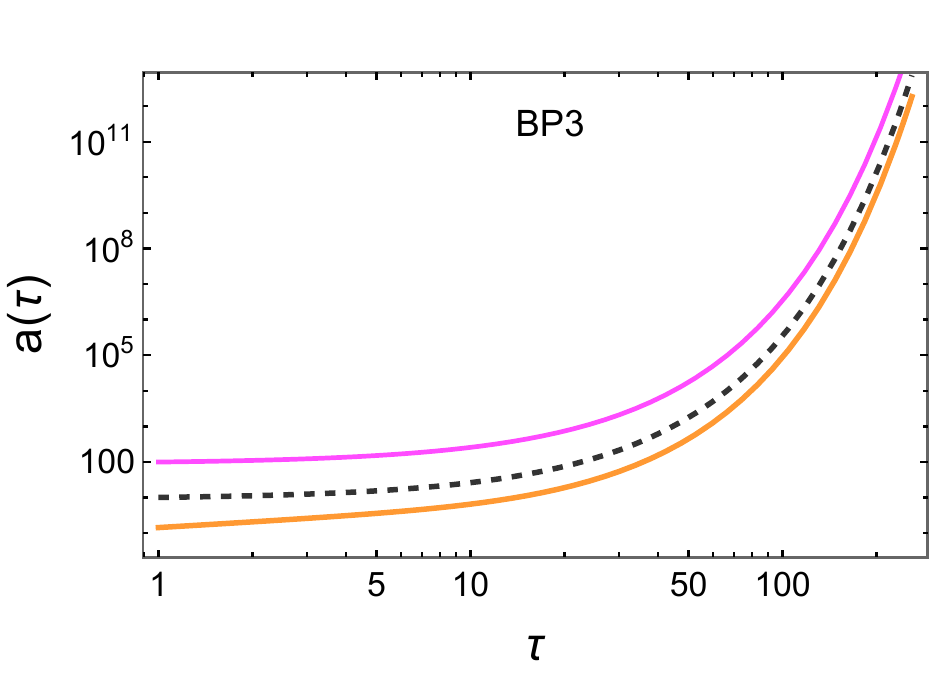}}\\
\subfloat[d) ]{\label{PlotMixBP1Xf}\includegraphics[width=5.3cm]{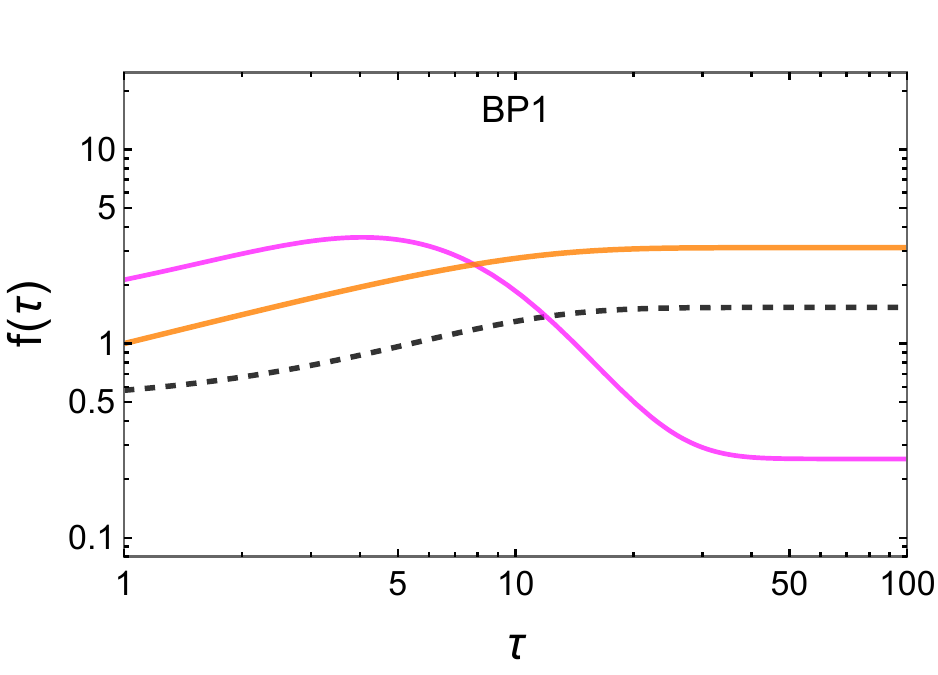}} \quad
\subfloat[e)]{\label{PlotMixBP2Xf}\includegraphics[width=5.3cm]{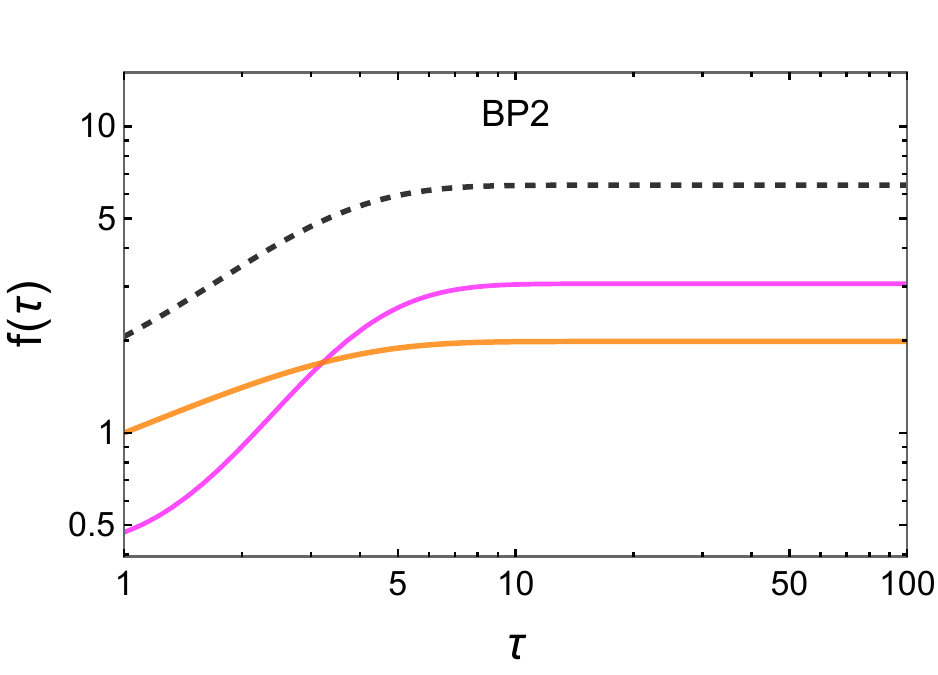}}\quad
\subfloat[f)]{\label{PlotMixBP3Xf}\includegraphics[width=5.3cm]{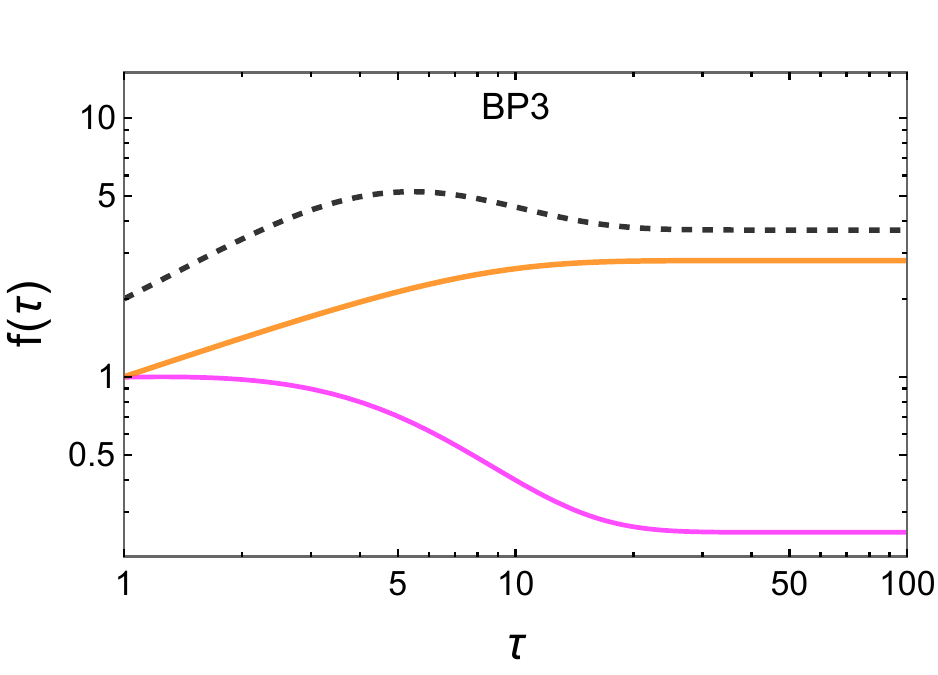}}\\
\subfloat[g) ]{\label{PlotMixBP1XHubble}\includegraphics[width=5.4cm]{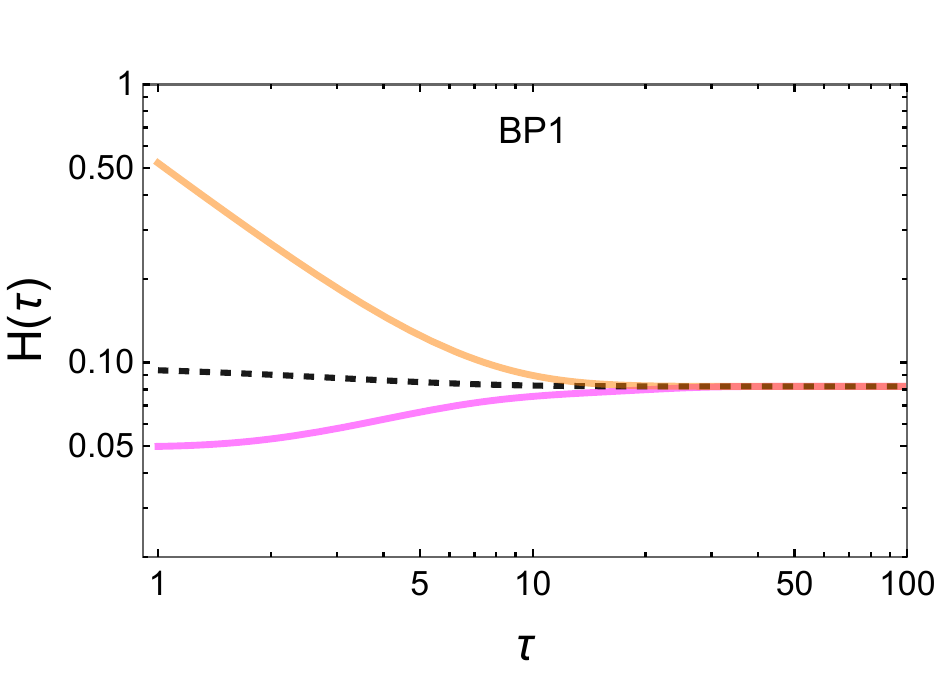}} \hspace{0.2cm}
\subfloat[h)]{\label{PlotMixBP2XHubble}\includegraphics[width=5.3cm]{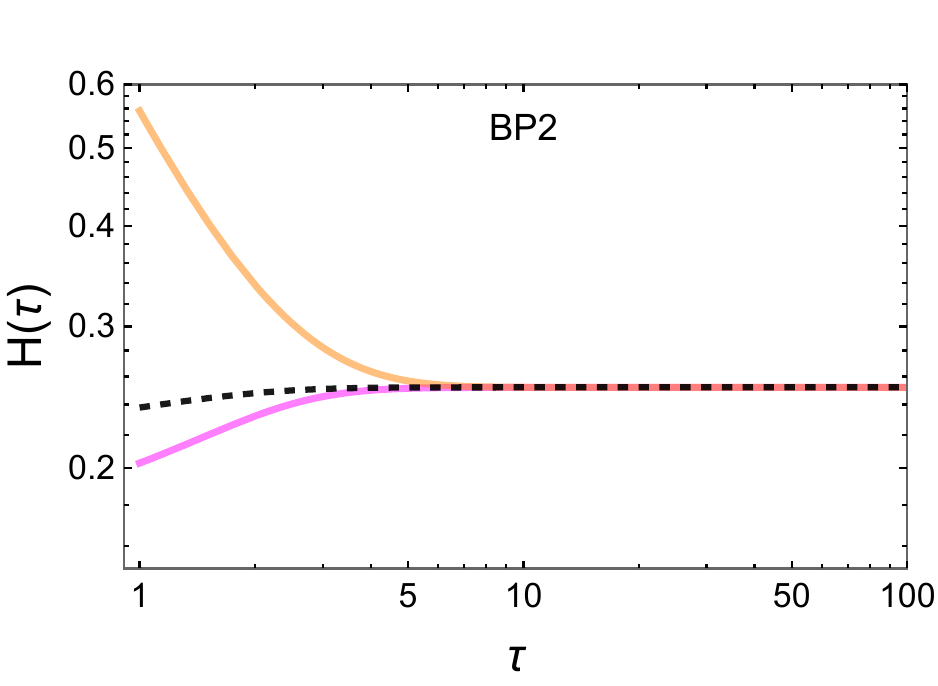}}\hspace{0.2cm}
\subfloat[i)]{\label{PlotMixBP3XHubble}\includegraphics[width=5.3cm]{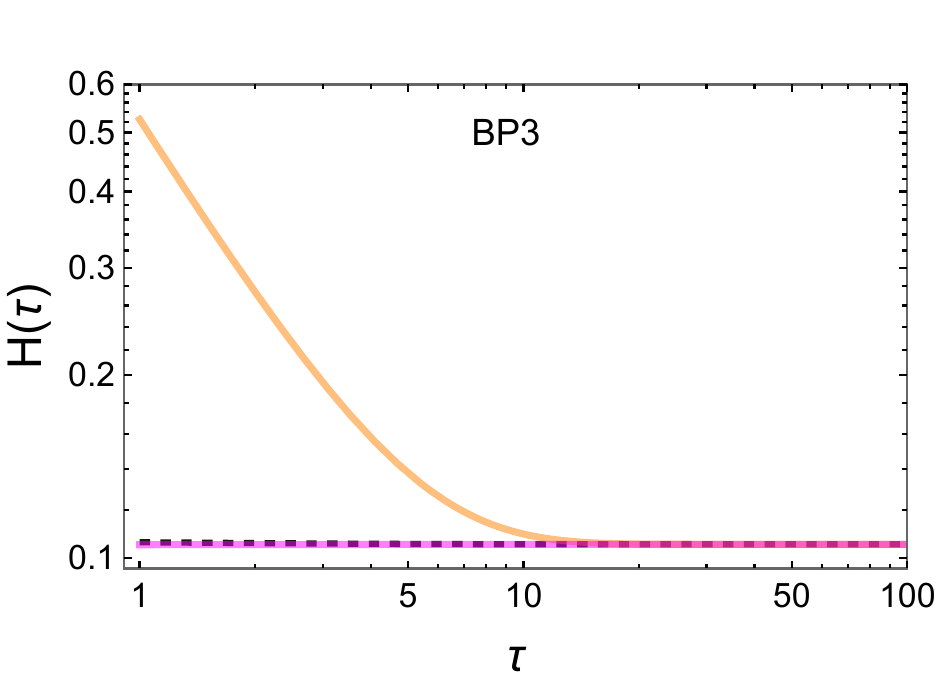}}
\end{tabular}
\caption{\label{Plotsmixcomparison} Comparison of $a(\tau)$,  $f(\tau)$, and  Hubble parameters for different initial conditions for the three-component (RMV) case, denoted by pink, orange, and dashed lines. The orange lines correspond to the same situation in the previous figure.}   
\end{figure}

Solution of $a(\tau)$ and $f(\tau)$ for BP1, given in Table~\ref{BPs}, is shown in Fig.~\ref{PlotmixBP1}. In Fig.~\ref{PlotMixBP1a}, the plot is given on a smaller scale to display the early behavior of the scale factor. Beginning with the initial conditions based on the exact radiation-only solution, given in Eq.~(\ref{radsolution}), the scale factor initially sits on the radiation-only solution, then exhibits matter-like behavior for a short interval and finally moves towards the exponential, vacuum-only solution, the last part of which is displayed in Fig.~\ref{PlotMixBP1b}. The comparison to GR behavior is also given in Fig.~\ref{PlotMixBP1b}, in which it is seen that the scale factor in the broken case begins close to the GR solution due to the initial condition selected and remains close up to the point where the vacuum behavior starts to dominate. The relatively big difference in the later stage is due to the rapid expansion of the vacuum-dominated stage. The coefficients $\alpha_a$ of BP1 (Table~\ref{BPs}) are not small enough to keep the scale factor close to the GR solution. For smaller coefficients, even several orders of magnitude,  the scale factor remains close to the GR case (as seen in Fig.~\ref{PlotMixGRBP3}, discussed below). The $f(\tau)$ solution is given in Fig.~\ref{PlotMixBP1c}, as well as the parametric plot in Fig.~\ref{PlotMixBP1d}. 

In Fig.~\ref{PlotsmixcomparisonGR}, we display the GR comparison. The deviation behavior is similar to the matter-vacuum case due to the late-time vacuum domination. For BP1 and BP2, the deviation grows with time since these $\alpha_a$ values are too large to keep $a(\tau)$ close to the GR counterpart at the beginning of the vacuum domination. For small enough $a(\tau)$, as for BP3, $a_{\tiny{\mbox{GR}}}(\tau)/a(\tau)$ eventually goes to a constant value, as desired. The Hubble parameter for each case is given in the bottom panel, which eventually settles into a constant value as expected from a late vacuum-dominated stage. In Fig.~\ref{PlotsmixcomparisonGR}, we display $a(\tau)$, $f(\tau)$, and $H(\tau)$ for substantially different sets of initial conditions. The situation is similar to the matter-vacuum case discussed above.
\newpage
\quad
\newpage
\section{Summary of the Results\label{sec:summary}}

 We have examined the cosmological evolution of the scale factor $a(t)$ in the diffemorphism-violating theory, given by the Lagrangian~(\ref{eqn:TotalLagrangian}), for different source configurations. Our purpose was to investigate whether there are solutions that continuously connect to GR, namely solutions that are nonsingular and yield the GR solutions in the $\{\alpha_n\}\rightarrow 0$ limit. Note that this is far from trivial (see footnote~\ref{example} for a simple example), as seen in the Pauli-Fierz massive gravity, where the diffeomorphism is broken at the linear level by the graviton mass term $m$~\cite{Hinterbichler:2011tt}; some of the physical implications of the massive theory differ from GR in the $m\rightarrow 0$ limit, indicating a discontinuity between the two theories and hence rendering the linear theory problematic. Similarly, in our case, such a behaviour would raise the question of the perturbativity of the theory and hence challenge the perturbative effective realization of the possible underlying diffeomorphism-violating theory.

Furthermore, we examined whether these solutions have the same time dependence as the GR counterparts, namely whether $a_{\tiny{\mbox{GR}}}(\tau)/a(\tau)$ is or approaches to a constant. This is important and should be addressed in addition to the question of $\{\alpha_n\}\rightarrow 0$ limit, mentioned above. Consider, for instance, a hypothetical solution for the radiation case in the form $a(\tau)\propto \frac{1}{1+J(\alpha_n)} (2 \tau)^{K(\alpha_m)+1/2}$, where $J$ and $K$ are some linear combinations of  $\alpha_n$. Such a scale factor would clearly be nonsingular and approach the GR solution, $a_{\tiny{\mbox{GR}}}(\tau)=(2 \tau)^{1/2}$, as $\{\alpha_n\}\rightarrow 0$. However, for finite $\{\alpha_n\}$, we would have some time-dependent deviation from the GR solution, namely time-dependent $a_{\tiny{\mbox{GR}}}(\tau)/a(\tau)$. Therefore, in that case, even minor modifications to GR may lead to substantial deviations in the observed cosmological evolution over time. 
 
We have found that no such discontinuity occurs for the theory given by the Lagrangian~(\ref{eqn:TotalLagrangian}).  In the radiation-only and vacuum-only cases, the solutions can be found analytically, allowing for the explicit demonstration of such good behavior. The vacuum-only case is an example of the behaviour described in the previous paragraph: the analytical solution approaches the GR counterpart as $\{\alpha_n\}$ values get smaller, but it always has some time-dependent deviation from GR (see Eq.~(\ref{ccsolution})), which will grow in time. That could have been a problem if it were to describe the evolution of our universe.  In the matter-only case, and in the more physically relevant double- and triple-fluid cases, the solutions are found numerically. It is observed that for cases where the vacuum is a subcomponent (including the most realistic, three-fluid case), we observed divergent behavior of the ratio $a_{\tiny{\mbox{GR}}}(\tau)/a(\tau)$ down to some small values of $\alpha_a$, which would indicate some potential issues for the observed evolution of the universe. However, for smaller values, even at the level of $10^{-3}$ for our example cases, we see controlled behavior of this ratio and hence consistent with our expectations (see the mid panels of Figs.~\ref{PlotscomparisonmatvacGR} and \ref{PlotsmixcomparisonGR}). Considering that the possible bounds on these diffemorphism-breaking terms are much smaller than this order of magnitude, as indicated by the results of Ref.~\cite{Anber:2009qp} (which found $\alpha_3 \lesssim 10^{-20}$), we conclude that diffemorphism-breaking effects are not expected to cause problems in the evolution of the scale factor. 

We also note that our results are independent of the conditions required for the linearized version of the theory to be free of ghosts and additional classical instabilities~\cite{Anber:2009qp}. These criteria, referred to as conditions I and II, are discussed below Eqs.~(\ref{constraint2}) and (\ref{ansatzmdu}).  The benchmark points BP1 and BP2 are given in Table~\ref{BPs},   with the former satisfying both conditions and the latter satisfying neither. We observe that a divergent behaviour for the scale factor occurs in both of these BPs, or neither. In cases where the divergent behaviour occurs for these BPs, it disappears for a smaller BP set, such as our BP3 given again in Table~\ref{BPs} (see the mid panels of Figs.~\ref{PlotscomparisonmatvacGR} and \ref{PlotsmixcomparisonGR}). This indicates that the issue lies with the order of magnitude of the parameters rather than with the satisfaction of the linearized-theory conditions. Consequently, our results suggest that the presence of ghosts does not induce instabilities in the evolution of the scale factor. 

\section{Conclusion\label{sec:conclusion}}
 
We investigated the cosmological evolution of the scale factor in gravity with explicitly broken diffeomorphism invariance, while keeping Lorentz invariance intact, in the effective theory framework. Our motivation was the following. If the diffeomorphism invariance of General Relativity (GR) emerges from an underlying theory that does not have this symmetry, the emergence may not be perfect, and there may be small violations of it at low energies, which can be realized as explicit symmetry-breaking terms in the effective Lagrangian. The PPN analysis performed in Ref.~\cite{Anber:2009qp} points to extremely stringent upper bounds on the possible violations. The question of whether such small violations yield a slight deviation from the outcome of the diffeomorphism invariant theory (GR) or lead to substantial differences is crucial. As explained in the previous section, this may or may not be related to whether the broken theory satisfies continuity conditions, namely, whether it smoothly connects to GR in the limit of vanishing symmetry-breaking effects. The answers to these questions can have implications regarding the validity of the perturbative effective description of the underlying theory (as in the case of Pauli-Fierz massive gravity). The cosmological evolution is a convenient ground to probe the corresponding classical effects due to its well-established status.

The effective Lagrangian contains all the possible symmetry-breaking terms up to two derivative order in the energy expansion, in addition to the usual symmetric term, namely the Einstein-Hilbert term (i.e., GR). The symmetry-breaking terms are different contractions of two factors of Christoffel connection coefficients, which are in the same order as the usual Ricci scalar. We have studied the systems with single-component and multicomponent fluids and found that small violations of general covariance do \textit{not} cause instabilities in the evolution of the scale factor, yielding close results to GR, except for the vacuum-only case. Furthermore, all the solutions (including the vacuum-only case)  show no signs of discontinuity with the diffeomorphism-invariant theory, GR, in the $\alpha\rightarrow 0$ limit. Another interesting question is whether the non-satisfaction of the conditions that prevent ghost-like and classical instabilities in the linearized theory would lead to problems in the nonlinear theory regarding the cosmological evolution. We haven't observed a significant difference in the behavior of the scale factor regarding the satisfaction of these conditions.

As a result, the emergence of General Relativity from an underlying theory without diffeomorphism invariance does \textit{not} receive a blow from the characteristics of the cosmological evolution of the scale factor.  Note that this itself does not present an argument in favor of or against either diffeomorphism-violating terms in the effective theory or the emergent gravity paradigm (recall that this is not the purpose of this paper), it just states that these ideas are not inconsistent with the evolution of the universe in the leading order in the energy expansion in the Lorentz invariant effective theory. Additional phenomenological studies on the effective theory considered in this paper could bring new challenges to the emergent gravity paradigm.  

\section*{Acknowledgment}
U.A.'s and partially M.E.'s works are funded by The Scientific and Technological Research Council of T\"urkiye (T\"UB\.ITAK) B\.{I}DEB 2232-A program under project number 121C067. M.E. is also supported by İstanbul Bilgi University research fund BAP with grant no: 2024.01.009 and by The Scientific and Technological Research Council of T\"urkiye (TÜBİTAK) under grant number 125F021. U.A. thanks Mohamed Anber and John F. Donoghue for their collaboration in the early stages of this work and their comments on the manuscript.

\appendix

\section{Possible but dependent diffeomorphism breaking terms}
\label{depapp}

Here, we will examine the two-derivative terms that can be written in terms of those we include in our Lagrangian~(\ref{diffLs}), $\mathcal{ L}_1...\mathcal{ L}_5$, and show that they do not lead to independent contributions.  In addition to the terms $\mathcal{ L}_1...\mathcal{ L}_5$, which include two factors of Christoffel connections, we may also have terms composed of the derivatives of a single connection. These terms actually exist in the definition of the Ricci scalar, and they can indeed be written in terms $\mathcal{ L}_1$ and $\mathcal{ L}_2$, as can be seen in obtaining Eq.~(\ref{Ricci}).  Let us examine them in detail. We have 2 possibilities in total: 
\begin{equation}
\mathcal{L}_6 = - g^{\rho\mu}\partial_\mu \Gamma^\nu_{\nu \rho},  \quad  \mathcal{ L}_7=- g^{\nu\rho} \partial_\mu \Gamma^\mu_{\nu\rho}.
\end{equation}
For $\mathcal{L}_6$, compute
\begin{eqnarray}
S_6 &=&  \int  d^4 x \sqrt{-g}\ \mathcal{L}_6     = - \int d^4 x \sqrt{-g}\ g^{\rho\mu} \partial_\mu \Gamma^\nu_{\nu \rho} \nonumber \\
&=& - \int d^4 x \ \partial_\nu    (  \sqrt{-g}  g^{\mu\nu} \Gamma^\lambda_{\mu\lambda} )+    \int d^4 x \  \Gamma^\lambda_{\mu\lambda}\partial_\nu (\sqrt{-g} g^{\mu\nu})\nonumber \\
&=&- \int d^4 x \ \partial_\nu    (  \sqrt{-g}  g^{\mu\nu} \Gamma^\lambda_{\mu\lambda} )+  \int d^4x \ \sqrt{-g} \Gamma^\lambda_{\mu\lambda}  \big(  g^{\mu\nu} \Gamma^\alpha_{\alpha\nu} - g^{\mu\alpha} \Gamma^\nu_{\alpha\nu} - g^{\nu\alpha} \Gamma^\mu_{\nu\alpha}    \big) \quad \nonumber \\
&=& \mathrm{surface \ terms}   +\int d^4x \sqrt{-g}\ \mathcal{L}_2 ,
\end{eqnarray}
where we have used the identities $\partial_\nu \sqrt{-g} =\sqrt{-g} \; \Gamma^\alpha_{\alpha\nu}$ and $\partial_\nu g^{\mu\lambda} = - g^{\mu\alpha}\Gamma^\lambda_{\nu\alpha} - g^{\lambda\alpha}\Gamma^\mu_{\nu\alpha}$. Similarly, for $\mathcal{L}_7$ we have
\begin{eqnarray}
S_7 &=& \int d^4 x \sqrt{-g}\ \mathcal{L}_7 = - \int d^4x\ \sqrt{-g} g^{\nu\rho}\partial_\mu\Gamma^\mu_{\nu\rho} \nonumber\\
&=& - \int d^4 x\  \partial_\mu \big( \sqrt{-g} g^{\nu\rho}\Gamma^\mu_{\nu\rho}  \big) + \int d^4 x\  \Gamma^\mu_{\nu\rho} \partial_\mu (\sqrt{-g} g^{\nu\rho}) \nonumber\\
&=& - \int d^4 x\  \partial_\mu \big( \sqrt{-g} g^{\nu\rho}\Gamma^\mu_{\nu\rho}  \big)  + \int d^4x  \sqrt{-g} \ \big( g^{\nu\rho}\Gamma^\mu_{\nu\rho} -g^{\nu\alpha}\Gamma^\mu_{\nu\rho} \Gamma^\rho_{\mu\alpha} -g^{\rho\alpha} \Gamma^\mu_{\nu\rho}\Gamma^\nu_{\mu\alpha}  \big)\nonumber\\
&=& \mathrm{surface  \ terms} - \int d^4 x \sqrt{-g}\ \mathcal{L}_2 + 2 \int d^4x \sqrt{-g}\ \mathcal{L}_1 . 
\label{S7}
\end{eqnarray}

Note that $\sqrt{-g} R=\sqrt{-g} (\mathcal{L}_1-\mathcal{L}_2+\mathcal{L}_6-\mathcal{L}_7)=\sqrt{-g} (\mathcal{L}_2-\mathcal{L}_1)\;+\; \mathrm{surface  \ terms}$, as given in Eq.~(\ref{Ricci}).

Other than the above terms that include derivatives of a single factor of Christoffel connections, one can also consider terms with two explicit derivatives. Consider e.g.,  $\sqrt{-g}\partial_\alpha\partial_\beta g^{\alpha\beta}$. The resulting action boils down to our set of Lagrange densities as 
\begin{eqnarray}
\int d^4 x \sqrt{-g}\partial_\alpha\partial_\beta g^{\alpha\beta} &=& \int d^4 x\ \partial_\alpha \Big(  \sqrt{-g} \partial_\beta g^{\alpha\beta} \Big) - \int d^4 x\ \big(\partial_\alpha \sqrt{-g}\big) \partial_\beta g^{\alpha\beta} \nonumber\\
&=&  \mathrm{surface  \ terms} + \int d^4 x\ \sqrt{-g}\ \Gamma^\nu_{\nu\alpha} \Big(  g^{\lambda\beta} \Gamma^\alpha_{\beta\lambda} +g^{\alpha\gamma}\Gamma^\beta_\beta\gamma \Big) \nonumber\\
&=& \mathrm{surface  \ terms} -\int d^4 x\ \sqrt{-g}( \mathcal{L}_2 + \mathcal{L}_5 ), 
\end{eqnarray} 
where we have used $\partial_\nu g^{\mu\lambda} = - g^{\mu\alpha}\Gamma^\lambda_{\nu\alpha} - g^{\lambda\alpha}\Gamma^\mu_{\nu\alpha}$. 
Similarly, another possible term, $\sqrt{-g} \partial^\gamma g_{\alpha\beta} \partial_\gamma g^{\alpha\beta}, $ can be expressed as a combination such that
\begin{equation}
\int d^4 x \ \sqrt{-g} \partial^\gamma g_{\alpha\beta} \partial_\gamma g^{\alpha\beta} = 2 \int d^4 x \sqrt{-g} \big( \mathcal{L}_1 + \mathcal{L}_3  \big)\;. 
\end{equation}

Therefore, we conclude that the densities $\mathcal{L}_a$ (\ref{diffLs}) are the only independent contributions.  

Let us also illustrate quickly how such terms break diffeomorphism invariance by considering the last contribution in (\ref{S7}), i.e., $\mathcal{L}_1$. According to (\ref{passive}), 
\begin{eqnarray}  
g^{\mu\nu} \to g^{\rho\sigma}\frac{\partial x^{\prime \mu} }{\partial x^\rho} \frac{\partial x^{\prime \nu}}{\partial x^\sigma},   \quad   \Gamma^{\rho}_{\mu\nu} \to 
\frac{\partial x^{\alpha}}{\partial x’^{\mu}}\,
\frac{\partial x^{\beta}}{\partial x’^{\nu}}\,
\frac{\partial x’^{\rho}}{\partial x^{\sigma}}\,
\Gamma^{\sigma}_{\alpha\beta}
\;-\;
\frac{\partial x^{\alpha}}{\partial x’^{\mu}}\,
\frac{\partial x^{\beta}}{\partial x’^{\nu}}\,
\frac{\partial^{2} x’^{\rho}}{\partial x^{\alpha}\,\partial x^{\beta}}\;.
\end{eqnarray}
The non-tensorial transformation rule of the Christoffel symbol can be clearly seen above. Since Christoffel symbols explicitly appear in our terms like $\mathcal{L}_1$ (\ref{diffLs}), these extra pieces are responsible for broken diffeomorphism invariance.

\section{Diffeomorphism invariance breaking terms in the field equations} 
\label{expMapp}
Here, we explicitly present our results of diffeomorphism-invariance-breaking contributions to Einstein's equation, $M_{\mu\nu}^{(a)}$, given in Eq.~(\ref{EE51}). 

\noindent {\bf{First term}}:
Let's consider the first diffeomorphism breaking Lagrangian term in (\ref{diffLs}), i.e.,
\begin{equation}
\label{L1}
L_1 = - g^{\mu\nu} \Gamma^\alpha_{\mu\lambda} \Gamma^\lambda_{\nu\alpha}. 
\end{equation}
Its variation results in the following contribution
\begin{equation}
\label{M1}
M_{\mu\nu}^{(1)} = \frac{1}{2} g_{\mu\nu} g^{\beta\gamma} \Gamma^\alpha_{\beta\lambda} \Gamma^\lambda_{\alpha\gamma} + \Gamma^\lambda_{\mu\alpha} \Gamma^\alpha_{\nu\lambda} - \Gamma^\beta_{\beta\alpha} \Gamma^\alpha_{\mu\nu} - \partial_\alpha\Gamma^{\alpha}_{\mu\nu}.
\end{equation}
\vspace{0.2cm}

\noindent {\bf{Second term}}:
We take the next one which is 
\begin{equation}
\label{L2}
L_2 = - g^{\mu\nu} \Gamma^\alpha_{\mu\nu} \Gamma^\lambda_{\lambda\alpha}\;,
\end{equation}
and find its following contribution as
\begin{eqnarray}
\label{M2}
M_{\mu\nu}^{(2)}  = &-&\frac{1}{2} g_{\mu\nu}  g^{\theta\beta}\Gamma^\lambda_{\lambda\alpha} \Gamma^\alpha_{\theta\beta} + g_{\mu\nu}g^{\beta\lambda} \Gamma^\gamma_{\lambda\theta} \Gamma^\theta_{\gamma\beta} 
+ \frac{1}{2} g_{\mu\nu} (g^{\alpha\gamma} \partial_\gamma \Gamma^\lambda_{\lambda\alpha} - g^{\theta\lambda} \partial_\gamma \Gamma^\gamma_{\lambda\theta})\nonumber\\
 &-& \frac{1}{2} (\partial_\mu \Gamma^\lambda_{\lambda\nu} +\partial_\nu \Gamma^\lambda_{\lambda\mu} ). \qquad 
\end{eqnarray} 
\vspace{0.2cm}

\noindent {\bf{Third term}}:
We continue with 
\begin{equation}
\label{L3}
L_3 = - g^{\alpha\gamma} g^{\beta\rho} g_{\mu\nu} \Gamma^\mu_{\alpha \beta} \Gamma^\nu_{\gamma\rho}.
\end{equation}
Varying the related action and after integration by parts we get
%
\begin{eqnarray}
\label{M3}
M_{\mu\nu}^{(3)} &=&\frac{1}{2} g_{\mu\nu} g_{\lambda\sigma} g^{\alpha\gamma} g^{\beta\rho} \Gamma^\lambda_{\alpha\beta}\Gamma^\sigma_{\rho\gamma} + g_{\mu\lambda} g_{\nu\sigma} g^{\alpha\gamma} g^{\beta\rho} \Gamma^\lambda_{\alpha\beta} \Gamma^\sigma_{\rho\gamma} 
-2 g^{\alpha\gamma} g_{\beta\lambda} \Gamma^\beta_{\mu\gamma} \Gamma^\lambda_{\nu\alpha}  \nonumber \\
&+&(g_{\mu\lambda} \Gamma^\lambda_{\nu\alpha} + g_{\nu\lambda} \Gamma^\lambda_{\mu\alpha})  g^{\beta\gamma}\Gamma^\alpha_{\gamma\beta} - g^{\alpha\gamma} (g_{\mu\lambda}\partial_{\gamma}\Gamma^\lambda_{\nu\alpha} + g_{\nu\lambda}\partial_{\gamma}\Gamma^\lambda_{\mu\alpha} )\nonumber\\
&-& 2 \Gamma^\lambda_{\mu\sigma}\Gamma^\sigma_{\nu\lambda} + \Gamma^\gamma_{\mu\nu} \Gamma^\beta_{\beta \gamma} + \partial_\gamma \Gamma^\gamma_{\mu\nu}\;.
\end{eqnarray}
Let us note that in \cite{Anber:2009qp}, the effects of $L_3$ and its contribution (\ref{M3}) were investigated within the Parametrized post-Newtonian (PPN) formalism. Note that in Ref.~\cite{Anber:2009qp}, there is an overall sign error in $M_{\mu\nu}^{(3)}$ and several missing terms, which do not affect the conclusion of Ref.~\cite{Anber:2009qp} regarding the bound on $\alpha_3$.
\vspace{0.7cm}

\noindent {\bf{Forth term}}: That fourth piece
\begin{equation}
\label{L4}
 L_4 = - g^{\alpha\gamma} g_{\beta\lambda} g^{\mu\nu} \Gamma^\lambda_{\mu\nu} \Gamma^\beta_{\gamma\alpha},
\end{equation} 
 results in the following expression.
\begin{eqnarray}
M_{\mu\nu}^{(4)} &=&\left( (\frac{1}{2} g_{\mu\nu} g_{\beta\lambda} + g_{\mu\beta} g_{\nu\lambda})g^{\eta\delta} \Gamma^\lambda_{\eta\delta} + g_{\mu\nu} \Gamma^\lambda_{\lambda\beta} \right) g^{\alpha\gamma} \Gamma^\beta_{\alpha\gamma} - 2 g_{\mu\nu} \Gamma^\beta_{\alpha\gamma} g^{\lambda\gamma} \Gamma^\alpha_{\beta\lambda} \nonumber \\  
&-& (g_{\mu\lambda} \Gamma^\lambda_{\nu\beta}
+ g_{\nu\lambda}\Gamma^\lambda_{\mu\beta}) g^{\alpha\gamma} \Gamma^\beta_{\alpha\gamma}+ 2 (g_{\mu\beta} \Gamma^\alpha_{\nu\lambda} + g_{\nu\beta} \Gamma^\alpha_{\mu\lambda})  \Gamma^\beta_{\alpha\gamma} g^{\lambda\gamma}\nonumber\\ 
&-& 2 g^{\alpha\gamma} g_{\lambda\beta} \Gamma^\beta_{\alpha\gamma} \Gamma^\lambda_{\mu\nu} + g^{\alpha\gamma} (g_{\mu\nu} \partial_\beta\Gamma^\beta_{\alpha\gamma} - g_{\mu\beta} \partial_\nu \Gamma^\beta_{\alpha\gamma} - g_{\nu\beta} \partial_\mu \Gamma^\beta_{\alpha\gamma})\;. 
\end{eqnarray}
%

\vspace{0.7cm}

\noindent {\bf{Fifth term}}:  When the last term 
\begin{equation}
\label{L5}
L_5 = - g^{\alpha\beta} \Gamma^\lambda_{\lambda\alpha} \Gamma^\mu_{\mu\beta},
\end{equation}
is taken into account one obtains 
\begin{equation}
 \label{M5}
M_{\mu\nu}^{(5)} = g_{\mu\nu} \left(  \frac{1}{2} g^{\alpha\beta} \Gamma^\lambda_{\lambda\alpha} \Gamma^\sigma_{\sigma\beta} + g^{\alpha\lambda}\Gamma^\beta_{\alpha\lambda}\Gamma^\sigma_{\sigma\beta} - g^{\alpha\beta} \partial_\alpha \Gamma^\lambda_{\lambda\beta}  \right) - \Gamma^\lambda_{\lambda \mu} \Gamma^\sigma_{\sigma\nu}.
\end{equation}
In \cite{Bello-Morales:2023btf}, only $L_5$ and its contribution to Einstein's equations (\ref{M5}) were considered.


\section{Constraint Equation}
\label{Appconstraint} 

Let us explicitly compute (\ref{con1}) term by term.

\noindent {\bf{First term}}: 
For the first contribution $M_{\mu\nu}^{(1)}$ (\ref{M1}), we have found 
\begin{eqnarray}
M_{00}^{(1)} = \frac{3}{2} \left( \frac{\dot{a}^2}{a^2} + \frac{\dot{f}^2}{f^2} \right) - \frac{3 \dot{a} \dot{f}}{a f} - \frac{\ddot{f}}{f}, \quad \quad 
M_{ij}^{(1)} = - \frac{a^2 \delta_{ij}}{2f^2} \left(  \frac{\dot{a}^2}{a^2} -  \frac{2\dot{a}\dot{f}}{a f} + \frac{2\ddot a}{a} + \frac{\dot{f}^2}{f^2}  \right), \quad
\end{eqnarray}
 which results in
\begin{equation}
\label{conM1}
\nabla^\mu M_{\mu 0}^{(1)} = - \frac{\dddot{f} }{f} + \frac{6 \ddot{f}}{f} \left( \frac{\dot{f}}{f} - \frac{\dot{a}}{a} \right) - \frac{3 \ddot{a} \dot{f}}{af} - \frac{6 \dot{f}^3}{f^3} - \frac{6 \dot{a}^2 \dot{f}}{a^2f} + \frac{12 \dot{a} \dot{f}^2}{af^2}. 
\end{equation}

\vspace{0.5cm}

\noindent {\bf{Second term}}: 
The components of $M_{\mu\nu}^{(2)}$ (\ref{M2}) are 
\begin{eqnarray}
M_{00}^{(2)} = \frac{3}{2} \left( \frac{3\dot{a}^2}{a^2} + \frac{\dot{f}^2}{f^2} \right) - \frac{3 \dot{a} \dot{f}}{a f} - \frac{\ddot{f}}{f}, \quad \quad 
M_{ij}^{(2)} = - \frac{a^2 \delta_{ij}}{2f^2} \left(  \frac{3\dot{a}^2}{a^2} -  \frac{6\dot{a}\dot{f}}{a f} + \frac{6\ddot a}{a} + \frac{\dot{f}^2}{f^2}  \right).\quad 
\end{eqnarray}
We realize that (\ref{con1}) yields the same equation as (\ref{conM1})
\begin{equation}
\label{conM2}
\nabla^\mu M_{\mu 0}^{(2)} = - \frac{\dddot{f} }{f} + \frac{6 \ddot{f}}{f} \left( \frac{\dot{f}}{f} - \frac{\dot{a}}{a} \right) - \frac{3 \ddot{a} \dot{f}}{af} - \frac{6 \dot{f}^3}{f^3} - \frac{6 \dot{a}^2 \dot{f}}{a^2f} + \frac{12 \dot{a} \dot{f}^2}{af^2} . 
\end{equation}

\vspace{0.5cm}

\noindent {\bf{Third term}}: 
The third term $M_{\mu\nu}^{(3)}$ (\ref{M3}) yields
\begin{eqnarray}
M_{00}^{(3)} = \frac{3}{2} \left( -\frac{3\dot{a}^2}{a^2} + \frac{\dot{f}^2}{f^2} \right) - \frac{3 \dot{a} \dot{f}}{a f} - \frac{\ddot{f}}{f}, \quad \quad
M_{ij}^{(3)} = - \frac{a^2 \delta_{ij}}{2f^2} \left( - \frac{3\dot{a}^2}{a^2} + \frac{6\dot{a}\dot{f}}{a f} - \frac{6\ddot a}{a} + \frac{\dot{f}^2}{f^2}  \right). \quad 
\end{eqnarray}
 When it comes to (\ref{con1}), again we have the same equation:
\begin{equation}
\label{conM3}
\nabla^\mu M_{\mu 0}^{(3)} = - \frac{\dddot{f} }{f} + \frac{6 \ddot{f}}{f} \left( \frac{\dot{f}}{f} - \frac{\dot{a}}{a} \right) - \frac{3 \ddot{a} \dot{f}}{af} - \frac{6 \dot{f}^3}{f^3} - \frac{6 \dot{a}^2 \dot{f}}{a^2f} + \frac{12 \dot{a} \dot{f}^2}{af^2}. 
\end{equation}

\vspace{0.5cm}

\noindent {\bf{Fourth term}}: 
Here are the results regarding the fourth term i.e., $M_{\mu\nu}^{(4)}$: 
\begin{eqnarray}
M_{00}^{(4)} &=& \frac{3}{2} \left( \frac{\dot{a}^2}{a^2} + \frac{\dot{f}^2}{f^2} \right) - \frac{3 \dot{a} \dot{f}}{a f} - \frac{\ddot{f}}{f} + 3 \frac{\ddot{a}}{a}\;, \nonumber\\  
M_{ij}^{(4)} &=& - \frac{a^2 \delta_{ij}}{2f^2} \left( - \frac{3\dot{a}^2}{a^2} + \frac{6\dot{a}\dot{f}}{a f} - \frac{6\ddot a}{a} - \frac{3\dot{f}^2}{f^2}  +  \frac{2\ddot{f}}{f^2}\right). \quad 
\end{eqnarray} 
Note that the extra $\ddot{a}$  and $\ddot{f}$ above. They lead to some modifications and this time (\ref{con1}) turns out to be
\begin{equation}
\label{conM4}
\nabla^\mu M_{\mu 0}^{(4)} = - \frac{\dddot{f} }{f} + \frac{6 \ddot{f}}{f} \left( \frac{\dot{f}}{f} - \frac{3}{2}\frac{\dot{a}}{a} \right) - \frac{9 \ddot{a} \dot{f}}{af} - \frac{6 \dot{f}^3}{f^3} - \frac{18 \dot{a}^2 \dot{f}}{a^2f} + \frac{18 \dot{a} \dot{f}^2}{af^2} +  \frac{3\dddot{a}}{a} + \frac{6\dot{a}^3}{a^3} + \frac{18 \ddot{a}\dot{a}}{a^2}\;,  
\end{equation}
which is different from the previous expressions (\ref{conM1}) etc.

\vspace{0.5cm}

\noindent {\bf{Fifth term}}: 
In the case of last term $M_{\mu\nu}^{(5)}$ we obtained
\begin{eqnarray}
M_{00}^{(5)} &=& \frac{3}{2} \left( -\frac{7\dot{a}^2}{a^2} + \frac{\dot{f}^2}{f^2} \right) - \frac{3 \dot{a} \dot{f}}{a f} - \frac{\ddot{f}}{f} - \frac{3\ddot{a}}{a}\;,\nonumber\\  
M_{ij}^{(5)} &=& - \frac{a^2 \delta_{ij}}{2f^2} \left( - \frac{3\dot{a}^2}{a^2} + \frac{6\dot{a}\dot{f}}{a f} - \frac{6\ddot a}{a} + \frac{5\dot{f}^2}{f^2}  - \frac{2\ddot{f}}{f}\right), \
\end{eqnarray}
again with extra $\ddot{a}$  and $\ddot{f}$ terms as above.
Then, (\ref{con1}) yields the following expression
\begin{equation}
\label{conM5}
\nabla^\mu M_{\mu 0}^{(5)} = - \frac{\dddot{f} }{f} + \frac{6 \ddot{f}}{f} \left( \frac{\dot{f}}{f} - \frac{\dot{a}}{2a} \right) + \frac{3 \ddot{a} \dot{f}}{af} - \frac{6 \dot{f}^3}{f^3} + \frac{6 \dot{a}^2 \dot{f}}{a^2f} + \frac{6 \dot{a} \dot{f}^2}{af^2} - \frac{3\dddot{a}}{a} - \frac{6\dot{a}^3}{a^3} - \frac{18\ddot{a}\dot{a}}{a^2}. 
\end{equation}

\section{Field equations in the modified FLRW spacetime} 
\label{EinsteinApp}

\subsection{Modified Einstein's equations }

Let us write down  Einstein's equations in the presence of diffeomorphism-breaking terms (\ref{EE51}). 
Treating $00$th and $ij$th components separately, field equations (\ref{EE51}) become 
\begin{equation}
\label{00gen}
\begin{split}
& 3 \frac{\dot{a}^2}{a^2} + \frac{3}{2}(\alpha_1 + 3 \alpha_2 - 3 \alpha_3 +\alpha_4 - 7 \alpha_5)   \frac{\dot{a}^2}{a^2} + 3 (\alpha_4 - \alpha_5) \frac{\ddot{a}}{a}  \\ & + (\alpha_1 + \alpha_2 +\alpha_3 +\alpha_4 +\alpha_5) \left( \frac{3}{2} \frac{\dot{f}^2}{f^2} - 3 \frac{\dot{a}\dot{f}}{af} - \frac{\ddot{f}}{f} \right) = 8 \pi G T_{00} 
\end{split}
\end{equation}
\begin{equation}
\label{ijgen}
\begin{split}
&\delta_{ij} \left(-1 - \frac{\alpha_1}{2} - \frac{3}{2}\alpha_2 + \frac{3}{2}\alpha_3 + \frac{3}{2}\alpha_4 + \frac{3}{2}\alpha_5 \right) \frac{\dot{a}^2}{f^2} + \delta_{ij} \left( 2 + \alpha_1 + 3\alpha_2 - 3\alpha_3 -3 \alpha_4 - 3 \alpha_5 \right) \frac{a\dot{a} \dot{f}}{f^3} \\
& +\delta_{ij} \left(-2 -\alpha_1 -3\alpha_2 + 3\alpha_3 +3 \alpha_4 + 3\alpha_5  \right) \frac{a\ddot{a}}{f^2} + \delta_{ij} (-\alpha_1 - \alpha_2 -\alpha_3 +3\alpha_4 -5\alpha_5) \frac{a^2 \dot{f}^2}{2 f^4} \\
& +\delta_{ij}(-\alpha_4 + \alpha_5) \frac{a^2 \ddot{f}}{f^3}   = 8 \pi G T_{ij}
\end{split}
\end{equation}
where we have used explicit form of $M_{\mu\nu}^{(\alpha)}$ presented in Appendix \ref{expMapp}. Together with the full consistency equation (\ref{congenx}), field equations  (\ref{00gen}) and (\ref{ijgen}) provide the most generic form of our diffeomorphism-breaking theory.

However, their complicated-looking form leads us to seek simplifications. Indeed, a particular case with an additional condition  
$\alpha_4 =\alpha_5$ was employed in \cite{Anber:2009qp}. Adopting this condition simplifies (\ref{00gen}) and (\ref{ijgen}) and yields our main equations (\ref{fieldeqsgeneral}).  

On the other hand, in \cite{Bello-Morales:2023btf} the authors only considered the effect of the fifth term, namely  $M_{\mu \nu}^{(5)}$. In order to obtain their equations, it is enough to switch off all $\alpha$ parameters except $\alpha_5$ in (\ref{00gen}) and (\ref{ijgen}). We emphasize that their theory can not be recovered from (\ref{fieldeqsgeneral}) because of the cancellation due to $\alpha_4 = \alpha_5$.

\subsection{Road to exact solutions: perturbative computations\label{perturbative}}
Adopting the cosmological time, the equations of motion become (\ref{EEc1}) and (\ref{EEc2}) supplemented with the constraint equation (\ref{cec}). Firstly, we would like to try perturbative solutions for them. Then, we will build our exact solution ansatzes on those perturbative results.   

Let us assume the following expansion of the metric components
\begin{subequations}
\label{perslns}
\begin{align}
a (\tau) &= a_0 + \epsilon a_1 + \epsilon^2 a_2+\cdots , \\
f (\tau )& = f_0 + \epsilon f_1 + \epsilon^2 f_2 + \cdots ,
\end{align}
\end{subequations}
where the couplings are rescaled as $\alpha_a \to \epsilon\alpha_a$ and $\epsilon << 1$. Using (\ref{perslns}), order by order we seek for perturbative solutions to each case.  Since the coupling constants are at the order $\mathcal{O} (\epsilon)$, the constant $K$ is first order but $L$ contains both zeroth and first order terms. With the change $ \tau\equiv \sqrt{\frac{8\pi G \rho_0}{3}}\; \tilde{t}$, the zeroth order equations  turn out to be
\begin{subequations}
\begin{align}
\label{EEc1zero}
\frac{(a_0')^2}{a_0^2} &= a_0^{-3(1+\omega)}, \\
 \frac{a'^2}{a^2} + 2 \frac{a''}{a}   &= -3 \omega\;  a_0^{-3(1+\omega_n)}.
\label{EEc2zero}
\end{align}
\end{subequations}
Below, we attempt to solve them for single component universes to get insight for our exact solutions. 

\vspace{0.5cm}

\noindent {\bf{Cosmological constant $\omega = -1$}}:  In that situation, first equation (\ref{EEc1zero}) can be solved via $a_0 (\tau) = e^{\pm \tau}$. However, the second equation does not provide any additional information. Therefore, in order to obtain $f_0(\tau)$, we need to expand either (\ref{EEc1zero}) or (\ref{EEc2zero}) to the next order and collect all the terms at $\mathcal{O}(\epsilon)$. After tedious but straightforward calculations, we observe that $f_1(\tau)$ terms drop out and we arrive a differential equation for $f_0(\tau)$ as
\begin{equation}
\frac{1}{3} \frac{f_0'''}{f_0} - \frac{2}{3} \frac{f_0' f_0''}{f_0^2} + \frac{1}{3} \frac{f_0'^3}{f_0^3} + 2 \frac{f_0''}{f_0} - \frac{f_0'^2}{f_0^2} + 3 \frac{f_0'}{f_0}= 0,
\end{equation}
where we have already used $a_0 = e^\tau$. The above equation is solved via $f_0(\tau) = C e^{-3\tau}$. Therefore, we have the following pair as the zeroth-order solutions.
\begin{equation}
\label{vacuumperturbative}
f_0(\tau) = C e^{-3\tau}\;, \qquad a_0(\tau) = e^\tau\;,
\end{equation}
where $C$ is an integration constant, cf. (\ref{ansatzcc}). In order to obtain $a_1(\tau)$ and $f_1(\tau)$ in (\ref{perslns}), we need to go to higher orders which will be quite complicated. However, even with the present solutions, we have enough physical insight for the exact solutions of the modified theory.  

\vspace{0.5cm}

\noindent {\bf{Radiation dominated universe $\omega = 1/3$}}: Substituting $\omega = \frac{1}{3}$ into (\ref{EEc1zero}), we again solve for $a_0$ as $a_0(\tau) = (2\tau)^{1/2}$. Going to the next step in perturbation, we obtain two solutions for $f_0(\tau)$ as
\begin{equation}
\label{radiationperturbative}
f_0 =A\;,   \qquad    a_0(\tau) = (2 A \tau)^{1/2},
\end{equation} 
where $A$ is a constant. The above set of solutions yield our ansatz (\ref{ansatzrdu}) with solution (\ref{radsolution}). 

\vspace{0.5cm}

\noindent {\bf{Matter dominated universe $\omega = 0$}}: Following the same steps outlined above with $\omega=0$, we obtained the following set of solutions at the zeroth order.
\begin{equation}
\label{matterperturbative}
f_0 = A\;,  \qquad a_0(\tau) = (3A/2  \tau)^{2/3}\;,
\end{equation}
where $A$ is again an integration constant. 
\\

\raggedright  
\bibliography{References_diff}{}
\bibliographystyle{JHEP}

\end{document}